\documentclass[twocolumn,secnumarabic,amssymb, nobibnotes, aps, prd]{revtex4-2}

\usepackage[dvipsnames]{xcolor}
\usepackage{amsmath}
\usepackage{amssymb}
\usepackage{graphicx}
\usepackage{comment}
\usepackage{bm}
\usepackage{braket}
\usepackage{ulem} 
\usepackage[caption=false]{subfig}
\usepackage{ragged2e}
\usepackage[colorlinks=true,bookmarks=false,citecolor=blue,linkcolor=red,hyperfootnotes=true,urlcolor=blue]{hyperref}
\usepackage{lipsum}
\DeclareCaptionJustification{justified}{\justifying}

\newcommand{\be}{\begin{equation}}
\newcommand{\ee}{\end{equation}}
\newcommand{\bea}{\begin{eqnarray}}
\newcommand{\eea}{\end{eqnarray}}
\newcommand{\Ht}{{\hat {\bf T}}}
\newcommand{\kts}[1]{\ket{t_{l,#1}}}
\newcommand{\bts}[1]{\bra{t_{l,#1}}}
\newcommand{\kTs}[1]{\ket{T_{l,#1}}}

\newcommand{\HT}[1]{\hat{\bf T}_{l,#1}}
\newcommand{\Hll}{\hat{\bf H}_{11}}
\newcommand{\Hhh}{\hat{\bf H}_{22}}
\newcommand{\Hhl}{\hat{\bf H}_{21}}
\newcommand{\Hlh}{\hat{\bf H}_{12}}
\newcommand{\Vhh}{\hat{\bf V}_{22}}
\newcommand{\Vhl}{\hat{\bf V}_{21}}
\newcommand{\Vlh}{\hat{\bf V}_{12}}
\newcommand{\V}{\hat{\bf V}}

\begin{document}

\title{Krylov Spaces for Truncated Spectrum Methodologies}%

\author{M\'arton K. L\'ajer}
\author{Robert  M. Konik}
\affiliation{ Division of  Condensed Matter Physics and Material Science,
Brookhaven National Laboratory, Upton, NY 11973-5000, USA}

%\email[REVTeX Support: ]{revtex@aps.org}

\begin{abstract}

We propose herein an extension of truncated spectrum methodologies (TSMs), a non-perturbative numerical approach able to elucidate the low energy properties of quantum field theories.  TSMs, in their various flavors, involve a division of a computational Hilbert space, $\mathcal{H}$, into two parts, one part, $\mathcal{H}_1$ that is `kept' for the numerical computations, and one part, $\mathcal{H}_2$, that is discarded or `truncated'.  Even though $\mathcal{H}_2$ is discarded, TSMs will often try to incorporate the effects of $\mathcal{H}_2$ in some effective way.  In these terms, we propose to keep the dimension of $\mathcal{H}_1$ small.  We pair this choice of $\mathcal{H}_1$ with a Krylov subspace iterative approach able to take into account the effects of $\mathcal{H}_2$.  This iterative approach can be taken to arbitrarily high order and so offers the ability to compute quantities to arbitrary precision.  
In many cases it also offers the advantage of not needing an explicit UV cutoff.
To compute the matrix elements that arise in the Krylov iterations, we employ a Feynman diagrammatic representation that is then evaluated with Monte Carlo techniques.  Each order of the Krylov iteration is variational and is guaranteed to improve upon the previous iteration.  The first Krylov iteration is akin to the NLO approach of Elias-Miró et al. \cite{Elias-Miro:2017tup}.
To demonstrate this approach, we focus on the $1+1d$ dimensional $\phi^4$ model and compute the bulk energy and mass gaps in both the $\mathbb{Z}_2$-broken and unbroken sectors.  We estimate the critical $\phi^4$ coupling in the broken phase to be $g_c=0.2645\pm0.002$.

\end{abstract}

\maketitle

\section{Introduction}
The non-perturbative study of quantum field theories (QFTs) is one of the great challenges of contemporary physics.
There are several approaches available to study nonperturbative aspects of QFTs, each having their own successes and limitations. Examples include lattice Monte Carlo \cite{Wilson:1974sk,Durr:2008zz}, the functional renormalization group \cite{Wetterich:1992yh,Morris:1993qb}, equal-time \cite{yurov1990truncated,James:2017cpc} and light cone Hamiltonian truncation methodologies \cite{Katz:2016hxp,anand2020introduction,https://doi.org/10.48550/arxiv.2209.14306},  tensor networks (matrix product states) in both their discrete form \cite{Milsted:2013rxa,Banuls:2013jaa,Kadoh:2018tis} and their continuous incarnation \cite{Verstraete:2010ft,Tilloy:2021yre,Tilloy:2021hhb,https://doi.org/10.48550/arxiv.2209.05341}, large-N expansions \cite{tHooft:1973alw}, Borel resummations and resurgence \cite{Serone:2018gjo,Serone:2019szm,Sberveglieri:2020eko,Abbott:2020qnl,Bajnok:2021dri}, the conformal bootstrap \cite{Belavin:1984vu, Rattazzi:2008pe, El-Showk:2014dwa} and integrability \cite{Karowski:1977th, Zamolodchikov:1978xm, Beisert:2010jr}.
The above methods have achieved spectacular successes, including the reproduction of the hadronic spectrum in quantum chromodynamics with lattice Monte Carlo \cite{Durr:2008zz}, the precise values for the critical exponents in the $3D$ Ising Conformal Field Theory from the conformal bootstrap \cite{El-Showk:2014dwa}, and an efficient algorithm for computing the anomalous dimensions of general operators in planar $\mathcal{N}=4$ Super Yang-Mills theory from integrability.

Nonetheless a large number of questions remain open, and difficulties remain to solve.  
To give a few examples, it is generally difficult to compute scattering amplitudes and other matrix elements in a fully non-perturbative way, especially when the asymptotic states involve bound states or topological excitations. This holds even for a simple model like the $1+1d$ $\phi^4$ theory. The control of such matrix elements is potentially important as baryons (and nuclei) appear as solitons in the low-energy chiral effective (e.g. Skyrme) models used in nuclear physics \cite{Skyrme:1961vq,Weinberg:1978kz,Witten:1983tx}. Progress is possible with Hamiltonian methods, where the L\"uscher formula provides direct access to elastic $2\rightarrow 2$ scattering below the inelastic threshold, while spectral sums of form factors can be used to obtain scattering information above threshold \cite{Henning:2022xlj}.
 
It is also hard to analyze systems out of thermodynamic equilibrium. Circumstances in which this occurs include relativistic heavy ion collisions, the nucleosynthesis of the early universe \cite{Berges:2015kfa}, as well as experiments involving ultracold quantum gases \cite{Kinoshita2006,Hofferberth2007}. A simple example when the real time dynamics of a 1+1d dimensional QFT can be studied directly with Hamiltonian methods is the calculation of overlaps and expectation values after a sudden quench\cite{Rakovszky:2016ugs,Horvath:2017wzf,Hodsagi:2018sul,Hodsagi:2019rcs,PhysRevLett.122.130603,Hodsagi:2020dqq,Horvath:2021vlx,Szasz-Schagrin:2021lal,Szasz-Schagrin:2022wkk}.

Finally, it is challenging to study systems at finite particle density and zero temperature. For example, the phase diagram of high-$T_c$ superconductor cuprates are often enriched by a quantum critical point at $T=0$, around which the long wavelength excitations are described by the spin-fermion model: an effective 2+1D field theory involving scalar bosons interacting with non-relativistic fermions in the presence of a Fermi sea. 
The study of such models by path integral methods is hindered by the so-called sign problem, while conventional Hamiltonian approaches face difficulties due to the enormous dimensionality of the truncated Hilbert space needed to reach sensible accuracy.

It is anticipated \cite{Jordan:2011ci,Jordan:2012xnu} that these difficulties will be overcome with the eventual development of quantum computers. However, despite rapid progress in this direction, a nontrivial simulation of any quantum field theory on a physical quantum computer is still someways in the future. Development of new methods serves a double purpose: on the one hand, they may break new paths towards the solution of the above challenges. On the other hand, they are interesting in the pursuit of finding the optimal classical algorithms to benchmark the evolving quantum computers on simpler toy models.

Of the methods listed above, truncated spectrum methodologies (TSMs) are perhaps the least explored, particularly for QFTs in two or more spatial dimensions.  
TSMs are a Hamiltonian approach that is essentially the Rayleigh-Ritz variational principle applied to a QFT. Two distinct variants of TSMs exist, one where the standard equal time quantization is employed \cite{yurov1990truncated,James:2017cpc}, and where lightcone quantization is employed \cite{Katz:2016hxp,anand2020introduction}.
We focus here on the first. TSM is a numerical approach well suited to study the low-energy spectrum as well as vacuum expectation values and matrix elements of operators. As such, it aims to be an alternative to lattice Monte Carlo, especially in situations where the latter is plagued by the sign problem. 

The method was originally proposed by Yurov and Zamolodchikov in \cite{yurov1990truncated} to study the relevant perturbation of the non-unitary Lee-Yang conformal minimal model. TSMs were soon extended to relevant perturbations of other conformal minimal models \cite{yurov1991truncated,Lassig:1990xy,Lassig:1990wc}, deformations of the $c=1$ boson \cite{Feverati:1998va,Feverati:1998dt}, and eventually to perturbations of a non-compact boson \cite{Coser:2014lla,Rychkov:2014eea,Rychkov:2015vap,Bajnok:2015bgw}.

TSMs have been used extensively in the context of $1+1$ dimensional field theories, where they are commonly employed to calculate the bound state spectrum \cite{Lassig:1990xy}, matrix elements \cite{Pozsgay:2007kn,Pozsgay:2007gx}, two-point correlation functions \cite{Kukuljan:2018whw}, elastic S-matrix phases \cite{Bajnok:2015bgw,sinhgordon} and even inelastic information \cite{Gabai2019}, entanglement entropy \cite{Palmai:2016act,Lencses:2018paa,Murciano:2021dga,Murciano:2021huy} and more. Recently, they have even been extended to study QFTs on an anti-de Sitter background \cite{Hogervorst:2021spa}.
Due to their finite volume formulation, equal-time TSMs provide a convenient way to compute the finite volume corrections to various observables \cite{Klassen:1990ub,Pozsgay:2007gx,Pozsgay:2007kn,Pozsgay:2014gza,Bajnok:2019cdf}.
Recently, TSMs have also been extended to $2+1d$ models in both the lightcone \cite{Anand:2020qnp,https://doi.org/10.48550/arxiv.2209.14306} and the equal-time \cite{Elias-Miro:2020qwz} framework.

In particular, TSMs are well-suited to study the spectrum and matrix elements of bound states and topological excitations. For example, the kink-antikink elastic scattering phase can be obtained in the the strongly coupled regime of the broken phase $\phi^4$ model \cite{Bajnok:2015bgw}.
TSMs are also an ideal tool to study QFTs subject to sudden (or gradual) changes of the couplings in the Hamiltonian or the environment, called quenches \cite{Rakovszky:2016ugs,Horvath:2017wzf,Hodsagi:2020dqq}. Real-time correlators in the presence of these quenches are available to be computed with TSMs \cite{Kukuljan:2018whw}. Periodic driving and Floquet dynamics was analyized in \cite{Bajnok:2021twm}. False vacuum decay was studied in \cite{Szasz-Schagrin:2022wkk}.

However, it is generally difficult to implement Hamiltonian truncations for systems featuring more than one space dimensions, multiple fields and/or gauge constraints.
In this paper, we present an incarnation of Hamiltonian truncation that aims to improve on the majority of the aforementioned issues.

TSMs are typically applied to a problem where the Hamiltonian comes in two parts:
\begin{equation}
     \hat{\bf H} = \hat{\bf H}_0 + \hat{\bf  V}.
\end{equation}
Here $\hat{\bf{H}}_0$ is some Hamiltonian over which one has complete control (i.e. $\hat{\bf H}_0$ is that of a conformal field theory with known structure constants or of a free boson or fermion).  The space of eigenstates of $\hat{\bf H}_0$ often provides the computational basis under which TSMs play out.  The second term of the Hamiltonian, ${\hat{\bf V}}$, is some interaction whose effects one is trying to understand.  
To begin to apply the TSM to understanding this problem, one first makes a division of the computational Hilbert space, $\mathcal{H}$, into two parts, 
\begin{equation}
    \mathcal{H} = \mathcal{H}_1\oplus \mathcal{H}_2.
\end{equation}
In the simplest (and original) form of TSMs, this division is predicated upon energy as determined by $\hat{\bf H}_0$.  In this particular case, $\mathcal{H}_1$ then consists of a finite basis of emphasized states and the effects of the (infinite dimensional) $\mathcal{H}_2$ are ignored entirely.  For example in the study of the relevant perturbation of the Lee-Yang conformal minimal model \cite{yurov1990truncated}, the computational Hilbert space,  $\mathcal{H}$ consists of the Lee-Yang CFT space of states, $\mathcal{H}_1$ consists of conformal states with energy (as defined by the conformal Hamiltonian) less than some cutoff, and $\mathcal{H}_2$ all states above this cutoff.  In this way the TSM reduces the solution of a quantum field theory to an exact diagonalization.  More sophisticated versions of the TSM build on this crude (but surprisingly often accurate) approximation by computing how the discarded part of the Hilbert space, $\mathcal{H}_2$, affects the physics.  Taking into account the effects of $\mathcal{H}_2$ can be done purely numerically in the spirit of a numerical renormalization group \cite{nrg} or analytically where the effects of integrating out $\mathcal{H}_2$ leads to a new effective Hamiltonian defined on $\mathcal{H}_1$ \cite{feverati2006renormalisation,giokas2011renormalisation, Lencses:2014tba, Hogervorst:2014rta,Elias-Miro:2017tup}. It is at this stage, in how to deal with $\mathcal{H}_2$, that we offer a novel approach.  

In most applications of TSMs, the dimension of $\mathcal{H}_1$ is on the order of $10^3$ to $10^7$.  Here we instead explore keeping the dimension of $\mathcal{H}_1$ small, less than $100$.  Keeping ${\rm dim}\mathcal{H}_1$ small mandates that we account for the effects of $\mathcal{H}_2$.  Here we propose to do so by using a Krylov subspace iterative approach.  In principle, this iterative approach can be taken to arbitrarily high order and so offers the ability to compute quantities to arbitrary precision. 
While ${\rm dim}\mathcal{H}_2=\infty$, at each Krylov order we do not need to introduce an explicit UV cutoff on the Hilbert space.  We are able to avoid doing so because we compute the matrix elements that arise at a given Krylov iteration by employing a Feynman diagrammatic representation that is then evaluated with Monte Carlo techniques. In general these diagrams can be regularized with conventional UV cutoffs from the perturbative literature. In particular, in the $1+1d$ $\phi^4$ model they are UV finite after normal ordering and do not require the introduction of a cutoff to be evaluated.
Importantly each Krylov iteration is variational and is guaranteed to improve upon the previous iteration.  The first Krylov iteration is akin to the NLO approach of Elias-Miró et al. \cite{Elias-Miro:2017tup}.

The paper is organized as follows.  In Section II we begin with an overview of the truncated spectrum methodology and its bipartition of the computational Hilbert space.  We show how the notion of a 'tail' state, first introduced in a TSM context by Ref. \cite{Elias-Miro:2017tup}, arises from this partition.  We then show how the tail states can be computed through an iterative, continued fraction approach.  This approach involves writing the tail states in terms of a Krylov subspace formed by the action of products of a resolvant operator $\bf{\hat T}$. 
We give a variational estimate to each iteration of the continued fraction, which is in turn
a variational estimate of the ground state energy of each symmetry sector, i.e., guaranteed to form an upper bound on the exact energy, and that each iteration of the continued fraction improves upon that before.    The discussion in Section II is model agnostic.  

In Section III we then turn to applying it to the $\phi^4$ model in $1+1d$.  There we provide a brief review of the properties of the model in both its unbroken and broken phase.  We then turn to defining the computational space of states and the division of the Hilbert space, $\mathcal{H}$ into two parts, $\mathcal{H} = \mathcal{H}_1\oplus \mathcal{H}_2$.  
We discuss in particular a division termed ``Zero-mode separation", where $\mathcal{H}_1$ consists of the zero-mode eigenstates that arise from a $\phi^4$ Hamiltonian shorn of its oscillator modes.

Having established the computational basis that we will use, we turn to numerical details on how the TSM Hamiltonian is evaluated in the Krylov subspace.  The first detail that we discuss is how matrix elements of the Hamilitonian are computed.  Here we show that these matrix elements can be written in terms of standard $\phi^4$ Feynman diagrams and that these diagrams in turn can be evaluated through Monte Carlo.
Formally, the exact ground state energy $E_*=E$ appears explicitly in the Krylov Hamiltonian. Thus in principle finding the eigenenergies of this Hamiltonian must be done in a self-consistent manner. However we will argue that $E_*$ is essentially a parameter of the Krylov basis and can be traded for another set of states with no explicit dependence on $E_*$.  This will allow us to choose a basis arising as a natural tensor product of the zero-mode states and the states with oscillators.

Having finished with how we will implement the method in the context of $\phi^4$, we then turn to results.  In Section IV we discuss results for the broken phase of the model.  There we present results for the kink mass and the value of the critical coupling.  In Section V we present results on $\phi^4$'s disordered phase, both the ground state energy and the mass of the first excited state.  In Section VI we demonstrate that our method likely provides a better estimate of the low-lying eigenvectors in the theory (as opposed to eigenenergies) than traditional truncated spectrum methods.

In the final section, Section VII, we turn to the discussion of the method.  We consider two points.  In the first, we can in principle compute the method at arbitrary high Krylov iterative order.  However in practice, we are limited by the ability to perform the numerical evaluation of n-point functions of the $\phi^4$ interaction.  In this paper the highest Krylov order that we go to is 3.  While we have some understanding of how to extrapolate in Krylov order, this issue remains in general open.  In the second, we discuss the prospects of applying the method to higher dimensional field theories.  We believe that the method, because it employs Feynman diagrammatic representations, has a good chance of being generalized to at least 
linear sigma models in $2+1d$.  We consider this to be an exciting possibility.  While applications of TSMs to $2+1d$ QFTs have only recently appeared, TSMs have been applied to 1+1d QFTs for over three decades.

\section{Truncated Spectrum Methods}
\subsection{Overview}
    Quantum field theory (QFT) is essentially quantum mechanics (QM) combined with Lorentz invariance.
More precisely, a QFT can be imagined as a limit of of a QM system as the number of
degrees of freedom is taken to infinity.
An immediate consequence of this is that numerical methods developed for quantum
mechanics are potentially applicable to QFT problems as well.

One such method is the Rayleigh-Ritz-L\"owdin approach coming from quantum chemistry \cite{lowdin,Klahn1977,Klahn1977a}.  In the context of quantum field theories with infinite dimensional Hilbert spaces, this approach often comes with the moniker `Truncated Spectrum Methods' or in its more narrow application to perturbed conformal field theories in $1+1d$, `Truncated Conformal Spectrum Approaches' \cite{yurov1990truncated,yurov1991truncated}. Let us write our infinite dimensional Hilbert space in which our quantum field theory operates as
\begin{equation}
    {\cal H} = \{\ket{c_i}\}^\infty_{i=1},
\end{equation}
where we suppose the $\ket{c_i}$ are orthonormal.
Now let $\hat{\bf H}$ be a Hamiltonian defined on this space.  Because ${\cal H}$ is infinite dimensional, it is not possible numerically to find the exact eigenstates $\{\ket{\psi_i}\}^\infty_{i=1}$ and their corresponding eigenenergies $E_i$:
\be
\hat {\bf H}\left|\psi_{i}\right\rangle=E_{i}\left|\psi_{i}\right\rangle.
\ee
To circumvent this problem, one can divide the Hilbert space into two parts:
\begin{equation}\label{hilbert_divide}
    {\cal H} = {\cal H}_1\oplus {\cal H}_2.
\end{equation}
Here ${\cal H}_1$ is a finite dimensional subspace and has been chosen to contain, in some sense, the most `important' states in the basis $\{\ket{c_i}\}^\infty_{i=1}$ for the problem.  We will return to this notion of `importance' later in the paper.

With this division, the Schr\"odinger equation for our Hamiltonian becomes, in block form:
\begin{equation}
    \left(
    \begin{array}{cc}
        \Hll & \Hlh \\
        \Hhl & \Hhh
    \end{array}
    \right)
    \left(
    \begin{array}{c}
    c_{1} \\
    c_{2}
    \end{array}
    \right) = E
    \left(
    \begin{array}{c}
    c_{1} \\
    c_{2}
    \end{array}
    \right)\label{HmatDSum}.
\end{equation}
Here the vector $(c_{1}~~c_{2})$ should be understood, schematically, as a vector in the bipartite Hilbert space:
\begin{equation}
    (c_{1}~~ c_{2}) = \sum_{i\in{\cal H}_1}\alpha_i \ket{c_i} +\sum_{i\in{\cal H}_2}\alpha_i \ket{c_i}.
\end{equation}
The simplest approximation that we can make is to suppose that $\Hlh$ and $\Hhl$ are small and thus can be ignored.  We then can solve the restricted, numerically finite problem, of finding the eigenvalues of $\Hll$:
\begin{equation}
    \Hll|\psi^{trunc}_{1i}\rangle = E^{trunc}_i\ket{\psi^{trunc}_{1i}}.
\end{equation}
Often this approximation is excellent.  This can be seen in problems where the Hamiltonian comes in two parts:
\begin{equation}\label{Ham_HV}
    \hat {\bf H} = \hat{\bf H_0} + \hat{\bf V}.
\end{equation}
Here $\hat {\bf H}_0$ is, typically, diagonal in terms of the basis $\{\ket{c_i}\}^\infty_{i=1}$, i.e.
\begin{equation}
    \hat {\bf H}_0\ket{c_i} = E_{0i}\ket{c_i},
\end{equation}
while $\hat {\bf V}$ is an operator that mixes this basis.  With the Hamiltonian in this form, it is natural to make the division between ${\cal H}_1$ and ${\cal  H}_2$ on the basis of the states $\ket{c_i}$ as determined by $\hat {\bf H}_0$: ${\cal H}_1$ is composed of a finite set of low energy states:
\begin{equation}
    {\cal H}_1 = {\rm span}\{\ket{c_i},E_{0i}<E_{trunc}\},
\end{equation}
while ${\cal H}_2$ consists of the complement: ${\cal H}_2=\cal H\setminus {\cal H}_1$.
This form of the Hamiltonian was present in two of the first uses of truncated spectrum methodologies to study properties of perturbed conformal field theories, \cite{yurov1990truncated,yurov1991truncated}.  Here the dimension of ${\cal H}_1$ was kept to be very small, a few tens, yet the answers generated were accurate to several significant figures.  Here the reason why the methodology worked so well was because the operator $\hat {\bf V}$ was highly relevant, i.e. $\Vlh$ was in some sense small and only mixed the two parts of the Hilbert space weakly.  But it will often be the case that one wants to understand theories where $\Vlh$ is not small.

In cases where it is not possible to ignore the mixing of ${\cal H}_1$ and ${\cal H}_2$ induced by $\Hlh/\Hhl$, we need a strategy to incorporate the effects of this mixing.
Such strategies can be numerical or they can be analytical.  Numerically, it is possible to adapt a Wilsonian numerical renormalization group to the problem \cite{nrg}. Analytically, a common way to improve precision is to perform perturbation theory to leading order in $\V$:
\begin{equation}
    \delta E^{trunc}_i = \bra{\psi^{trunc}_{1i}}\Vlh\frac{1}{E^{trunc}_{i}-H_{0,22}}\Vhl\ket{\psi^{trunc}_{1i}}.
\end{equation}
This is of course not standard perturbation theory in $\V$.  This operator's effects in mixing states in ${\cal H}_1$ are being accounted for exactly.  What is instead being treated perturbatively in $\V$ is the mixing between ${\cal H}_1$ and ${\cal H}_2$.

While computing this correction can sometimes improve the results, often it does not. In response, one could imagine going further and computing higher corrections in the mixing between the two parts of the Hilbert space.  However, it can be that the perturbative series in $\V$ is asymptotic.  This is the case in studying the $\phi^4$ perturbation of a scalar real field in $1+1d$ \cite{Elias-Miro:2017tup}.  It is also the case in applying truncated spectrum methods to the sinh-Gordon model \cite{sinhgordon}. We thus want to develop a framework where these corrections can be analyzed systematically and sense made of any asymptotic series.

\newcommand{\DlH}{\hat {\bf \Delta H}}
The first step is to recast the exact problem so that it a finite dimensional one.  The full Schr\"odinger equation reads when broken into components as
\begin{eqnarray}
    \Hll\ket{c_1} + \Hlh\ket{c_2} &=& E\ket{c_1}\cr\cr
    \Hhl\ket{c_1} + \Hhh\ket{c_2} &=& E\ket{c_2}.
\end{eqnarray}
We can eliminate $\ket{c_2}$ from this equation leaving us with a single Schr\"odinger equation for the first $N_1\equiv {\rm dim}~{\cal H}_1$ eigenvalues and eigenstates of the problem:
\begin{eqnarray}
    E\ket{c_1} &=& (\Hll + \DlH)\ket{c_1};\cr\cr
  \DlH &\equiv& \Hlh\frac{1}{E-\Hhh}\Hhl.
\end{eqnarray}
While we have recast the problem into a finite dimensional one, it is still in general intractable because we cannot compute the matrix elements of $\DlH$.

To begin to attack this issue, we take the stance of Ref. \cite{Elias-Miro:2017tup}: we recognize that if one is interested in the obtaining the exact values of the first $N_1$ eigenvalues of the problem, one does not need to deal with the infinite part of Hilbert space, ${\cal H}_2$, in its entirety.  Rather the subspace of ${\cal H}_2$ defined by 
\begin{equation}
    {\cal H}_t \equiv \rm span\{\frac{1}{E-\Hhh}\Hhl\ket{c_i}\}^{N_1}_{i=1} \subset {\cal H}_2
\end{equation}
is sufficient.  Following Ref. \cite{Elias-Miro:2017tup}, we call the states in this span, `tail' states.

\newcommand{\Pt}{\hat {\bf P}_t}
Considering the problem in the expanded (but still finite dimensional) space ${\cal H}_1 \bigcup {\cal H}_t$, we can write the Hamiltonian in the form
\begin{eqnarray}
    \left(
    \begin{array}{cc}
        \Hll & \hat {\bf H}_{1t} \\
        \hat {\bf H}_{t1} & \hat {\bf H}_{tt}
    \end{array}
    \right)
    \left(
    \begin{array}{c}
    c_1 \\
    c_t
    \end{array}
    \right) &=& E
    \hat{\mathbf{G}}
    \left(
    \begin{array}{c}
    c_1 \\
    c_t
    \end{array}
    \right)\label{HmatLO};\cr\cr
    \hat{\mathbf{G}}&=&\left(
    \begin{array}{cc}
        1 & 0  \\
        0 & \hat {\bf G}_{tt}
    \end{array}
    \right)
  \cr\cr
    \hat {\bf H}_{1t} = \Hlh \Pt; &&~~ \hat {\bf H}_{t1} = \Pt\Hhl;\cr\cr
    \hat {\bf H}_{tt} &=& \Pt\Hhh \Pt,
\end{eqnarray}
where $\Pt$ is a projector onto ${\cal H}_t$ and $\hat {\bf G}_{tt}$ is the inner product matrix of the tail states -- while the basis in ${\cal H}_1$ is orthonormal, the tail states are not so.  While this recasting of the problem does not in itself solve our problem -- not being able to compute the matrix elements of $\DlH$ amounts to be unable to write down the exact form of the tail states --  it does provide us with a basis for finding systematic, iterative approximations for the tail states.  This we do in the next section.

\subsection{Tail States as Continued Fractions}

Here we construct in an iterative fashion expressions for the tail states that have a continued fraction form.  The basis on which we develop the continued fraction will be to exploit the division of the Hamiltonian into a `free' part $\hat {\bf H}_0$ and an interacting part $\V$ as in Eqn. \eqref{Ham_HV}.  Our presentation here is based on Ref. \cite{PhysRevA.28.2151}.

The exact tail states,
\begin{equation}
    \ket{T_l} \equiv \frac{1}{E-\Hhh}\Vhl\ket{c_l}, ~~l=1,\cdots,N_1,
\end{equation}
can be rewritten implicitly as
\begin{eqnarray}
 \ket{T_l} &=& \frac{1}{E-\hat {\bf H}_{0,22}}\Vhl\ket{c_l}+\frac{1}{E-\hat {\bf H}_{0,22}}\Vhh\ket{T_l}\cr\cr
 &\equiv & \ket{t_{l,0}} + \Ht\ket{T_l}.\label{IterLinEq}
\end{eqnarray}
We can think of $\ket{t_{l,0}}$ as a zero-order in $\Vhh$ approximation to the full tail state $\ket{T_l}$ while 
$\Ht \ket{T_l}$ encapsulates the first and higher order terms of $\ket{T_l}$ in $\Vhh$.
Our goal here is to develop such terms systematically.

To do so, we want to define an operator, $\HT{1}$, that separates from $\ket{T_l}$ the contribution of $\kts{0}$, i.e.
\begin{equation}
    \HT{1}\ket{t_{l,0}} = 0.
\end{equation}
The operator that does this is
\begin{eqnarray}
\HT{1} = \Ht - \frac{\Ht \kts{0}\bts{0} \Ht}{\bts{0} \Ht \kts{0}}.
\end{eqnarray}
We can use this definition to write the exact tail state as $\ket{T_l}$ as
\begin{equation}\label{almost}
    \ket{T_l} = \kts{0} + \frac{1}{1-\HT{1}}\Ht\kts{0}\frac{\bts{0}\Ht\ket{T_l}}{\bts{0}\Ht\kts{0}}.
\end{equation}
Eq. \eqref{almost} is almost what we are looking for in terms of isolating the next correction beyond $\kts{0}$ to the tail state $\ket{T_l}$ in the form of a continued fraction. The only problem with it is that $\ket{T_l}$ appears on both sides of this equation. To evade this, we multiply eq. \eqref{almost} with $\bts{0}\Ht$ from the left and rearrange terms to yield
\begin{equation} \label{ts1}
    \ket{T_l} = \kts{0} + \frac{\bts{0}\Ht\kts{0}}{\bts{0}\Ht\kts{0}-\bts{0}\Ht\kTs{1}}\kTs{1},
\end{equation}
where we have defined $\kTs{1}$ as
\begin{eqnarray}
\kTs{1} &=& \frac{1}{1-\HT{1}}\kts{1}; \cr\cr
\kts{1} &\equiv& \Ht\kts{0}.
\end{eqnarray}
At the next order of approximation, we can write
\begin{equation}
    \kTs{1} \approx \kts{1} ~{\rm if}~\lVert\HT{1}\Ht\kts{0}\rVert \ll 1.
\end{equation}
giving us an approximate for the exact tail state $\ket{T_l}$:
\begin{equation}
       \ket{T_l} = \kts{0} + \frac{\bts{0}\Ht\kts{0}}{\bts{0}\Ht\kts{0}-\bts{0}\Ht\kts{1}}\kts{1}.
\end{equation}
We stress that the correction to $\ket{T_l}$ beyond $\kts{0}$ is not merely one higher order in $\V$ but is some resummation of terms in powers of $\V$.  Unlike a naive perturbation theory, this expansion is well-behaved, i.e. is not asymptotic.  

We can continue this expansion by repeating the process with $\kTs{1}$.  We define an operator $\HT{2}$ that projects from $\kTs{1}$ the contribution of $\kts{1}$:
\begin{eqnarray}
\HT{2} &=& \HT{1} -\frac{\HT{1}\kts{1}\bts{1}\HT{1}}{\bts{1}\HT{1}\kts{1}};\cr\cr
\HT{2}\kts{1} &=& 0.
\end{eqnarray}
As before we can then write
\begin{eqnarray}
  \kTs{1} &=& \kts{1} + \frac{\bts{1}\HT{1}\kts{1}}{\bts{1}\HT{1}\kts{1}-\bts{1}\HT{1}\kTs{1}}\kTs{2},\cr\cr
  \kTs{2} &=& \frac{1}{1-\HT{2}}\kts{2},\cr\cr
  \kts{2} &=& \HT{1}\kts{1}.
\end{eqnarray}
Here we have the option of approximating $\kTs{2}$ as $\kts{2}$, valid if
$$
\lVert \HT{2}\kts{2}\rVert \ll 1.
$$
We can continue this process iteratively, defining a sequence of states $\kTs{n},\kts{n}$ and operators, $\HT{n}$ defined as
\begin{eqnarray}\label{tsn}
\HT{n} &=& \HT{n-1} -\frac{\HT{n-1}\kts{n-1}\bts{n-1}\HT{n-1}}{\bts{n-1}\HT{n-1}\kts{n-1}};\cr\cr
0 &=& \HT{n+1}\kts{n};\cr\cr
  \kTs{n} &=& \kts{n} \!+\! \frac{\bts{n}\HT{n}\kts{n}}{\bts{n}\HT{n}\kts{n}\!-\!\bts{n}\HT{n}\kTs{n}}\kTs{n+1},\cr\cr
  \kTs{n} &=& \frac{1}{1-\HT{n}}\kts{n},\cr\cr
  \kts{n} &=& \HT{n-1}\kts{n-1}.
\end{eqnarray}
This sequence describing the exact tail states can be terminated at finite order by making the approximation
\begin{equation}
    \kTs{n+1} \approx \kts{n+1}=\HT{n}\kts{n}. \label{IterTerminateGen}
\end{equation}

\newcommand{\HV}{\hat {\bf H}_V}
\newcommand{\HG}{\hat {\bf G}_V}
\newcommand{\Hlt}[1]{{\rm H}_{1t}^{(#1)}}
\newcommand{\Htl}[1]{{\rm H}_{t1}^{(#1)}}
\newcommand{\Htt}[1]{{\rm H}_{tt}^{(#1)}}
\newcommand{\G}[1]{{\rm G}^{(#1)}}

\subsection{A Variational Improvement}

The takeaway point of the above calculation is that we have recast the exact tail states $\ket{T_l}$ as a linear combination of states
\begin{equation}
    \ket{T_l} = \sum_{k=1}^{N_T}\tau_{lk} \ket{\tilde t_{lk}}
\end{equation}
with basis vectors
\begin{equation}
    \ket{\tilde t_{lk}}=\Ht^{k-1}\frac{1}{E-{\bf \hat H_{0,22}}}{\bf \hat V_{21}}\ket{c_l}.\label{tildet}
\end{equation}
In the above, we have truncated our approximation for the tail states at order $N_T$.
The iterative procedure described in the previous section gives the form of the nonperturbative coefficients $\tau_{lk}$.
However rather than compute the coefficients out to some order, we are going to treat them as variational parameters that are chosen so as to minimize the energy.

We do so by starting from a tail-extended basis given by the combination of the original low energy Hilbert space, ${\cal H}_1$, with $\ket{\tilde t_{lk}}$:
\begin{equation}\label{comp_space}
    {\cal H}_1~\bigcup~\{\ket{\tilde t_{lk}};l=1,\cdots,N_1;k=1,\cdots, N_T\}.
\end{equation}
This computational basis has dimension $N_1(N_T+1)$.  Because these basis states are not necessarily orthogonal, we set
up a generalized eigenvalue problem
in block matrix form as
\begin{widetext}
\begin{equation}\label{eigen}
\begin{pmatrix}
   {\rm H}_{11} & \Hlt{1} & \Hlt{2} &\dots & \Hlt{N_T} \\
   \Htl{1} & \Htt{11} & \Htt{12} &  & \vdots\\
   \Htl{2} & \Htt{21} & \Htt{22} & & \vdots \\
   \vdots & & & \ddots & \vdots  \\
   \Htl{N_T} & \dots & \dots & \dots & \Htt{N_TN_T}
    \end{pmatrix}
   \begin{pmatrix}
    c_1 \\[.1in]
    \tilde t_1 \\[.1in]
    \tilde t_2 \\[.1in]
    \vdots \\
    \tilde t_{N_T}
\end{pmatrix}
   = E
    \begin{pmatrix}
        1 & 0 & 0 &\dots & 0 \\[.1in]
        0 & \G{11} & \G{12} & \dots & \G{1N_T} \\[.1in]
        0 & \G{21} & \G{22} & \dots & \\[.1in]
        \vdots & & & \ddots & \\
        0 & \G{N_{T1}} & \cdots & \cdots & \G{N_TN_T}
  \end{pmatrix}
  \begin{pmatrix}
    c_l \\[.1in]
    \tilde t_1 \\[.1in]
    \tilde t_2 \\[.1in]
    \vdots \\
    \tilde t_{N_T}
  \end{pmatrix}
 \end{equation}
\end{widetext}

Recalling that the operator $\hat{\mathbf{T}}$ was defined in eq. \eqref{IterLinEq} as
\be
\hat{\mathbf{T}}=\frac{1}{E-\hat {\bf H}_{0,22}}\Vhh,
\ee
the matrix elements of the block matrices ${\rm H}_{11}$, $\Hlt{n}$, $\Htt{nm}$ are defined as
\begin{eqnarray}
{\rm H}_{11,ll'} &=& \bra{c_l}\Hll \ket{c_l};\cr\cr
{\rm H}^{(n)}_{1t,ll'} &=& \bra{c_l}\Vlh \Ht^{n-1}\frac{1}{E-{\bf H}_{0,22}}\Vhl\ket{c_l};\cr\cr
{\rm H}^{(nm)}_{tt,ll'} &=& \bra{c_l}\Vlh \frac{1}{E-{\bf H}_{0,22}}(\Ht^{n-1})^\dagger \Hhh\cr\cr
&& \hskip .4in \times \Ht^{m-1} \frac{1}{E-{\bf H}_{0,22}}\Vhl\ket{c_l};\cr\cr
&=& {\rm H}^{(n+m)}_{1t,ll'} - {\rm H}^{(n+m-1)}_{1t,ll'} + E \G{nm}_{ll'},
\label{GenHtails}\end{eqnarray}
and the non-trivial elements of the Gram matrices are defined by 
\begin{eqnarray}
\G{nm}_{ll'} &=& \bra{c_l}\Vlh \Ht^{n-1} \left(\frac{1}{E-{\bf H}_{0,22}}\right)^2(\Ht^{m-1})^\dagger \nonumber\\
&&\Vhl\ket{c_l'}.
\end{eqnarray}

\subsection{The NLO Approximation of Ref. \cite{Elias-Miro:2017tup} and Beyond}

As we have stated, Ref. \cite{Elias-Miro:2017tup} introduced the idea of working with a basis formed from ${\cal H}_1$ and the span of the tail states, ${\cal H}_t$.  However Ref. \cite{Elias-Miro:2017tup} did not attempt to construct the exact tail states of eqs. \eqref{ts1} and \eqref{tsn}. Instead they examined the leading approximation to the tail states of the form
\begin{equation}
    \ket{\tilde t_{l1}}=\frac{1}{E-H_{0,22} }H_{21}\ket{c_l} \label{NLOtail}
\end{equation}
and set up the truncated Hamiltonian as
\begin{equation}\label{1st_tail_order}
\begin{pmatrix}
   {\rm H}_{11} & \Hlt{1} \\
   \Htl{1} & \Htt{11} & \\
    \end{pmatrix}
   \begin{pmatrix}
    c_1 \\[.1in]
    \tilde t_1 \\
\end{pmatrix}
   = E
    \begin{pmatrix}
        1 & 0  \\[.1in]
        0 & \G{11} \\
  \end{pmatrix}
  \begin{pmatrix}
    c_l \\[.1in]
    \tilde t_1 \\
  \end{pmatrix}
 \end{equation}
They termed this approximation 'next-leader order' (NLO).  In this paper we will also explore variational approximations that go two orders beyond this - taking the eigenvalue problem in \eqref{eigen} with $N=2$ - we term this 2NLO.  
\begin{widetext}
\begin{equation}\label{eigen2}
    \left(
    \begin{array}{ccccc}
        {\rm H}_{11} & \Hlt{1} & \Hlt{2} \\
        \Htl{1} & \Htt{11} & \Htt{12}\\
        \Htl{2} & \Htt{21} &  \Htt{22}\\
    \end{array}
    \right)
    \left(
    \begin{array}{c}
    c_l \\[.05in]
    \tilde t_1 \\[.05in]
    \tilde t_2 \\
    \end{array}
    \right) = E
    \left(
    \begin{array}{ccccc}
        1 & 0 & 0  \\[.05in]
        0 & \G{11} & \G{12}  \\[.05in]
        0 & \G{21} & \G{22}\\
    \end{array}
    \right)
    \left(
    \begin{array}{c}
    c_l \\[.05in]
    \tilde t_1 \\[.05in]
    \tilde t_2 \\
    \end{array}
    \right)
\end{equation}
and $N=3$ (we term this 3NLO)
\begin{equation}\label{eigen3}
    \left(
    \begin{array}{ccccc}
        {\rm H}_{11} & \Hlt{1} & \Hlt{2} & \Hlt{3} \\
        \Htl{1} & \Htt{11} & \Htt{12} & \Htt{13} \\
        \Htl{2} & \Htt{12} & \Htt{22} & \Htt{23} \\
        \Htl{3} & \Htt{13} & \Htt{32} & \Htt{33} \\
    \end{array}
    \right)
    \left(
    \begin{array}{c}
    c_l \\[.05in]
    \tilde t_1 \\[.05in]
    \tilde t_2 \\[.05in]
    \tilde t_3
    \end{array}
    \right) = E
    \left(
    \begin{array}{ccccc}
        1 & 0 & 0 & 0 \\[.05in]
        0 & \G{11} & \G{12} & \G{13} \\[.05in]
        0 & \G{21} & \G{22} & \G{23} \\[.05in]
        0 & \G{31} & \G{32} & \G{33} \\
    \end{array}
    \right)
    \left(
    \begin{array}{c}
    c_l \\[.05in]
    \tilde t_1 \\[.05in]
    \tilde t_2 \\[.05in]
    \tilde t_3
    \end{array}
    \right)
\end{equation}
\end{widetext}
To this point, this section has presented a general formalism independent of particular models and particular choices of ${\cal H}_1$ and ${\cal H}_2$.   For a model with the simple structure of eq. \eqref{Ham_HV}, matrix elements of $H_{tt,ll\prime}^{(nm)}$ involve $n+m+1$ powers of the interaction: at most $3$ at NLO order, $5$ at 2NLO and $7$ at 3NLO. In the next section, we turn to these issues in the context of the $\phi^4$ scalar field theory.

\section{Implementing the Methodology on the $\phi^4$ Theory}

The discussion in Section 2 was completely general and was presented independent of any model.  It also did not specify certain crucial implementation details that would be needed in any actual application of the methodology.  In particular, it did not indicate how we intended to divide the computational Hilbert space into ${\cal H}_1$ and ${\cal H}_2$. And we did not discuss how we are going to compute the matrix elements of $\Hlt{n}$, $\Htl{n}$, and $\Htt{nm}$.  We take up these tasks in this section in the context of the $\phi^4$ theory.

\subsection{Review of $\phi^4$ Theory}

The two-dimensional $\phi^4$ theory corresponds to the normal ordered Hamiltonian
\bea
\hat{\mathbf{H}}&=&\hat{\mathbf{H}}_0+\hat{\mathbf{V}},\nonumber\\
\hat{\mathbf{H}}_0&=&\intop dx :\frac{\hat{\boldsymbol{\pi}}(x)^2}{2}+\frac{(\partial_x \hat{\boldsymbol{\phi}})^2}{2}+\frac{m_0^2}{2} \hat{\boldsymbol{\phi}}(x)^2:_{m_0,\infty},\cr\cr
\hat{\mathbf{V}}&=&\intop dx :\left(g_2-\frac{m_0^2}{2}\right)\hat{\boldsymbol{\phi}}(x)^2+g_4 \hat{\boldsymbol{\phi}}(x)^4:_{m_0,\infty} \label{Vdef}
\eea
where $g_4>0$ and the normal ordering is understood with respect to the mass scale $m_0$ and at infinite volume. 
The Hamiltonian is invariant against the global $Z_2$ transformation $\hat{\boldsymbol{\phi}}(x)\rightarrow-\hat{\boldsymbol{\phi}}(x)$ in addition to Poincar\'e invariance and spatial parity symmetry.
Classically, separating the quadratic term into two pieces is of course redundant. In the quantum theory, different choices for $\hat{\mathbf{H}}_0$ are connected by finite a renormalization of the couplings.

In finite volume, $L$, Lorentz symmetry is lost but translation invariance persists. The periodic finite volume mode expansion of the field $\hat{\boldsymbol{\phi}}$ takes the form
\be
\hat{\boldsymbol{\phi}}(\tau,x)=\sum_{n=-\infty}^{\infty} \frac{1}{\sqrt{2L\omega_n}}\left(\hat{\mathbf{a}}_n e^{i k_n x-\omega_n \tau}+\hat{\mathbf{a}}_n^\dagger e^{-i k_n x+\omega_n \tau}\right),
\ee
with wavenumber $k_n=2\pi n L^{-1}$ and frequency $\omega_n=\sqrt{m_0^2+k_n^2}$, $n\in\mathbb{Z}$.
The finite volume oscillators are subject to the usual Fock commutation relations,
\be
\left[\hat{\mathbf{a}}_{n},\hat{\mathbf{a}}_{m}^{\dagger}\right] = \delta_{nm}.
\ee
The Hilbert space in which $\hat{\mathbf{H}}$ acts can be built up from the vacuum $\ket{0}$ using the creation operators $\hat{\mathbf{a}}_n^\dagger$. A general Fock basis vector takes the form
\be
\ket{\psi_{\dots,n_{-1},n_0,n_1,\dots}}=\prod_{k=-\infty}^{\infty} \left(\hat{\mathbf{a}}_k^{\dagger}\right)^{n_k}\ket{0}.
\ee
In practice it is more feasible to work with operators normal ordered at finite volume $L$. Strictly speaking this yields a different regularization scheme. The schemes can be connected by applying finite, volume-dependent corrections to the coefficients $g_2$ and $g_4$. A derivation for these corrections is provided in Appendix \ref{SubsecNormOrdSchemes}.
The explicit form of the Hamiltonian, expressed with finite volume normal ordering,
reads
\bea
\hat{\mathbf{H}}&=&\intop dx :\frac{\hat{\boldsymbol{\pi}}^2}{2}+\frac{(\partial_x \hat{\boldsymbol{\phi}})^2}{2}+\hat{g}_2 \hat{\boldsymbol{\phi}}(x)^2+g_4 \hat{\boldsymbol{\phi}}(x)^4:_{m_0,L}\nonumber\\
&+&E_0(m_0,L)\label{FiniteLH}
\eea
where we introduced
\bea
g'_2&=&g_2+6g_4 z(m_0L),\cr\cr
E_0(m_0,L)&=&m_0 e_0(m_0L)+L\left(g_2-\frac{m_0^2}{2}\right)z(m_0L) \nonumber \\
& & +3Lg_4 z(m_0L)^2. \label{E0def}
\eea
The functions $z(x)$ and $e_0(x)$ are defined in eqs. \eqref{zfuncdef} and \eqref{e0funcdef}, respectively.

The model possesses two phases. Classically, for $g_2>0$, the ground state is invariant
with respect to the parity $Z_2$ symmetry, while for $g_2<0$ the symmetry is
spontaneously broken.
In the quantum model, the situation is more complicated due to the presence of the additional energy scale $m$ introduced by normal ordering.
In particular, a duality emerges between a theory with $g_2^{(1)} = \frac{m_1^2}{2}$ and another one with $g_2^{(2)}=-\frac{m_2^2}{4}$, the normal ordering scales fixed to $m_1$ and $m_2$, respectively, provided that the dimensionful quartic coupling agrees $g_4^{(1)}=g_4^{(2)}$ and the relation
\be
\log \frac{g_4}{m_1^2}-\frac{\pi m_1^2}{3 g_4}=\log \frac{g_4}{m_2^2}+\frac{\pi m_2^2}{6 g_4}
\ee
holds \cite{Rychkov:2015vap}. This is called the Chang duality and is a weak-strong duality in the quartic coupling. A consequence is that the Ising critical point and the broken phase can be reached starting from the symmetric phase and increasing the quartic coupling $g_4$ at fixed $g_2$. We provide a collection of estimates for the location of the critical point in the literature in Table \ref{tab:CritRefTab}. 

\begin{table}[]
\centering
\begin{tabular}{|c|c|c|c|}
\hline
Ref. &  Method & broken $g_c$  & unbroken $g_c$\\     \hline
\cite{Serone:2018gjo}  & PT+Borel resum & 0.2620(45)  & 2.807(34)          \\
\cite{Serone:2019szm} & $\langle\phi\rangle$ PT+Borel & .290(20) & 2.64(11) \\ 
\cite{Bajnok:2015bgw}  & Fock space TSM & .2656(86) & 2.780(60)          \\
\cite{Elias-Miro:2017tup}  & Fock space TSM+RG& .2683(44) & 2.760(30)          \\
\cite{Milsted:2013rxa}  & MPS I & .26707(28) & 2.7690(20)          \\
\cite{Milsted:2013rxa}  & MPS II & .26797(11) & 2.7625(8)          \\
\cite{Bosetti:2015lsa} & LMC & .2645(20)(11) & 2.788(15)(8) \\
\cite{Pelissetto:2015yha} & LMC+Borel & .2697(15) & 2.750(10) \\
\cite{Bronzin:2018tqz} & LMC & .26779(49)& 2.7638(35) \\
\cite{Kadoh:2018tis} & TRG & .2730(21) & 2.728(14) \\
\cite{Heymans:2021rqo} & OPT & .2657(35) & 2.779(25) \\
\cite{Delcamp:2020hzo} & Gilt-TNR &.26672(32) & 2.7715(23) \\
\cite{Vanhecke:2021noi} & Boundary MPS & .266343(11)&  2.774250(78)\\
& This work & .2645(20) & 2.788(15) \\
\hline
\end{tabular}

\caption{Results for the critical point both in broken and unbroken phases from the literature.}
\label{tab:CritRefTab}
\end{table}

\subsubsection{Symmetry preserving phase}
In the symmetry preserving phase, there is a single, self-interacting excitation that correspond to the elementary fluctuations of the field.
The bulk energy density, $\mathcal{E}_0$, has been calculated to
$8$ loops in \cite{Serone:2018gjo}, yielding ($g=g_4/(m_0^2)$)
\bea
\frac{\mathcal{E}_0}{m_0}&=&-\frac{21 \zeta(3)}{16\pi^3}g^2+\frac{27\zeta(3)}{8\pi^4}g^3\nonumber\\
&&-0.116125964(91)g^4+0.3949534(18))g^5\nonumber\\
&&-1.629794(22)g^6+7.85404(21)g^7\nonumber\\
&&-43.02(21)g^8 +O(g^9)\label{phiE0pertser}
\eea
This is an asymptotic series, which can be resummed using Borel techniques. 

The physical mass $M$  
is given by the energy difference of the lowest-energy one-particle state and the vacuum. It can be obtained from the position of the pole in the interacting two-point function, which, in turn, is accessible through a coupling expansion.The expansion is given as \cite{Serone:2018gjo}
\bea
\frac{M^2}{m_0^2}&=&1-\frac32 g^2+\left(\frac{9}{\pi}+\frac{63\zeta(3)}{2\pi^3}\right)g^3 - 14.655869(22)g^4 \nonumber\\
&& + 65.97308(43)g^5 - 347.8881(28) g^6 \nonumber \\
&&  +2077.703(36)g^7 - 13771.04(54) g^8 + O(g^9)
\eea
which is again a Borel resummable asymptotic series.

\subsubsection{Symmetry broken phase}
In infinite volume, where symmetry breaking is exact, there is a doubly degenerate ground state. In finite volume, instanton effects restore the symmetry, but the energy splitting of the two vacua remains on the order of $e^{-M_K L}$, where $M_K$ is the kink mass, given semiclassically ($g_2<0$, $|g_2|>>g_4$) as \cite{Dashen:1975hd,Evslin:2021gxs}
\be
M_K=\frac{2|g_2|^{3/2}}{3g_4}-2|g_2|^{1/2}\left(\frac{3}{2\pi}-\frac{1}{4\sqrt{3}}\right)+O\left(\frac{g_4}{|g_2|^{1/2}}\right)\label{kinkmassSC}
\ee

The broken phase is enriched by a number of mesons, resonances and topological kinks: The exact structure of the spectrum also depends on the coupling.
The bulk energy density in the broken sector can be calculated by expanding in the fluctuations around either vacua. Its perturbative expansion in the coupling starts as \cite{Serone:2019szm}

\bea
\frac{\Lambda}{m_0^2}&=&-\left(\frac{\psi^{(1)}(1/3)}{4\pi^2}-\frac16\right)g-0.042182971(51)g^2\nonumber \\
&&- 0.0138715(74)g^3-0.01158(19)g^4\nonumber\\
&&+O(g^5)
\eea

This series is also Borel resummable, but the naive series provides a robust estimate in the weakly coupled branch ($g_2<0$, $|g_2|>>g_4$) of the broken phase.

\subsection{The Computational Space of States}

We now turn to how we will set up our computational space of states.  The first step in doing so is specify how we are dividing the Hilbert space of the $\phi^4$ theory into two, i.e., how we are choosing $\mathcal{H}_1$ and $\mathcal{H}_2$ in eq. \ref{hilbert_divide}.  

$\mathcal{H}_1$ consists of states in the zero mode sector of the theory, i.e., states involving no oscillator modes, $\hat{\mathbf{a}}_n, n\neq 0$.  To explain this further, we follow \cite{Rychkov:2015vap, Bajnok:2015bgw} and single out the zero mode part of the field $\hat{\boldsymbol{\phi}}$ and its conjugate $\hat{\boldsymbol{\pi}}$:
\be
\hat{\boldsymbol{\phi}}(x)=\boldsymbol{\phi}_0+\widetilde{\boldsymbol{\phi}}(x),\quad \hat{\boldsymbol{\pi}}=\boldsymbol{\pi}_0+\widetilde{\boldsymbol{\pi}}(x). \label{phisepar}
\ee
In eq. \eqref{phisepar} and in the following, we apply the following notation. For an operator $\mathbf{A}$ acting only on the zero mode subspace, we omit the hat. For an operator $\widetilde{\mathbf{B}}$ acting only in the oscillator subspace, we apply a wide tilde. Whenever a hatless operator is added to a wide tilded operator, it is understood that they are tensored with the unity in the other subspace, so that the addition is meaningful.

We then divide 
the Hamiltonian $\hat{\mathbf{H}}$ into a zero mode $\mathbf{H}_{ZM}$ and non-zero mode
$\hat{\mathbf{H}}_{NZM}$ part:
\begin{eqnarray}%\label{eIIIiii}
\hat{\mathbf{H}} &=& \mathbf{H}_{ZM} + \hat{\mathbf{H}}_{NZM}\cr\cr
\mathbf{H}_{ZM}&=&\mathbf{H}_{0,ZM}+\mathbf{V}_{ZM}\cr\cr
\hat{\mathbf{H}}_{NZM}&=&\widetilde{\mathbf{H}}_{0,osc}^{(m)}+ \hat{\mathbf{ V'}}\cr\cr
\hat{\mathbf{V}}&=&\hat{\mathbf{V'}}+\mathbf{V}_{ZM},\label{Vseparation}
\end{eqnarray}
with the free parts defined by
\begin{eqnarray}
  \mathbf{H}_{0,ZM}&=&m_0 \mathbf{a}_0^\dagger \mathbf{a}_0\cr\cr
  \widetilde{\mathbf{H}}_{0,osc}^{(m_0)}&=&\intop dx :\frac{\widetilde{\boldsymbol{\pi}}^2}{2}+\frac{(\partial_x \widetilde{\boldsymbol{\phi}})^2}{2}+\frac{m_0^2}{2}\widetilde{\boldsymbol{\phi}}^2:_{m_0,L}\cr\cr
&=&\sum_{n\neq 0} \omega_n \widetilde{\mathbf{a}}_n^\dagger \widetilde{\mathbf{a}}_n\label{Hnzmdef}  
\end{eqnarray}
There is some freedom in how one distributes $\hat{\mathbf{V}}$ between $\mathbf{V}_{ZM}$ and $\hat{\mathbf{V'}}$ which we exploit in this work.  One choice for this division is given in Section III.2.1.

We first describe our choice for $\mathcal{H}_1$ and $\mathcal{H}_2$ as described in Eq.~\eqref{hilbert_divide}.  Our choice for $\mathcal{H}_1$ is given by
\begin{eqnarray}
\mathcal{H}_1 &=& \{\ket{m}\otimes\ket{\tilde{0}}\}^{N_{ZM}}_{m=1};\cr\cr
\mathbf{H}_{ZM}|m\rangle &=& E_m^{(ZM)}|m\rangle. \label{ZMEVprob}
\end{eqnarray}
In forming $\mathcal{H}_1$, we use a finite dimensional subspace spanned by the first $N_{ZM}$ lowest energy eigenvectors of 
$\mathbf{H}_{ZM}$. $\mathbf{H}_{ZM}$ is an interacting Hamiltonian with a discrete spectrum. The determination of the eigenstates $|m\rangle$ is easily done numerically. To do so, we need to choose a basis in which to represent $|m\rangle$.  Here we choose a massive oscillator basis. Having specified $\mathcal{H}_1$, $\mathcal{H}_2$ then consists of all states with non-trival oscillator content, any states involving a non-zero number of $\widetilde{\mathbf{a}}^\dagger_n$.

We now turn to specifying our tail state basis.  For each $\ket{m}\in \mathcal{H}_1$, we define a sequence of tails given by (see Eq.~\ref{tildet})
\begin{eqnarray}
      \ket{\tilde t_{mk}} &=& \Ht^{k-1}\frac{1}{E-{\bf\hat H_{0,22}}}{\bf\hat V_{21}}\ket{m};\label{OrigTailBasis}\cr\cr
      \hat{\mathbf{T}} &=& \frac{1}{E-\hat {\bf H}_{0,22}}\Vhh \cr\cr
      {\bf\hat H_{0,22}} &=& \hat{\mathbf{P}}_{\perp}\left(\mathbf H_{ZM}+ \widetilde{\mathbf{H}}_{0,osc}^{(m)}\right)\hat{\mathbf{P}}_{\perp} ;\cr\cr
      {\bf\hat H_{21}} &=& \hat{\mathbf{P}}_{\perp}\hat{\mathbf H} (1-\hat{\mathbf{P}}_\perp),\cr\cr
      {\bf\hat V_{22}} &=& \hat{\mathbf{P}}_{\perp}\hat{\mathbf{V'}}\hat{\mathbf{P}}_{\perp}.
\end{eqnarray}
where $P_\perp$ is a projector onto $\mathcal{H}_2$.

\subsubsection{``Conventional" tails}

An obvious choice for separation of $\hat{\mathbf{V}}$ into $\mathbf{V}_{ZM}$ and $\hat{\mathbf{V'}}$ is to take 
\bea
\mathbf{V}_{ZM} &=&\mathcal{G}_2L :\boldsymbol{\phi}_0^2:_{m_0}\nonumber \\
& &+g_4 L :\boldsymbol{\phi}_0^4:_{m_0}+E_0(m_0,L);\label{VZMOrigChoice}\cr\cr
\hat{\mathbf{ V'}} &=& \sum_{n=2}^4\mathbf{g}_n\intop dx  :\widetilde{\boldsymbol{\phi}}(x)^n:_{m_0,L} 
 \label{V1OrigChoice}
\eea
where we used the notations
\bea
\mathbf{g}_4&=&g_4\mathbf{1},\quad\mathbf{g}_3=4g_4 \boldsymbol{\phi}_0,\quad \mathbf{g}_2=\mathcal{G}_2\mathbf{1}+6g_4 :\boldsymbol{\phi}_0^2:_{m_0};\cr\cr
\mathcal{G}_2&=&g_2^\prime-\frac{m_0^2}{2}\label{boldgdef}
\eea
In the above $g^\prime_2$ and  $E_0(m_0,L)$ are
defined in Eq. \eqref{E0def}. 
In order to complete the computation we need to specify the matrix elements of our basis.  At first tail order (see Eq.~\ref{1st_tail_order}), we can substitute the projected operators with the unprojected ones in the relevant matrix elements. 
This is because the extra terms induced by the projectors always involve one-point functions of operators $:\widetilde{\boldsymbol{\phi}}^n:$ for some $n\geq1$, which are all zero. We obtain the following matrix elements  
\begin{widetext}
    \begin{eqnarray}\label{1st_tail_mes}
        (H_{11})_{mm'} &=& \langle m | \mathbf{H}_{ZM} | m'\rangle =E_m^{(ZM)} \delta_{mm^\prime};\cr\cr
        (\Hlt{1})_{m\tilde t_{m'1}} &=& \langle m |\mathbf{\hat V}_{12}   \ket{\tilde t_{m'1}} = \braket{m|\mathbf{\hat V'}\frac{1}{E-\mathbf H_{ZM}- \widetilde{\mathbf{H}}_{0,osc}^{(m)}}\mathbf{\hat V'}|m'} ;\cr\cr
        (\Htl{1})_{\tilde t_{m1}m'} &=& \bra{\tilde t_{m1}}\mathbf{\hat V}_{21} \ket{m'} = (\Hlt{1})_{m'\tilde t_{m1}};\cr\cr
        (\Htt{1})_{\tilde t_{m1} \tilde t_{m'1}} &=& \bra{\tilde t_{m1}}\mathbf{\hat V}_{22} \ket{\tilde t_{m'1}} = \braket{m|\mathbf{\hat V'}\frac{1}{E-\mathbf H_{ZM}- \widetilde{\mathbf{H}}_{0,osc}^{(m)}}\mathbf{\hat V'}\frac{1}{E-\mathbf H_{ZM}- \widetilde{\mathbf{H}}_{0,osc}^{(m)}}\mathbf{\hat V'}|m'};\cr\cr
        (\G{11})_{\tilde t_{m1} \tilde t_{m'1}} &=& \bra{\tilde t_{m1}} \tilde t_{m'1}\rangle = \braket{m|\mathbf{\hat V'}\frac{1}{(E-\mathbf H_{ZM}- \widetilde{\mathbf{H}}_{0,osc}^{(m)})^2}\mathbf{\hat V'}|m'}.
    \end{eqnarray}
\end{widetext}
 We provide similar expressions for the matrix elements involved in the eigensystem at second tail order in Appendix \ref{AppZMExDetails} (see Eq.~\ref{eigen2}). 
In turn, these expressions can be transformed into Euclidean multipoint functions of the interaction $\mathbf{\hat V'}$ by using the integral representation 
\be
\intop_0^\infty \tau^r e^{-\hat{\mathbf{M}} \tau} =\frac{r!}{\hat{\mathbf{M}}^{r+1}}.\label{DenomToMultipoint}
\ee
In particular, for the matrix element $(\Hlt{1})_{m\tilde t_{m'1}}$ we obtain
\bea
&&(\Hlt{1})_{m\tilde t_{m'1}}=-L\intop_{0}^{\infty}d\tau e^{-E_m^{(ZM)})\tau}\intop_{0}^L dx\nonumber\\
&&\left(\left\langle:\widetilde{\boldsymbol\phi}^2(\tau,x)::\widetilde{\boldsymbol\phi}^2(0):\right\rangle \braket{m|\mathbf{g}_2(\tau)\mathbf{g}_2(0)|m'}+\right.\nonumber\\
&&\left\langle:\widetilde{\boldsymbol\phi}^3(\tau,x)::\widetilde{\boldsymbol\phi}^3(0):\right\rangle \braket{m|\mathbf{g}_3(\tau)\mathbf{g}_3(0)|m'}+\nonumber\\
&&\left.g_4^2\left\langle:\widetilde{\boldsymbol\phi}^4(\tau,x)::\widetilde{\boldsymbol\phi}^4(0):\right\rangle\right).
\eea

In order to implement our scheme we need to make a choice for $N_{ZM}$. This choice is unproblematic.  The lowest eigenvalues of the full problem converge quickly in the number, $N_{ZM}$, of kept zero-mode states. Typically $N_{ZM}$ can be kept to a number less than 10.  And at least in certain cases, the zero-mode Hamiltonian, $\mathbf{H}_{ZM}$, encodes much of the physics of the full problem.  In such cases the effects of the oscillator modes on the physics of the problem can be considered as perturbative.  We have found that this choice of separation of $\mathcal{H}_1$ and $\mathcal{H}_2$ works in both the unbroken and broken phases of the theory.

\begin{figure}
    \centering
     \includegraphics[draft=false,width=\columnwidth]{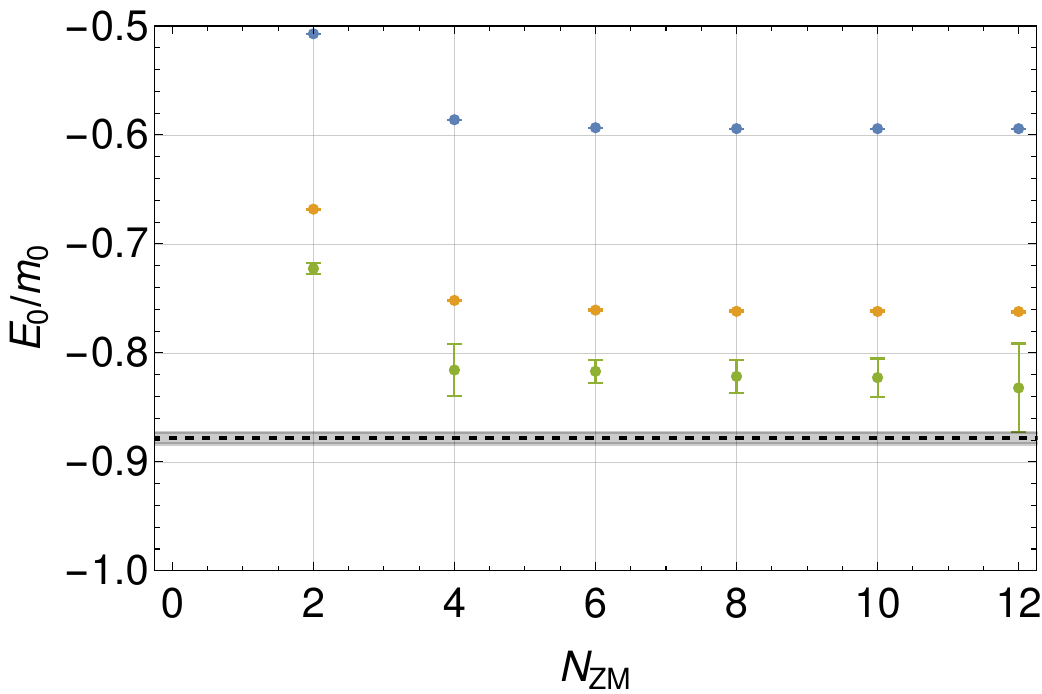}

    \caption{
    The values of the ground state energy for the first three tail orders computed at $g_4=1.5$, $L=10$ for different choices of $N_{ZM}$.
    Blue, orange, green dots correspond to the first, second, and third Krylov orders, respectively. We see the results rapidly converge as a function of $N_{ZM}$.  We compare our results to data provided to us by the authors of \cite{Elias-Miro:2017tup} (here shown as a dashed black line with error bounds shown in gray).
\label{unapproximated_tails}}
\end{figure}

In Fig.~\ref{unapproximated_tails} we provide an example of our use of these tails. There we compute the ground state energy in the unbroken phase up to the third tail order for a non-trivial value $g_4$ as a function of $N_{ZM}$.\footnote{We extrapolate the reference data from \cite{Elias-Miro:2017tup} using two extrapolation functions, $f_4(E_{cut})=a+b E_{cut}^{-3}+ c E_{cut}^{-4}$ and $f_3(E_{cut})=a+b E_{cut}^{-3}+ c E_{cut}^{-4}$. We accept the result of the extrapolation with $f_4$ as the central value, while the difference betweem fitting with $f_3$ and $f_4$ estimates the error. We opted for this to emphasize the uncertainty arising from the unknown exact form of the  fitting function. Note that following the original error analysis in \cite{Elias-Miro:2017tup} yields an even sharper estimate.}  We see that after $N_{ZM}$ exceeds 6, the result reaches an asymptotic value.

\subsubsection{Using numerically efficient tail states}
The vanilla method described in the first part of this section suffers from three technical drawbacks. Firstly, the computational basis is not a simple tensor product of states in $\mathcal{H}_1$ and states built by acting with oscillator modes above the oscillator vacuum.  The evaluation of matrix elements using Monte Carlo results in a linear algebra that scales cubically with the number, $N_{ZM}$ of zero mode states kept. Secondly, the tail states depend on the coupling $g_4$ in a complicated way and so a separate computation is needed for each value of $g_4$ to be investigated. Thirdly, the tails depend explicitly on the exact energy level $E_*$, which has to be tuned self-consistently, and is in principle different for the ground state and for excited states. These three drawbacks make the method as described computationally expensive. 

We can avoid these issues if we define our tails with the following choice for $\mathbf{V}_{ZM}^{(U)}$ and $\hat{\mathbf{V'}}^{(U)}$.  The superscript $(U)$ appears here to emphasize a different separation of $\hat{\mathbf{V}}$ as compared to eqs. \eqref{VZMOrigChoice}-\eqref{V1OrigChoice}.
We choose 
\bea
\mathbf{V}_{ZM}^{(U)} &=&-\mathbf{H}_{0,ZM}+E_*\label{NumEffVZM};\cr\cr
\hat{\mathbf{ V'}}^{(U)} &=& \intop dx :\mathcal{G}_2 \hat{\boldsymbol{\phi}}(x)^2 
\cr\cr 
&& + g_4 \hat{\boldsymbol{\phi}}(x)^4:_{m_0,L} +\mathbf{H}_{0,ZM}-E_*.
\eea
We also alter our choice for our computational basis.  We change ${\cal H}_1$ to 
\begin{eqnarray}
\mathcal{H}_1 &=& \{\ket{p}\otimes\ket{\tilde{0}}\}^{N_{ZM}}_{p=1};\cr\cr
\mathbf{H}_{0,ZM}|p\rangle &=& E_p^{(ZM)}|p\rangle.
\end{eqnarray}
Here $E^{ZM}_p = m_0p$.
Now ${\cal H}_1$ consists of a basis of free oscillator states $\ket{p}$.

Neglecting for a moment the cutoff, $N_{ZM}$, in the zero-mode subspace, tails of order $K$ obtained with $\mathbf{V}_{ZM}^{(U)}$ of eq. \eqref{NumEffVZM} can be written as a linear combination ($l_j\geq1$)
\begin{eqnarray}
\ket{\widetilde t_{pK}} &=& \sum_{p'=1}^{\infty}\sum_{Q=1}^K\sum_{\sum_{j=1}^Q l_j=K}\sum^4_{k_j=2} c^{pK}_{p'Q,\{k\},\{l\}}\cdot\cr\cr
&&\ket{p'}\otimes\ket{\widetilde  t_{Q,\{k\},\{l\}}}\cr\cr
\ket{\widetilde  t_{Q,\{k\},\{l\}}}&\equiv& \prod_{j=1}^Q\frac{1}
{(-\widetilde{\mathbf{H}}_{0,osc})^{l_j}}:\widetilde{\boldsymbol\phi}^{k_j}:\ket{\widetilde{0}}\label{CompTailBasis}
\end{eqnarray}
where $c^{pK}_{p'Q,\{k\},\{l\}}$ are numerical coefficients and $\ket{\widetilde{0}}$ is the ground state of $\widetilde{\mathbf{H}}_{0,osc}^{(m)}$.
Rather than work with the tail states $\ket{\tilde t_{pK}}$, we will use the larger set $\ket{p'}\otimes\ket{\widetilde  t_{Q,\{k\},\{l\}}}$ directly as a basis.  One can see that this basis choice is a tensor product of the zero-mode space and the space of oscillator states.  This will lead to a dramatic simplification in our numerical evaluation of matrix elements.

Using states of the form $\ket{p'}\otimes\ket{\widetilde  t_{Q,\{k\},\{l\}}}$ represents a change of basis but it is not expected to spoil the convergence of the method (see also Appendix \ref{OtherKrylovRel}). In fact this extension of the basis improves on the accuracy through the increase of the dimension of the variational basis. However one issue seen with this choice of basis is that the Gram matrix rapidly becomes degenerate, particularly at higher Krylov orders. To alleviate this issue we will impose an extra restriction on the tails and use only $l_j=1$ states (i.e. states that involve only single inverse powers of $\widetilde{\mathbf{H}}_{0,osc}^{(m)}$ in our numerical basis). We will refer to the resulting subspace as the \textit{universal} tail space. 

The further reduction of this tail space to one spanned by tails with at most $K$ denominators will be called the universal tail space of order $K$. Accordingly, when it does not lead to confusion, henceforth we simplify the notation of the $l_j=1$ tails, specifying them by the list of operator powers $k_l$ only: $\ket{\widetilde  t_{Q,\{k\},\{l\}}}\equiv\ket{\widetilde  t_{\{k_1\dots k_K\}}}$. %
Note that the resulting basis still has significantly higher dimensionality than the number of conventional tails. At order $K$, the total number of conventional tails is $KN_{ZM}$, while the dimension of the full universal tail space truncated at order $K$ is $N_{ZM}\sum_{q=0}^K 3^q=N_{ZM}(3^{K+1}-1)/2$. %

\begin{figure}
    \centering
    \includegraphics[draft=false,width=\columnwidth]{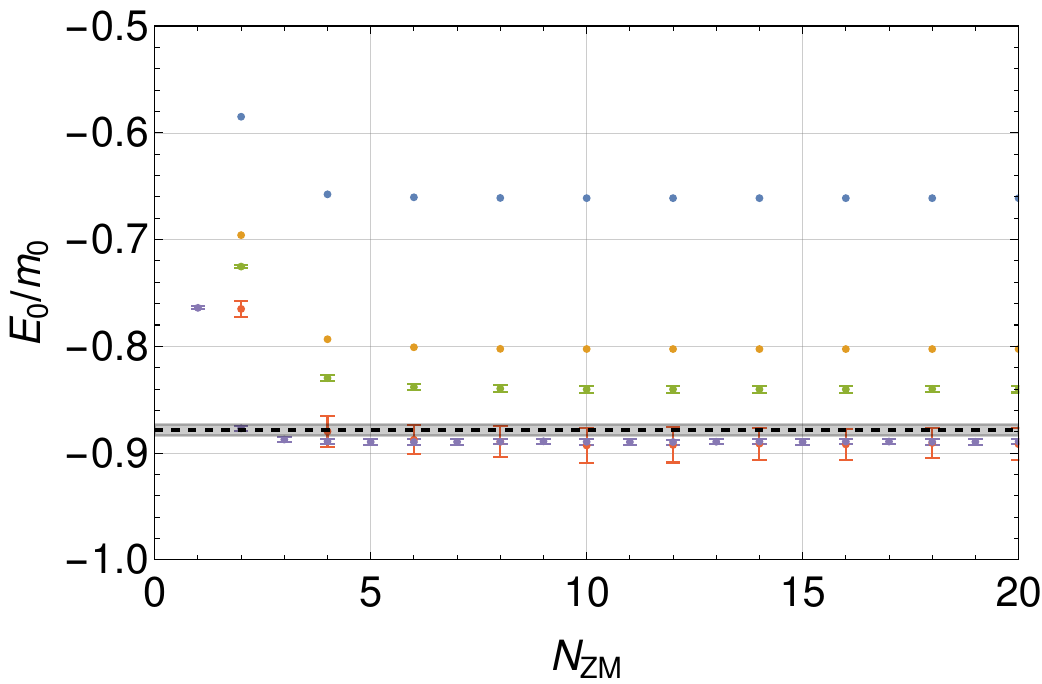}%
    \caption{
    The values of the ground state energy for the first three tail orders computed at $g_4=1.5$, $L=10$ using the approximate tails for different choices of $N_{ZM}$.  We again compare our results with those found in \cite{Elias-Miro:2017tup}.  Blue, orange, green dots correspond to $1$, $2$ and $3$ Krylov orders, respectively. Also shown is the power law extrapolation using statistically generated eigenvalues (red) and its empirical improvement (purple).  We see the results rapidly converge as a function of $N_{kept}$. We compare our results to data provided to us by the authors of \cite{Elias-Miro:2017tup} (here shown as a dashed black line with error bounds shown as a thick gray line around it). Here $N_{MC,0}=10^8$, $N_{iter}=50$.}
    \label{approximated_tails}
\end{figure}
\begin{figure}
    \centering
    \includegraphics[width=\columnwidth]{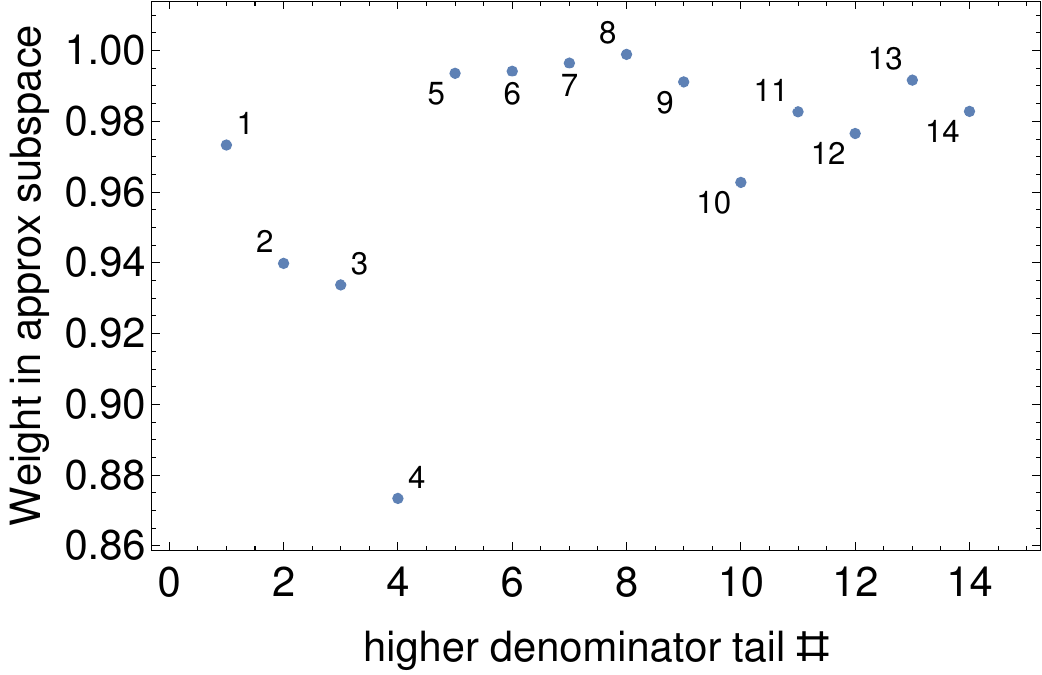}
    \caption{
    Overlaps of normalized higher denominator tails with the universal tail space of order 2 (see main text). Even parity sector, $m_0L=10$. Tails $\ket{\widetilde t_{Q,\{k\},\{l\}}}$ corresponding to numbered points are as follows: 1) $\ket{\widetilde t_{1,\{2\},\{2\}}}$, 2) $\ket{\widetilde t_{1,\{4\},\{2\}}}$, 3) $\ket{\widetilde t_{1,\{2\},\{3\}}}$ , 4) $\ket{\widetilde t_{1,\{4\},\{3\}}}$, 5) $\ket{\widetilde t_{2,\{22\},\{12\}}}$, 6) $\ket{\widetilde t_{2,\{24\},\{12\}}}$, 7) $\ket{\widetilde t_{2,\{33\},\{12\}}}$, 8) $\ket{\widetilde t_{2,\{42\},\{12\}}}$, 9) $\ket{\widetilde t_{2,\{44\},\{12\}}}$, 10) $\ket{\widetilde t_{2,\{22\},\{21\}}}$, 11) $\ket{\widetilde t_{2,\{24\},\{21\}}}$, 12) $\ket{\widetilde t_{2,\{33\},\{21\}}}$, 13) $\ket{\widetilde t_{2,\{42\},\{21\}}}$, 14) $\ket{\widetilde t_{2,\{44\},\{21\}}}$. Note that points 3 through 14 only appear in $K=3$ tails.}
    \label{FigHighOrdTail_overlaps}
\end{figure}

We reproduced the quantity depicted in Fig. \ref{unapproximated_tails} with universal tails in Figure \ref{approximated_tails}. The error bars are smaller in the latter plot, as more integrand evaluations were possible due to the simpler numerics. In turn, this made possible to explore various extrapolation approaches in Krylov orders, which we discuss further in Sections \ref{Extrap1Sec} and \ref{ExtrapSec2}.

With higher denominator tails omitted, the method still remains variational, but in principle prone to a small systematic error due to the neglected basis states.
We show in Figure \ref{FigHighOrdTail_overlaps} that, after normalizing to unity, the overlaps of all $l_j>1$ tails appearing in eq. \eqref{CompTailBasis} up to $K=3$ on the universal tail space of order $2$ is close to 1. While Fig. \ref{FigHighOrdTail_overlaps} includes tails from only the even parity sector, the situation is entirely analogous in the odd sector. Thus the neglected states lie almost entirely in the space of states that we do keep. Further, since the tails $\ket{\widetilde  t_{Q,\{k\},\{l\}}}$ are independent of the couplings, so are their overlaps. Therefore, no significant errors are expected to arise from omitting the higher power denominators.

\begin{figure}[t]
    \centering
    \includegraphics[width=\columnwidth]{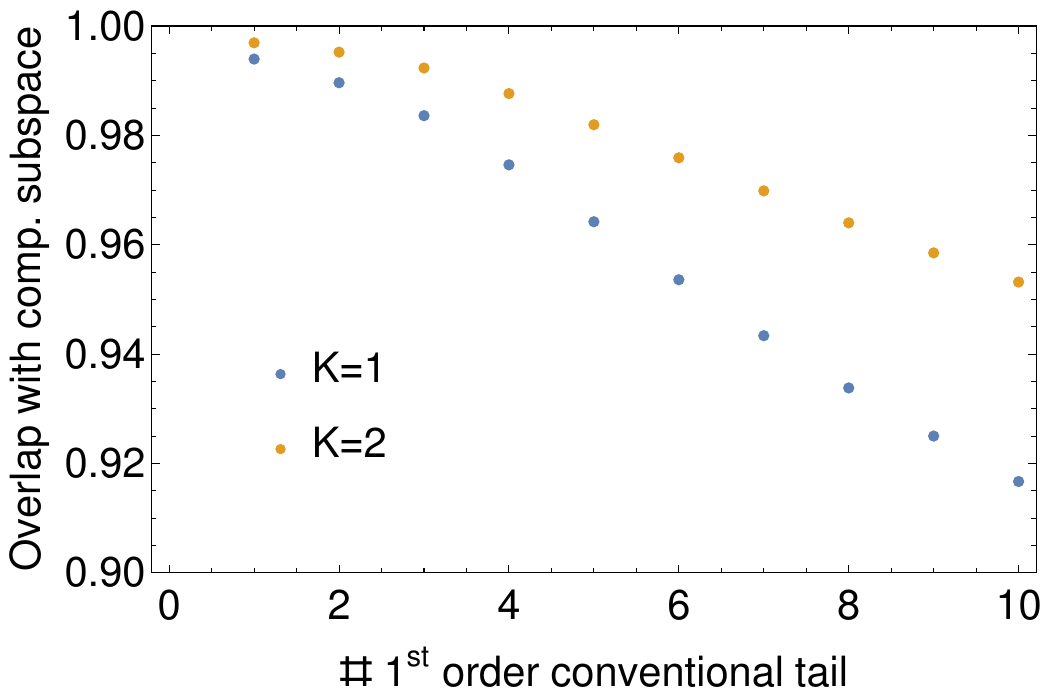}
    \caption{Overlaps of normalized conventional first order tails in the broken sector with $g_4=1.5$, $mL=10$, and $E_*=-0.878$, with the universal tail space of orders $K=1$ (blue dots) and $K=2$ (orange dots).  Tails built upon the $10$ zero-mode eigenstates with lowest energy are shown sorted by increasing ZM energy.}
    \label{FigConvTail_overlaps}
\end{figure}

As another test of this modified computation basis, we test our ability to represent the ``conventional" tails in \eqref{OrigTailBasis} (with $\mathbf{V}_{ZM}$ defined in eq. \eqref{VZMOrigChoice}) in our new basis. In Figure \ref{FigConvTail_overlaps}, we see that the overlaps of first order ``conventional" tails are close to one. These overlaps are defined as,
\bea
\braket{\tilde t_{m1}|\hat{\mathbf{P}}_U|\tilde t_{m1}}&=&\sum_{I,J,p,q}o_{m1;p,I}(G^{-1}_U)_{IJ}\delta_{pq}o_{m1;q,J}\\
\label{ome}o_{m1;p,\{k_1,\dots k_K\}}&=&\bra{\tilde t_{m1}}\left(\ket{p}\otimes\ket{\widetilde t_{\{k_1,\dots k_K\}}}\right)
\eea
where the indices $I,J$ run over all universal tails up to order $K$, and $G_U$ is the inner product matrix between universal tails,
\be
(G_U)_{\{k_1,\dots k_{K_1}\},\{k^\prime_1,\dots k^\prime_{K_2}\}}=\braket{\widetilde{t}_{\{k_1,\dots k_{K_1}\}}|\widetilde{t}_{\{k^\prime_1,\dots k^\prime_{K_2}\}}}.
\ee
This remains true for a wide range of couplings, $E_*$ parameters and volumes.
This completes a numerical demonstration that replacing our original tail states with $\ket{p}\otimes\ket{\widetilde  t_{\{k\}}}$ is a very good approximation.

A particularly suitable form of the Hamiltonian is gained by emphasizing its tensor product structure relative to this new basis. Introducing the shorthand $\widetilde{\mathbf{V}}_n = \intop_0^L :\widetilde{\boldsymbol\phi}(x)^n:_{m_0,L} dx$, the Hamiltonian $\hat{\mathbf H}$ can be written as 
\begin{eqnarray}
\mathbf{\hat{H}}=(\mathbf{H}_{0,ZM}+\mathbf{V}_{ZM})\otimes \widetilde{\mathbf{1}}+\mathbf{1}\otimes\widetilde{\mathbf{H}}_{0,osc}
+\sum_{n=2}^4\mathbf{g}_n\otimes\widetilde{\mathbf{V}}_n,\label{HamNumeric}\cr\cr&&
\end{eqnarray}
with $\mathbf{V}_{ZM}$ as defined in eq. \eqref{VZMOrigChoice} and $\mathbf{g}_n$ defined in eq. \eqref{boldgdef}.
In our new basis, the needed matrix elements can be reduced into matrix elements of $\widetilde{\mathbf{V}}_n$ and $\widetilde{\mathbf{1}}$ between the universal tails:
\begin{widetext}
    \begin{eqnarray}\label{1st_approx_tail_mes}
        (H_{11})_{pp'} &=&\delta_{pp'}(E_p^{(0)}+E_0(m_0,L))+\mathcal{G}_2L (:\phi_0^2:)_{pp^\prime}+g_4 L (:\phi_0^4:)_{pp^\prime} ;\cr\cr
        (\Hlt{1})_{p,p'\otimes\widetilde t_{k}} &=&\delta_{pp^\prime}\left(\mathcal{G}_2 \braket{\widetilde{0}|\widetilde{\mathbf V}_2|\widetilde t_{k}}+g_4 \braket{\widetilde{0}|\widetilde{\mathbf V}_4|\widetilde  t_{k}}\right)+4g_4(\phi_0)_{pp^\prime}\braket{\widetilde{0}|\widetilde{\mathbf V}_3|\widetilde  t_{k}}+6g_4^2(:\phi_0^2:)_{pp^\prime}\braket{\widetilde{0}|\widetilde{\mathbf V}_2|\widetilde  t_{k}};\cr\cr
        (\Htl{1})_{p\otimes\widetilde t_{k}, p'} &=& (\Hlt{1})_{p',p\otimes\widetilde t_{k}}; \cr\cr
        (\Htt{11})_{p\otimes\widetilde t_{k_1},p'\otimes\widetilde t_{k_2}} &=&(H_{11})_{pp'}\braket{\widetilde t_{k_1}|\widetilde t_{k_2}}+\delta_{pp^\prime}\left(\mathcal{G}_2 \braket{\widetilde t_{k_1}|\widetilde{\mathbf V}_2|\widetilde t_{k_2}}+g_4 \braket{\widetilde t_{k_1}|\widetilde{\mathbf V}_4|\widetilde t_{k_2}}+\braket{\widetilde t_{k_1}|\widetilde{\mathbf H}_{0,osc}^{(m)}|\widetilde t_{k_2}}\right)\cr\cr
        &+&4g_4(\phi_0)_{pp^\prime}\braket{\widetilde t_{k_1}|\widetilde{\mathbf V}_3|\widetilde t_{k_2}}+6g_4^2(:\phi_0^2:)_{pp^\prime}\braket{\widetilde t_{k_1}|\widetilde{\mathbf V}_2|\widetilde t_{k_2}};
    \end{eqnarray}
\end{widetext}
where the oscillator space matrix elements take the form
\bea
&&\braket{\widetilde{0}|\widetilde{\mathbf V}_n|\widetilde t_{k}} = -\braket{\widetilde t_{n}|\widetilde{\mathbf H}_{0,osc}^{(m)}|\widetilde t_{k}}\cr\cr
&=&L\intop_{0}^Ldx\braket{\widetilde{0}|:\widetilde{\phi}^n:(x)\frac{1}{-\widetilde{\mathbf H}_{0,osc}^{(m)}}:\widetilde{\phi}^k:(0)|\widetilde{0}}\cr\cr
&&\braket{\widetilde t_{n}|\widetilde t_{k}}=L\intop_{0}^Ldx\braket{\widetilde{0}|:\widetilde{\phi}^n:(x)\frac{1}{(\widetilde{\mathbf H}_{0,osc}^{(m)})^2}:\widetilde{\phi}^k:|\widetilde{0}}\cr\cr
&&\braket{\widetilde t_{l}|\widetilde{\mathbf V}_n|\widetilde t_{k}}=L\intop_{0}^\infty d^2\tau \intop_{0}^L d^2x\cr\cr
&&\braket{\widetilde{0}|:\widetilde{\boldsymbol\phi}^l:(\tau_2+\tau_1,x_1):\widetilde{\boldsymbol\phi}^n:(\tau_1,x_2):\widetilde{\boldsymbol\phi}^k:|\widetilde{0}}.
\eea
In the last line we transformed the energy denominators into multi-point functions with the aid of eq. \eqref{DenomToMultipoint}. Note that the oscillator space matrix representation of $\widetilde{\mathbf H}_{0,osc}^{(m)}$ and the inner product matrix are actually diagonal between first order universal tails.

The use of the basis vectors \eqref{CompTailBasis} in calculating the matrix elements of the effective Hamiltonian reduces to calculating general n-point functions of $:\widetilde{\boldsymbol \phi}^n:$. Matrix elements involving higher tail orders contain disconnected pieces. As with the original tail states, we provide the explicit expressions for the matrix elements needed to carry out the computation with second order tail states in Appendix \ref{AppExplicitDiscTerms}.

We show a demonstration of the precision that we obtain with these approximated tail states in Fig. \ref{approximated_tails}. One can see that both the energy estimates and the error bars have improved compared to Fig. \ref{unapproximated_tails}. The improvement in computed energies is due to the larger variational basis. The error bars have decreased in part due to auspicious error propagation from matrix elements to the lowest eigenvalues, in part due to cheaper evaluation costs as we were able to increase the number of integral evaluations by an order of magnitude.

The matrix elements that we need to evaluate can be represented in terms of Feynman diagrams.  To see this, we
introduce the multi-index quantity
\bea
&&V_{i_n,\dots,i_0}(\tau_n,\dots,\tau_1,\tau_0)=L\intop_{0}^L d^nx\cr\cr
&&\braket{\widetilde{0}|:\widetilde{\boldsymbol\phi}^{i_n}(\tau_n,x_n):\dots  :\widetilde{\boldsymbol\phi}^{i_1}(\tau_1,x_1)::\widetilde{\boldsymbol\phi}^{i_0}(\tau_0,0):|\tilde{0}}\cr\cr\label{multiV}
&&
\eea
for an (n+1)-point function ($n\geq1$) of the integrated field powers.  In terms of $V_{i_n,\cdots,i_0}$,
the matrix elements can be written as
\bea
\braket{\widetilde t_{i}|\widetilde{\mathbf H}_{0,osc}^{(m)}|\widetilde t_{j}}&=&\intop_0^\infty d\tau V_{ij}(\tau,0),\cr\cr
\braket{\widetilde t_{i}|\widetilde{\mathbf V}_j|\widetilde t_{k}}&=&\intop_0^\infty d^2\tau V_{ijk}(\tau_2+\tau_1,\tau_1,0),\cr\cr
\braket{\widetilde t_{i}|\widetilde t_{j}}&=&\intop_0^\infty d\tau \tau V_{ij}(\tau,0).
\eea
In turn, the multi-point functions can be evaluated with Wick's theorem, yielding a diagrammatic structure similar to a perturbative calculation.  
We list the simplest matrix elements below, while the general Feynman rules are discussed in Appendix \ref{FeynApp}.
\bea
V_{NN}(\tau)&=&N! L\intop_0^L dx \left(\tilde \Delta^{(2)}_{L}(\tau,x)\right)^N\cr\cr
V_{222}(\tau_2,\tau_1)&=&8  L\intop_0^L d^2x \tilde \Delta^{(2)}_{L}(\tau_2-\tau_1,x_2-x_1)\cdot\nonumber\\
&&\tilde \Delta^{(2)}_{L}(\tau_2,x_2)\tilde \Delta^{(2)}_{L}(\tau_1,x_1)\cr\cr
V_{444}(\tau_2,\tau_1)&=&1728  L\intop_0^L d^2x \left(\tilde \Delta^{(2)}_{L}(\tau_2-\tau_1,x_2-x_1)\right)^2\cr\cr
&&\left(\tilde \Delta^{(2)}_{L}(\tau_2,x_2)\right)^2\left(\tilde \Delta^{(2)}_{L}(\tau_1,x_1)\right)^2 .\label{eqV444}
\eea
In the next subsection, we examine the explicit form of the propagator $\tilde{\Delta}^{(2)}_L$, given in eq. \eqref{ModFVProp} below.
\begin{figure}[]
    \centering
     \subfloat[]{\includegraphics[draft=false,width=0.4\columnwidth]{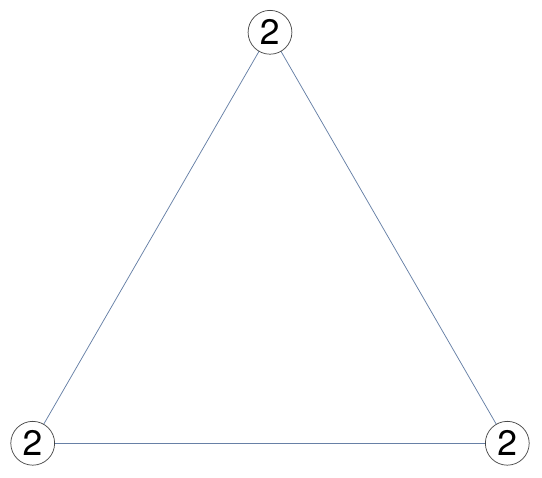}}%
 \subfloat[]{\includegraphics[draft=false,width=0.4\columnwidth]{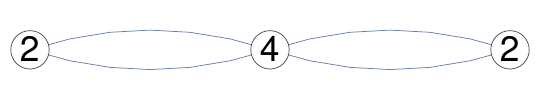}}\\
 \subfloat[]{\includegraphics[draft=false,width=0.4\columnwidth]{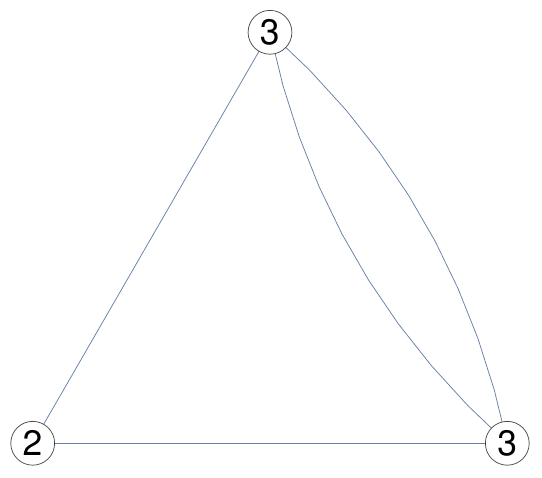}}%
 \subfloat[]{\includegraphics[draft=false,width=0.4\columnwidth]{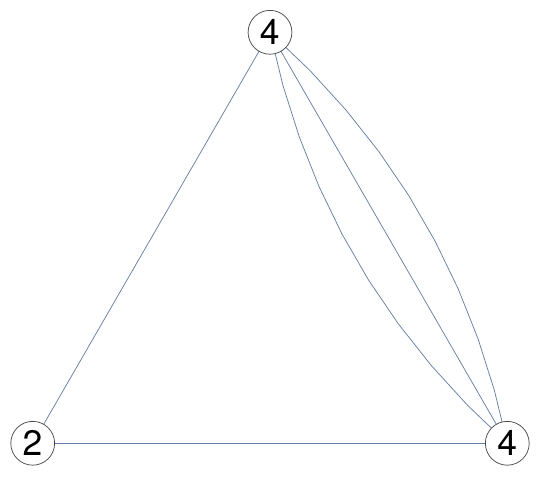}}\\
  \subfloat[]{\includegraphics[draft=false,width=0.4\columnwidth]{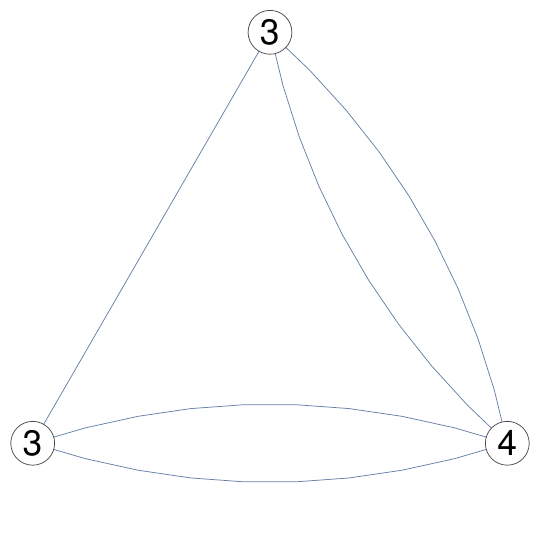}}%
 \subfloat[]{\includegraphics[draft=false,width=0.4\columnwidth]{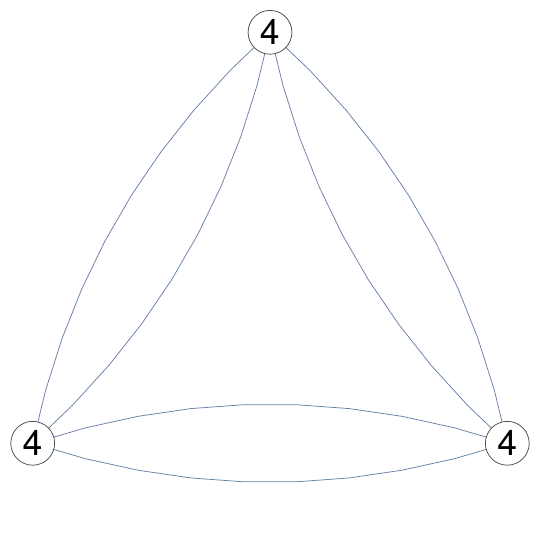}}

    \caption{Feynman diagrams with 3 insertions in the oscillator sector at first Krylov order. (a): $V_{222}$  (b):  $V_{224}$, (c): $V_{233}$, (d): $V_{244}$, (e): $V_{334}$, (f): $V_{444}$ (see also eq. \eqref{eqV444}). Permutations of the indices lead to the same diagrams with permuted $\tau$ arguments.}
    \label{FigFeynNLO}
\end{figure}
The diagrammatic representations of certain representative $V_{i_n,\cdots,i_0}$ are given in Fig. \ref{FigFeynNLO}.

\subsubsection{Finite Volume Propagator of $\phi^4$}
In our implementation of Krylov space methods for $\phi^4$, we simultaneously use both the Hamiltonian formulation of $\phi^4$ together with a Feynman diagrammatic description.  We need to set the conventions of the the finite volume propagator that appears in such diagrams.

For the moment we will work in general space-time dimension.  
Our action for the Lagrangian of a free massive scalar field is
\be
S_m^{(D)}=\frac{1}{2}\intop d^D x \left(\partial\phi\right)^2+m_0^2\phi^2. \label{GenScalarAct}
\ee
The corresponding two-point function in infinite volume then takes the form:
\bea
\Delta^{(D)}(r)&=&\left\langle T_{\tau}\phi(\mathbf{x}_1)\phi(\mathbf{x}_2)\right\rangle\cr\cr\label{GenDProp}
&=&(2\pi)^{-\frac{D}{2}}\left(\frac{m_0}{r}\right)^{\frac{D}{2}-1}K_{-\frac{D}{2}+1}(m_0r),
\eea
where $T_\tau$ indicates ordering with respect to an arbitrarily chosen Euclidean time direction $\tau$, $r=|\mathbf{x}_1-\mathbf{x}_2|$ is the Euclidean distance, and $K_\nu(x)$ is the modified Bessel function of the second kind.

In the special case $D=1$, eq. \eqref{GenScalarAct} describes a single harmonic oscillator with frequency $m_0$ and mass $1$. The propagator $D^{(D)}$ from eq. \eqref{GenDProp} reduces to
\be
\Delta^{(1)}(\tau)=\frac{1}{2m_0}e^{-m_0 \tau}.
\ee
As our computations for $D=2$ will be done in finite volume, we need the corresponding finite volume propagator.  We will assume that we have either periodic (P=0) or anti-periodic (P=1) boundary conditions.  As derived in Appendix \ref{AppFVprop}, in $D=2$ the finite volume mode expansion leads to the propagator
\bea
\Delta^{(2)}_L&=&\Delta^{(2)}(\tau,x)+\Delta^{(2)}_{L,\Delta}(\tau,x)\label{FVMassiveProp}\cr\cr
&=&\frac{1}{2\pi}\left[K_0\left(m_0\sqrt{x^2+\tau^2}\right)\right.+\cr\cr
& &\left.(-1)^{P}K_0\left(m_0\sqrt{(L-x)^2+\tau^2}\right)\right]+\sum_{n=1}^{\infty}\delta \Delta_n,\cr\cr
\delta \Delta_n&=&\frac{(-1)^{Pn}}{2\pi}\left[K_0\left(m_0\sqrt{(nL+x)^2+\tau^2}\right)+\right.\cr\cr
& &\left.(-1)^{P}K_0\left(m_0\sqrt{((n+1)L-x)^2+\tau^2}\right)\right].
\eea
As we can see, in finite volume, the propagator $\Delta^{(2)}$ acquires corrections that are suppressed as $e^{-m_0L}$.
When the zero mode is separated and the multipoint functions are restricted to the oscillator sector, we are required to substitute the complete propagator in eq. \eqref{FVMassiveProp} with a modified one,
\bea
\tilde{\Delta}^{(2)}_L(r)&=&\left\langle T_{\tau}\tilde{\phi}(\mathbf{x}_1)\tilde{\phi}(\mathbf{x}_2)\right\rangle\cr\cr
&=&\Delta^{(2)}_{L}(r)-\frac{1}{2m_0L}e^{-m_0\tau}, \label{ModFVProp}
\eea
In practical computations, we truncate the sum in eq. \eqref{FVMassiveProp} to $n\leq3$, which is a good approximation for $L>3$.

Finally, we note that for small volumes, it is possible to work in a different scheme where the unperturbed oscillators are massless. In this case the oscillator mass term is considered part of the perturbation and the modified two-point function is given by the closed formula
\be
\tilde{\Delta}^{(2)}_{0,L}(r)=-\frac{1}{4\pi}\log\left[1+e^{\frac{-4\pi}{L} t}-2e^{\frac{-2\pi}{L} t}\cos\left(\frac{2\pi}{L} x\right)\right]
\ee
The benefit of the massless scheme is that the volume dependence of the matrix elements factors out, so the numerical integrations do not have to be repeated for each studied volume. However, we found the precision of massless setup strongly volume-dependent and suboptimal for the volume range of our interest.  Thus we do not elaborate upon this direction further.

\subsection{Numerical Implementation} \label{SubsecNumImp}

Now that we have specified the computational bases that we will use in evaluating $\phi^4$, we turn to a discussion of the details of the numerical evaluation of the eigenvalues of the Hamiltonian. 

The numerical method proceeds as follows. We first
compute the matrix elements of $\tilde{\mathbf V}_n$ and $\widetilde{\mathbf{1}}$ of eq. \eqref{HamNumeric} between order one universal tails. We then assemble the full Hamiltonian with a zero mode cutoff $N_{ZM}$ and solve the generalized eigenproblem in this $4N_{ZM}$ dimensional basis. In the next step, we calculate all matrix elements involving universal tails of second order, and repeat the diagonalization in this larger, $13N_{ZM}$ dimensional basis. Finally we also include third order tails and obtain the eigenpairs of the resulting matrices. We thus get a series of eigenvalues as the function of the largest order of tails kept. This set of points, possibly supplemented by a ``zero-order” approximation with all tails neglected, serves as basis to extrapolate to infinite Krylov order. It is useful to note that both the conventional and universal tails are $\mathbb{Z}_2$ parity eigenstates, so the even-odd particle number separation can be done even in the tail Hamiltonian.

For all couplings examined, eigenvalues converge exponentially quickly in increasing $N_{ZM}$. This is exemplified for $g=1.5$ in Figure \ref{approximated_tails}. We found that the regimes most sensitive to the choice of $N_{ZM}$ are when either $g_2$ is large and negative and $g_4$ is small such that $\phi$ acquires a large vacuum expectation value, or when $g_4$ is very large.  On the basis of these observation, in the following we fix $N_{ZM}=40$.

The details of this method can be grouped into three areas: i) the evaluation of the matrix elements through Monte Carlo; 
ii) generating a statistics of eigenvalues to estimate their errors,
iii) the procedure by which we extrapolate to infinite Krylov order $K=\infty$.  
We take each in turn.

\subsubsection{Evaluation of Matrix Elements} \label{MCMatelSection}

Matrix elements involving two--dimensional integrals are evaluated using the numerical global adaptive method \texttt{Cuhre}, in the implementation of the \textsc{Cuba} package \cite{Hahn:2004fe}. These matrix elements consist of $H_{1t}^{(1)}$ and $G^{11}$. Their estimated relative numerical errors are negligible, on the order of $10^{-8}$. In turn, matrix elements requiring the evaluation of at least four-dimensional integrals are computed with the importance-sampling \texttt{Vegas} method of the \textsc{Cuba} library. This is basically a Monte Carlo method, in which the sampling ensemble is updated self-consistently in a number $N_{iter}$ of iterations. Inside each iteration, a Monte Carlo-type probing of the integrand is performed with $N_{MC,0}$ evaluations. The total number of integrand evaluations is thus $N_{eval}=N_{MC,0}\cdot N_{iter}$. It is generally efficient to choose the value of $N_{iter}$ to be a few dozen \cite{Lepage:2020tgj}. We fix $N_{iter}=50$. As a general rule, larger dimensionality of the integrand comes with slower convergence with respect to Monte Carlo evaluations. 

The integrations are take place over a finite domain. The restriction in space occurs naturally because of our working in finite volume.  We however introduce  a "temporal" cutoff, fixed to be $\tau_{max}=50m_0$ for $V_{i_1\dots i_n}$  with up to $3$ indices, and $tau_{max}=40m_0$ for those $V_{i_1\dots i_n}$'s with more than $3$ indices. These choices are made so that the results are insensitive to further increasing $\tau_{max}$. This is reasonable as the integrand vanishes exponentially for large values of each of its $\tau$ arguments, which in turn follows from the asymptotics of the propagators and the presence of disconnected terms. 

\begin{figure}[h]
    \centering
 \includegraphics[draft=false,width=\columnwidth]{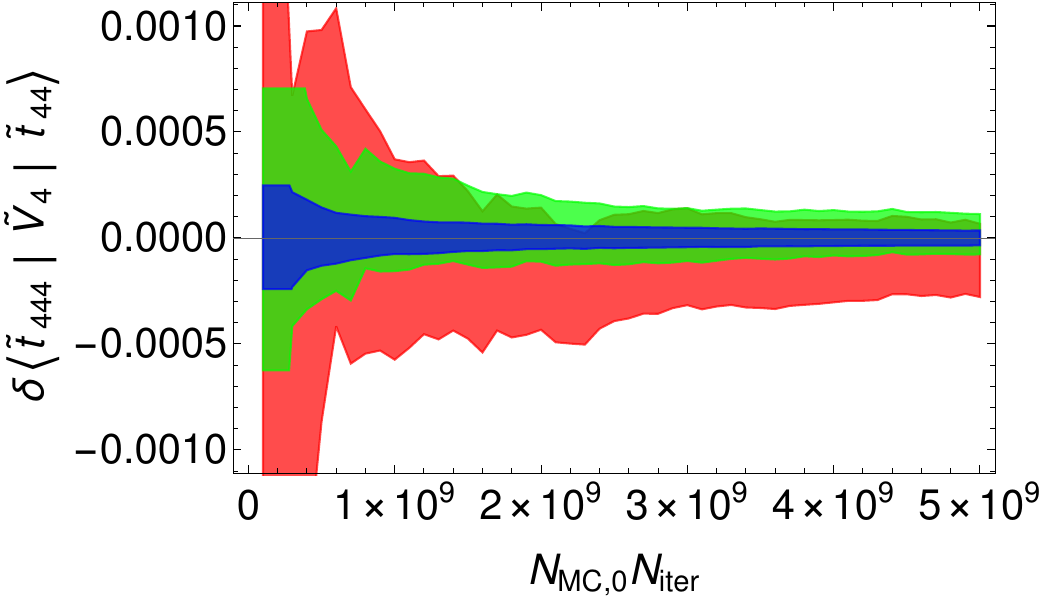}%
    \caption{
    Relative error of the matrix element $\braket{\widetilde t_{444}|\widetilde{\mathbf V}_j|\widetilde t_{44}}$ as function of the numerical integrand evaluations, in the presence of small momentum cutoffs:  $p_{max}=2\pi L^{-1}$ (blue), $p_{max}=4\pi L^{-1}$ (green), $p_{max}=8\pi L^{-1}$ (red), at volume $m_0L=10$. The differences are measured and normalized with respect to the exact matrix element obtained by linear algebra in the explicit Fock basis for these cutoffs. The importance sampling method uses $N_{iter}=50$ iterations of updating of the sampling ensemble with $N_{MC,0}=10^8$ evaluations inside each iteration. }
    \label{FigH5interr}
\end{figure}

It is easy to check the precision of numeric integrals by imposing an artificial momentum cutoff $p_{max}$. In the presence of a small enough $p_{max}$, the required matrix elements can be exactly computed in an alternative way: building up the matrix representations of the operators in eq. $\eqref{GenHtails}$ directly in a truncated Fock space with appropriate particle number and one-particle momentum cutoffs. On the other hand, such a cutoff is easily implemented in the integrals by replacing the propagator via restricting the sum in the mode expansion eq. \eqref{massivePropMode}. The dependence of numerical error on the number of evaluations for the matrix element $\braket{\widetilde t_{444}|\widetilde{\mathbf V}_j|\widetilde t_{44}}$ for various values of $p_{max}$ is shown in Figure \ref{FigH5interr}.  Other matrix elements behave similarly (or better).

\subsubsection{Statistics for third order tail states} \label{StatSubsec}
With the introduction of higher order tails, the inner product matrix becomes approximately degenerate. This issue is especially pressing in the case of universal tails. At the third tail order, the numerical precision of matrix elements is not sufficient for a direct computation, which manifests in (small) negative eigenvalues of the Gram matrix.

\begin{figure}[h]
    \centering
 \includegraphics[draft=false,width=\columnwidth]{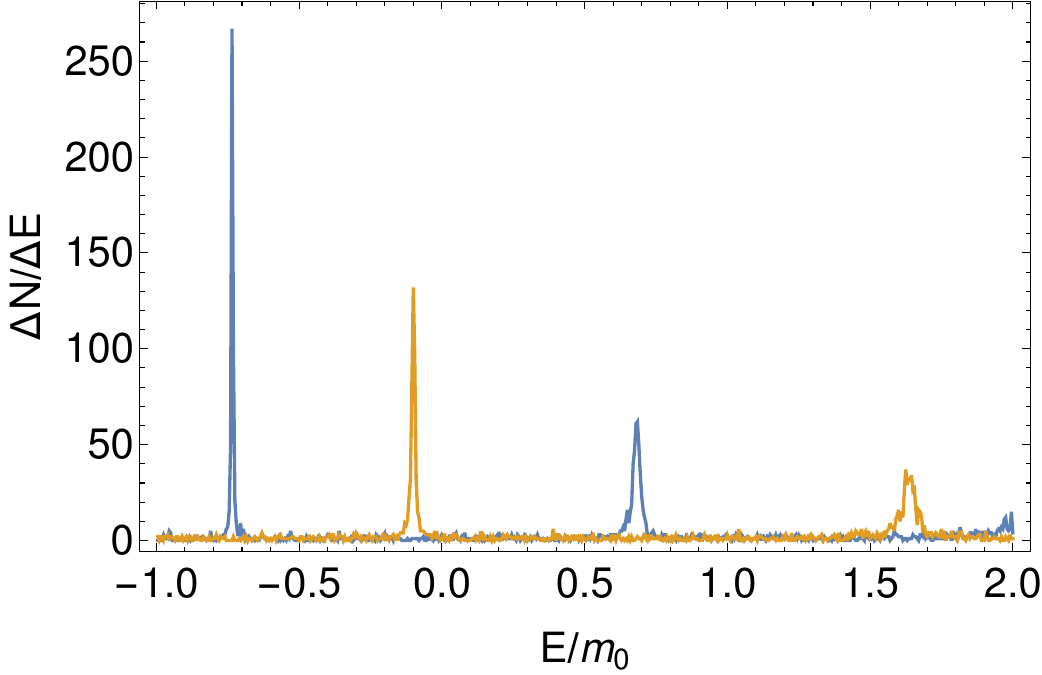}%

    \caption{
   Distribution of the lowest eigenvalues for $3$ tail orders, $g=1.4$ $m_0L=10$, based on an ensemble of $N_s=500$ Hamiltonians, sampled assuming normal distributions for the matrix elements with standard deviations from errors derived from the Monte Carlo integration routines. The histogram is based on calculating the lowest $10$ eigenvalues per parity sector for each matrix in the ensemble. Blue dots correspond to the even $\mathbb{Z}_2$ parity sector, while orange dots correspond to the odd sector. Energy bins of size $\Delta E=0.005m_0^2$ were used. }
    \label{FigEvalStats}
\end{figure}

To alleviate this problem, we generate an ensemble of $\hat{\mathbf{H}}$ and $\hat{\mathbf{G}}$ matrices using the numerical integration errors of matrix elements and repeat the eigenvalue calculation several times, projecting out the directions corresponding to numerically negative eigenvalues of the Gram matrix. The result of this analysis for a sample size $N_s=500$ is shown in Figure \ref{FigEvalStats}. For each peak, we then refine the energy bins until the maximum of the peak becomes less than $\left(\frac{\Delta N}{\Delta E}\right)_{\mathrm{treshold}}=60$. Separate Gaussian fits are then applied to the peaks. The standard deviation of these subsequent fits provide the errors of the eigenvalues at third order.

\subsubsection{Extrapolations in $K$} \label{Extrap1Sec}

After the eigenvalues from the three Krylov orders are obtained, we perform a power-law of the form
\be
f(x)=a+\frac{b}{x^c} \label{ExpolFunc}
\ee
As the input has an error, we generate a new ensemble of eigenvalues with the previously obtained errors. The fit is repeated for each member $i$ of the ensemble, providing a set of extrapolations $a_i$. $a_i$ are then averaged and their computed standard deviation provides the final error estimate (red error bars in Figure \ref{approximated_tails}). The result of this extrapolation fits nicely to previous TSM results \cite{Elias-Miro:2017tup}.

\section{Broken phase: "weak" coupling}

We start our detailed presentation of our results for the $\phi^4$ theory by focusing on its broken phase. We fix $g_2/m_0^2=-0.25$ and vary $g_4$ so that an Ising type quantum critical point is present at a relatively small quartic coupling. The spectrum and various physical quantities are well described by the perturbative treatment of fluctuations around the degenerate vacua. We consider the ground state energy, the meson mass gap, and the vacuum expectation value of the field, $\langle \phi \rangle$, before turning to a finite volume scaling analysis to determine the location of the critical point.

\subsection{Ground state energy}
\begin{figure}
\centering
\includegraphics[width=0.45\textwidth]{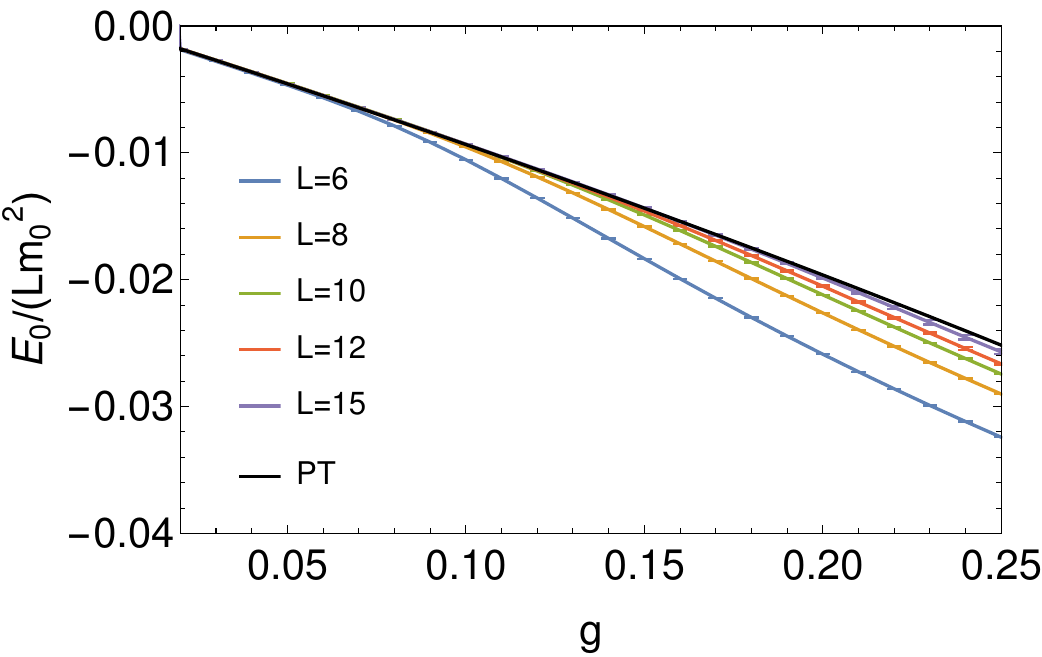}
\caption{Bulk energy measurement in the broken sector from different volumes and compared to 4 loop $L\rightarrow\infty$ perturbation theory (black).}
\label{FigBrokenBulk}
\end{figure}

Figure \ref{FigBrokenBulk} shows the coupling dependence of the bulk energy (ground state energy per unit length) as a function of the quartic coupling $g_4=g m_0^2$ for different system sizes $L$ compared to the $4$-loop perturbative result in infinite volume.
There are two kinds of physical finite volume effects that alter the finite volume eigenvalues from their $L\rightarrow\infty$ asymptotics. On the one hand, the doubly degenerate vacuum is split due to finite volume instanton effects, which is semiclassically exponentially small in the parameter $M_{K}L$, where $M_K$ is the kink mass (eq. \eqref{kinkmassSC}). On the other hand, there are finite volume corrections corresponding to the Casimir effect, which are exponentially small in the parameter $\bar{M}L$, $\bar{M}$ being the smallest mass in the spectrum. At small quartic coupling, $\bar{M}$ is the lightest meson mass $M_1$ and is about unity. In the regime $g>0.1$, $\bar{M}$ is again the kink mass and is approximately order $1/g$, before it vanishes at the critical point, giving rise to a stronger, inverse-power volume dependence. In the entire broken sector, the numerics significantly outperforms raw Fock space truncation if no extrapolations are taken into account. On the other hand, we obtain the bulk energy in agreement with perturbation theory, up to finite volume effects. 

Since the method does not rely on building up explicit Fock states, it is expected to perform especially well compared to raw truncation at larger volumes. This is indeed true. However, the numerical precision of matrix elements are observed to decrease in increasing the volume, which eventually restricts the applicability of the method in the infinite volume limit.

\subsection{Meson mass}
\begin{figure}
\centering
\includegraphics[width=0.45\textwidth]{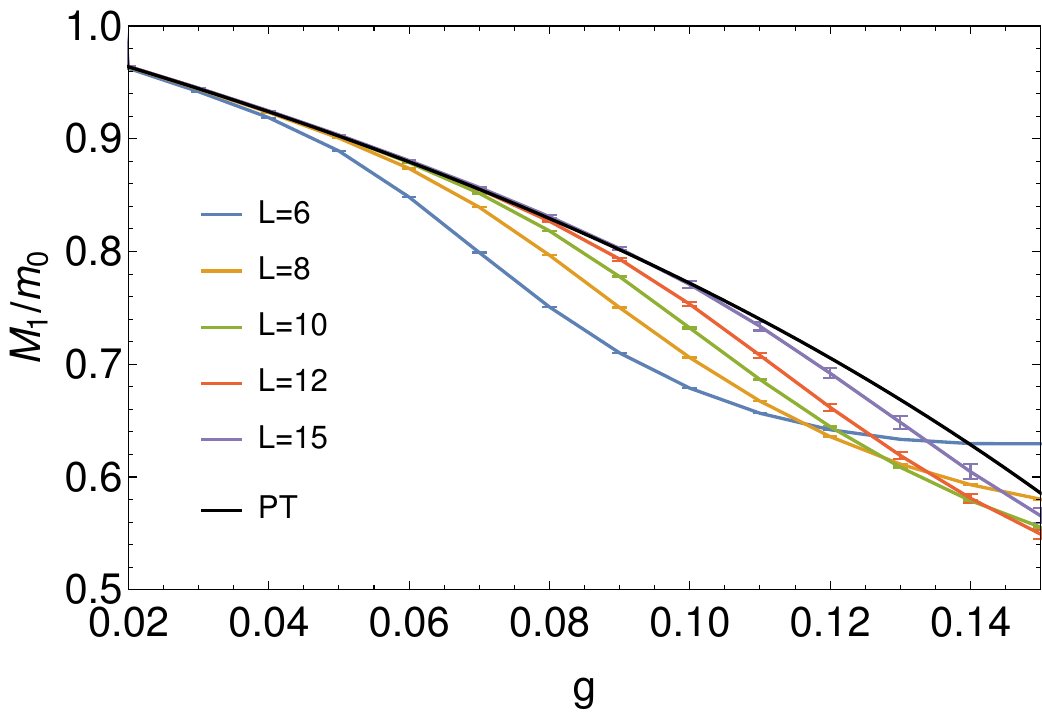}
\caption{Lowest energy gap measurement in the broken sector from different volumes and compared to 4 loop $L\rightarrow\infty$ perturbation theory (black). At couplings $g<0.15$ it corresponds to the $B1$ one-particle state, while for }
%\ref{tablech} .) }
\label{FigBrokenMesonMass}
\end{figure}

Turning to the lightest meson, $M_1$, it is again natural to compare to available perturbation theory results.  Here the finite volume effects are largely analogous to the ground state energy, apart from additional exponentially small effects due to virtual particle pairs.  $F$-terms arise due to vacuum polarisation: creation-annihilation of a virtual pair with a nontrivially winding trajectory on the cylinder, thereby scattering on the physical particle. $\mu$-terms correspond to the dissolution of the particle to virtual constituents, which reassemble into the original particle after traveling around the cylinder \cite{Luscher:1985dn,Klassen:1990ub}. Numerical results are shown in Figure \ref{FigBrokenMesonMass}. The meson disappears from the spectrum at around $g=0.15$, as then it becomes kinematically allowed to decay into a kink-antikink pair. Thus for $g>0.15$ in the broken phase, the first excited state is actually a kink-antikink scattering state. \cite{Bajnok:2015bgw}

\subsection{$\phi$ VEV}
\begin{figure}
\centering
\includegraphics[width=0.45\textwidth]{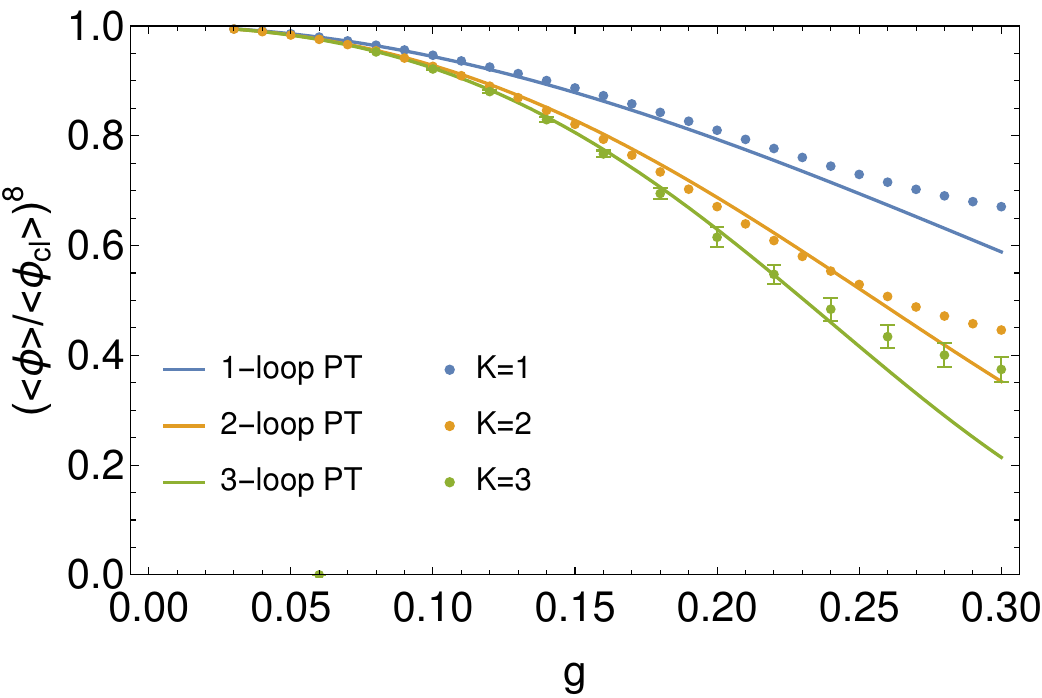}
\caption{$\left\langle\phi\right\rangle$ measured from TSM at $L=15$ and compared to $L\rightarrow\infty$ perturbation theory.}
\label{FigBrokenVEV}
\end{figure}

Since the method provides approximate eigenvectors, matrix elements of operators can be evaluated directly. Here we focus on the vacuum expectation value of the field $\phi$ in the broken sector. In finite volume, this is actually the matrix element of $\hat{\boldsymbol{\phi}}$ between the even and odd parity ground states. To single out the nontrivial quantum corrections, we normalize with the classical expectation value $\phi_{cl}=(8g)^{-1/2}$. As the critical point is in the Ising universality class,  $\langle\phi\rangle$ is expected to vanish as $(g-g_c)^{1/8}$ at the critical point, we plotted its eighth power in Figure \ref{FigBrokenVEV}, against one-, two-, and three-loop perturbation theory. 
The third order results and the corresponding error bars are again obtained by creating an ensemble of $N_S=500$ Hamiltonians and Gram matrices as described in Section \ref{StatSubsec}, projecting out any negative $\hat{\mathbf{G}}$ eigenvalues, and computing averages and standard deviations of $\langle\phi\rangle$ over the ensemble. 
We do not attempt to extrapolate the vacuum expectation value through Krylov orders. The outcome of the analysis is that universal tail sets of orders $K=1,2,3$ provide results consistent with a perturbative expansion to $1,2,3$ loops, respectively, provided the coupling is not in the vicinity of the critical point. Around the critical point, finite volume effects dominate.

\subsection{Critical point}

The broken and unbroken phases of $\phi^4$ model are separated by a second order phase transition. Classically, the phase transition happens when the parameter $g_2$ changes sign. In the quantum model, fluctuations drive the transition and it can thus be reached via tuning $g_4=g m_0^2$ for a fixed $g_2$. Despite working in finite volume, where there is no phase transition per se, truncated spectrum methods are effective in measuring the critical $g$ coupling. This is because finite volume effects in the vicinity of a critical point are well understood.

It is well known that the critical point of $\phi^4$ theory lies in the Ising universality class. Therefore, in the vicinity of the critical coupling  $g_{c}$, we can model the system as described by the Ising Hamiltonian on the cylinder, modified by relevant as well as irrelevant perturbations:
\bea
\hat{\mathbf{H}}_{eff}&=&\frac{2\pi}{L}\left(L_0+\bar{L}_0-\frac{1}{24}\right)\cr\cr
&+&L\alpha_1(g-g_{c})\left(\frac{L}{2\pi}\right)^{-1}\epsilon(z=1,\bar{z}=1)\cr\cr
&+&L(\alpha_2 g_{c}+\alpha_3(g-g_{c}))\left(\frac{L}{2\pi}\right)^{-3}\partial\bar\partial\epsilon(1,1)\cr\cr
&+&(\text{more irrelevant terms})
\eea
When the $g$ coupling is fine-tuned to the critical point $g_{c}$, there are still irrelevant perturbations affecting the finite volume spectrum. Fortunately, the irrelevant perturbations can be filtered by their symmetry properties and classified by their relevance. 

\begin{figure}
\centering
\includegraphics[width=0.45\textwidth]{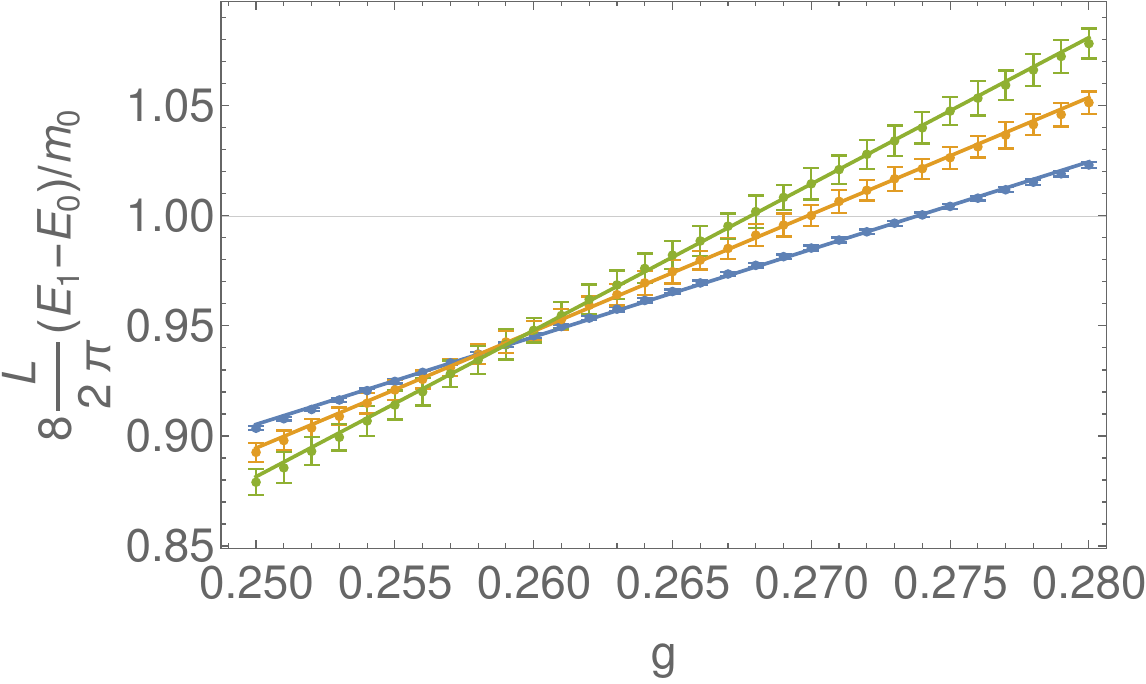}
\caption{We present here data for the determination of the critical value of $g_4$. Rescaled $E_1-E_0$ energy differences ("vacuum splitting") for volumes $L=6$ (blue), $L=8$ (orange) and $L=10$ (green). The corresponding value in the Ising model is shown with a horizontal black line.}
%\ref{tablech} .) }
\label{FigCritPoint}
\end{figure}

In the first order of conformal perturbation theory, it is simple to estimate the corrections by the irrelevant terms. Since the operators are multiplied by $L^{-2h+1}$, with $h$ being the chiral dimension, the least irrelevant operators dominate the large-volume spectrum, while more irrelevant ones are suppressed by powers of $L$. In our analysis, we approximate the perturbed Ising model by keeping the relevant thermal perturbation $\epsilon$ as well as the descendant $\partial\bar\partial\epsilon$. We restrict our attention to the energy difference between the even and the odd ground states and consider the volume range $L=6$ to $L=10$. The rescaled volume-dependent gaps as function of the coupling, for different volumes $L$, are depicted in Figure \ref{FigCritPoint}. The linear coupling dependence of the gap is consistent with the assumption that in the vicinity of the critical point, the effective Ising couplings can be approximated as linear functions of the quartic  $\phi^4$ coupling $g$ .

In the absence of irrelevant terms, one would conclude that the transition occurs at the point where the rescaled lines cross. But this has to be corrected by the effect of the irrelevant term.
To improve the accuracy, we fit linear functions $f_L(g_4)=a_L + b_L g$
to the datasets in Figure \ref{FigCritPoint}. Equating these empirical gap functions with their theoretical approximations from conformal perturbation theory, and separating coefficients of powers of $g_4$, we obtain a (generally overdetermined) system of linear equations for the critical coupling $g_{4c}$ as well as the parameters $\alpha_i$. There are multiple ways to reduce the resulting system into a uniquely determined one, which provides slightly different results for $g_c$. The variation in these results provides an error estimate.
We so obtain an estimate of the critical point to be $g_c=0.2645\pm0.002$.
 The order of performing the extrapolation in tail orders versus taking the difference of eigenvalues does not affect this final estimate.
 
\section{Unbroken Phase and strong coupling regime}
Now we turn to the unbroken phase. Here we choose $g_2=m_0^2/2$ and vary $g_4$. As per the Chang duality, the critical point appears at $g\approx2.7$, but higher orders of perturbation theory breaks down noticeably for below the critical point at $g\approx0.3$. In this phase, there is a single bosonic excitation in the spectrum.

\subsection{Empirical study of cutoff dependence and tentative improvement on errors } \label{ExtrapSec2}
\begin{figure}[h]
    \centering
     \includegraphics[draft=false,width=\columnwidth]{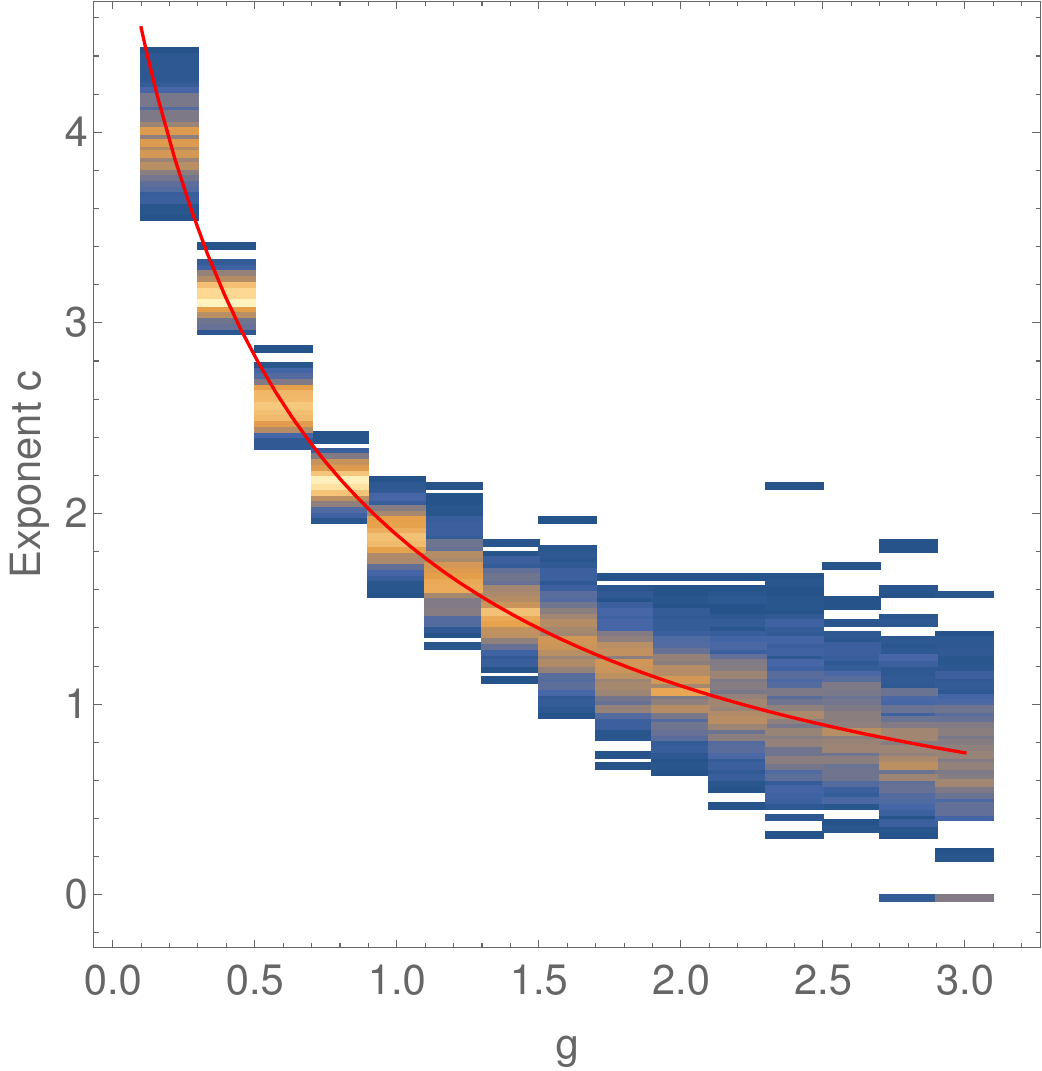}

    \caption{Fitted power law exponent to statistical ensemble of ground state eigenvalues}
    \label{FigFitStat}
\end{figure}

Before turning to the measurement of the ground state energy and the mass gap, let us study the tail order extrapolation in more detail. It is difficult to determine the extrapolating function eq. \eqref{ExpolFunc} from first principles. However, it is interesting to see how the fitted power law exponent $c$ varies as function of the coupling. We find that performing the fit without any restrictions on the parameters $a,b,c$, the resulting function $c(g)$ is well described by a simple function form
\be
c(g_4)=c_1+ \frac{c_2}{g+c_3}. \label{FitCForm}
\ee
We fitted the parameters $c_i$, $i=1,2,3$ to the distribution of exponents in Figure \ref{FigFitStat}, obtained by performing the fit \eqref{ExpolFunc} on a set of $N_S=500$ samples generated on the basis of the eigenvalue uncertainties.

The proposed functional form eq. \eqref{FitCForm} is empirical and likely model-dependent. At the same time, we find that using the exponent $c(g)$ determined from eq. \eqref{FitCForm} in fitting our data with eq. \eqref{ExpolFunc} results in good agreement with the data of Ref.\cite{Elias-Miro:2017tup}.
In the following plots, we report the results and errors both by fitting directly every sample with eq. \eqref{FitCForm} and alternatively, performing the fit with $c$ constrained to be its eq. \eqref{FitCForm} value.

\subsection{Ground state energy}
\begin{figure}[h]
    \centering
 \subfloat[Coupling $g_4=0.4m_0^2$ ($M=0.937$, $\mathcal{E}=-0.00682$).  The corresponding result for the free boson with mass $m_0$ is shown for reference with a gray dashed line.]{\includegraphics[draft=false,width=\columnwidth]{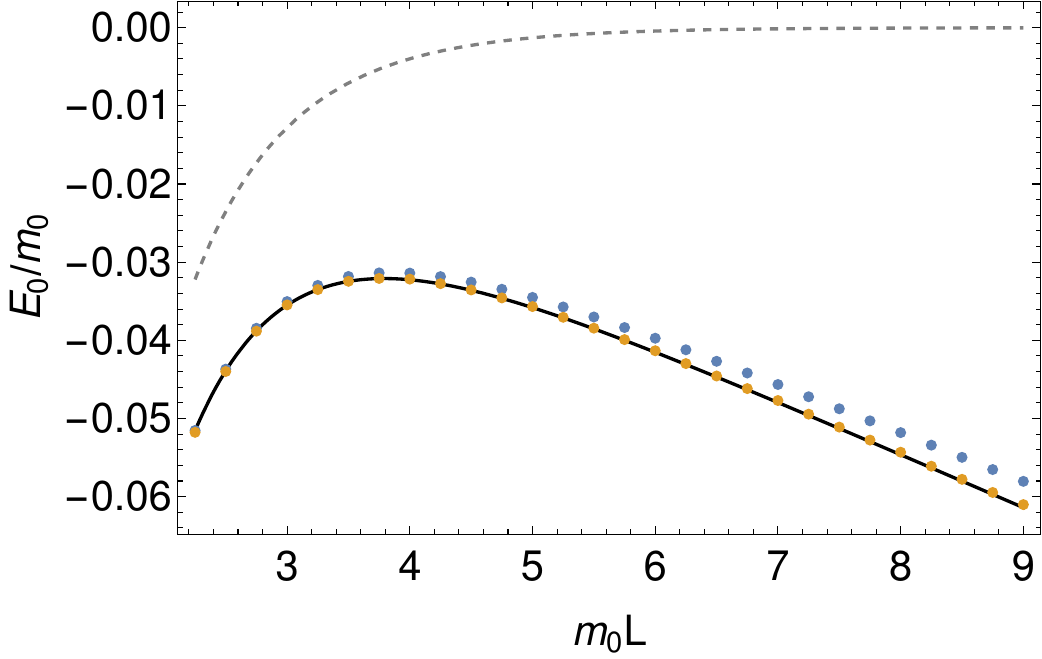}}\\
 \subfloat[Coupling $g_4=0.8m_0^2$ ($M=$, $\mathcal{E}=$). $K=3$ results shown with green dots]{\includegraphics[draft=false,width=\columnwidth]{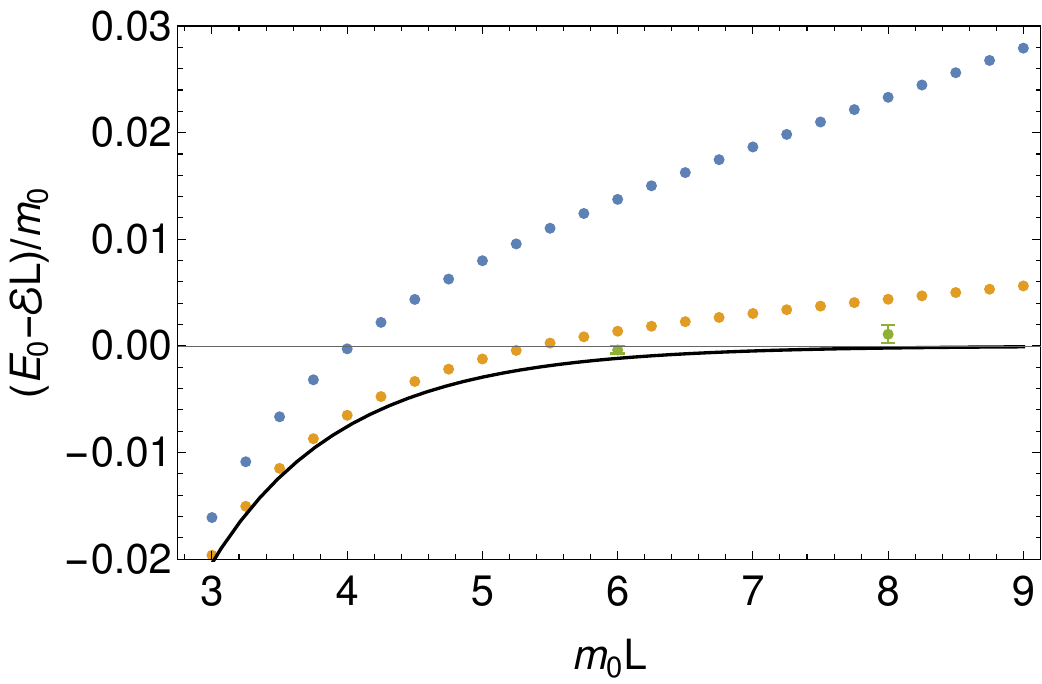}}

    \caption{Volume dependence of the ground state energy in the unbroken sector, at Theoretical behavior including the leading Lüscher correction is shown with a continuous black curve . The results for the $K=1$ and $K=2$ universal tail basis computations are shown with blue and orange dots, respectively.}
    \label{FigVoldep}
\end{figure}

The truncated spectrum method (TSM) in its equal time formulation is essentially a finite volume method. The form of leading terms in the large volume expansion of the ground state energy are known. In case of an unbroken vacuum and a single type of massive particle, It has the structure
\be
E_0(L)=\mathcal{E}_0 L-\frac{M}{2\pi}\intop_{-\infty}^\infty \cosh u\: e^{-ML\cosh u}du + O(e^{-2ML}) \label{Firstluschereq}
\ee
In eq. \eqref{Firstluschereq}, $\mathcal{E}$ is the bulk energy (density), $M$ is the mass of the excitation, while the integral term is the first Lüscher (leading exponential) correction. The value of the bulk energy and the mass is taken from the numerics, but knowledge of these parameters provides the volume dependence of the ground state energy $E_0$ for intermediate to large volumes. This is checked against TSM numerics in Figure \ref{FigVoldep}. 

\begin{figure}[h]
    \centering

\includegraphics[draft=false,width=\columnwidth]{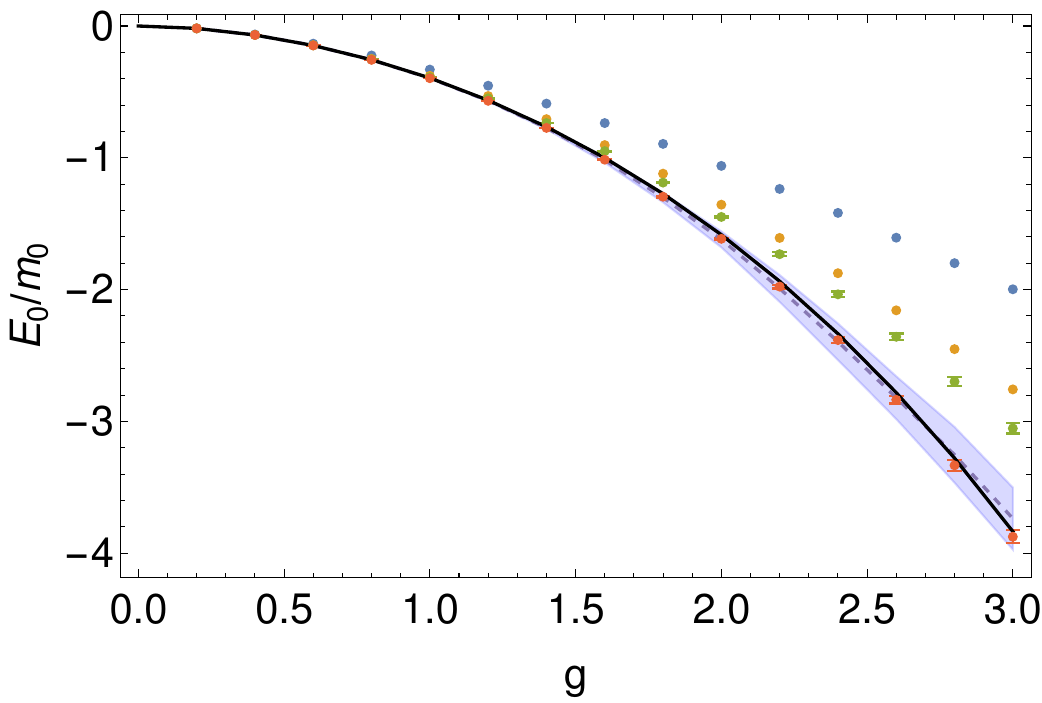}\\%

\caption{Ground state energy in the unbroken sector, $m_0L=10$. 
Show are our computations of $E_0/m_0$ as function of coupling $g$ compared to \cite{Elias-Miro:2017tup} (black curve). The numerical results corresponding to $K=1,2,3$ universal tail sets are marked by blue, orange and green dots, respectively. The power-law extrapolation is shown with a dashed purple line, with errors given by the purple shaded area. Extrapolation instead by constraining the exponent $c(g_4)$ to its value given by eq. \eqref{FitCForm} is marked by the data with red dots.}
    \label{FigE0Unbroken}
\end{figure}

We show the computed ground state energy at volume $m_0L=10$ in the unbroken sector, as the function of the coupling $g$, in Fig. \ref{FigE0Unbroken}.

\subsection{Mass gap}
\begin{figure}[h]
    %\centering
     \subfloat[][perturbative series truncated at various orders, compared to the Borel resummation of the perturbative series (\cite{Serone:2018gjo}). ]{\includegraphics[draft=false,width=\columnwidth]{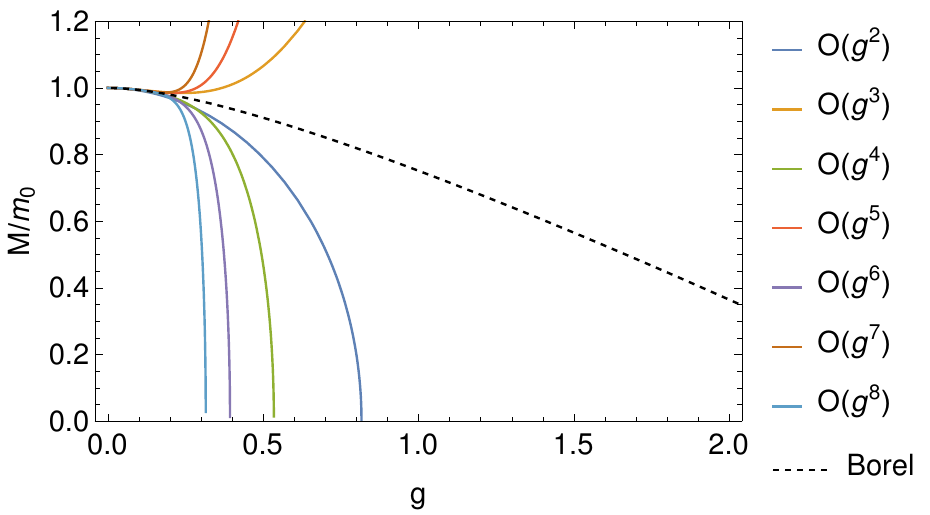} }\\
 \subfloat[][Krylov TSM numerical data at $m_0L=10$, compared to \cite{Elias-Miro:2017tup} (reference data shown as black curve).  Numerical results corresponding to $K=1,2,3$ universal tail sets correspond to blue, orange and green dots, respectively. The power-law extrapolation is shown with a dashed purple line, with errors given by the shaded area. Extrapolation with the exponent $c$ fixed by eq. \eqref{FitCForm} results in the red dots.]{\includegraphics[draft=false,width=\columnwidth]{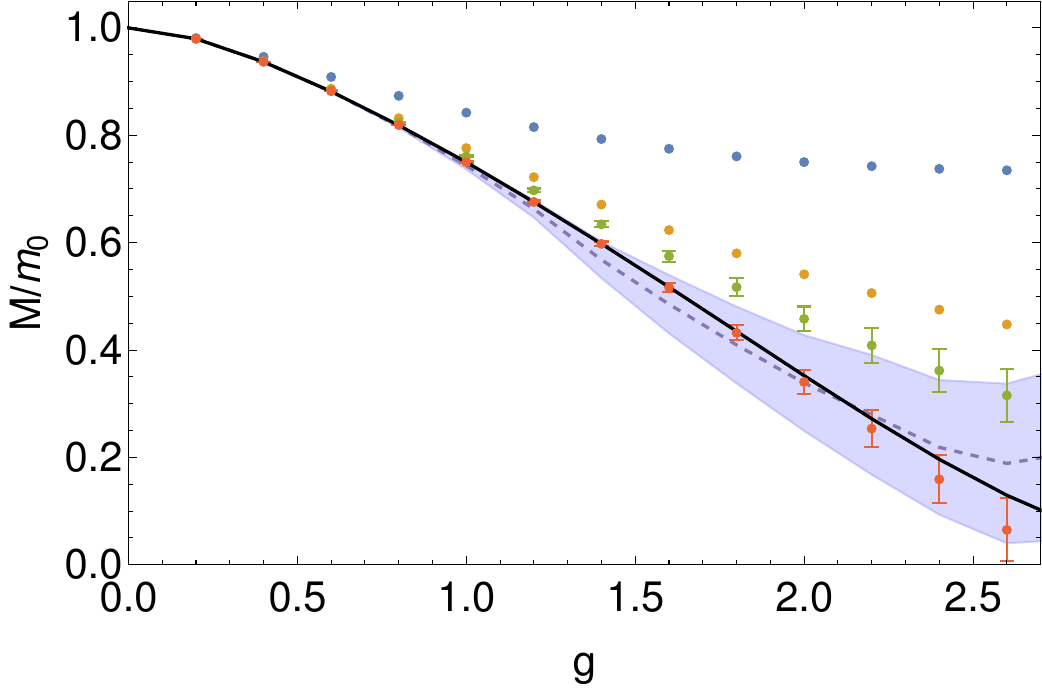}}%

    \caption{Mass gap in the unbroken sector, as function of coupling $g$. }
    \label{FigMassdep}
\end{figure}

In the top panel of Figure \ref{FigMassdep} we show the validity of the perturbative expansion eq. \eqref{phiE0pertser}. Consistently with the asymptotic nature of the perturbative expansion, higher-order results come with a more restricted domain of validity.  It is interesting to see how this compares to Borel resumming the perturbative series. While conventional perturbation theory breaks down spectacularly at around $g=0.2$, our Krylov-enabled TSM is able to follow the resummed result over an extended range of couplings, as shown on the bottom panel.

\section{Residual error}
\begin{figure}[h]
\centering
\includegraphics[width=0.45\textwidth]{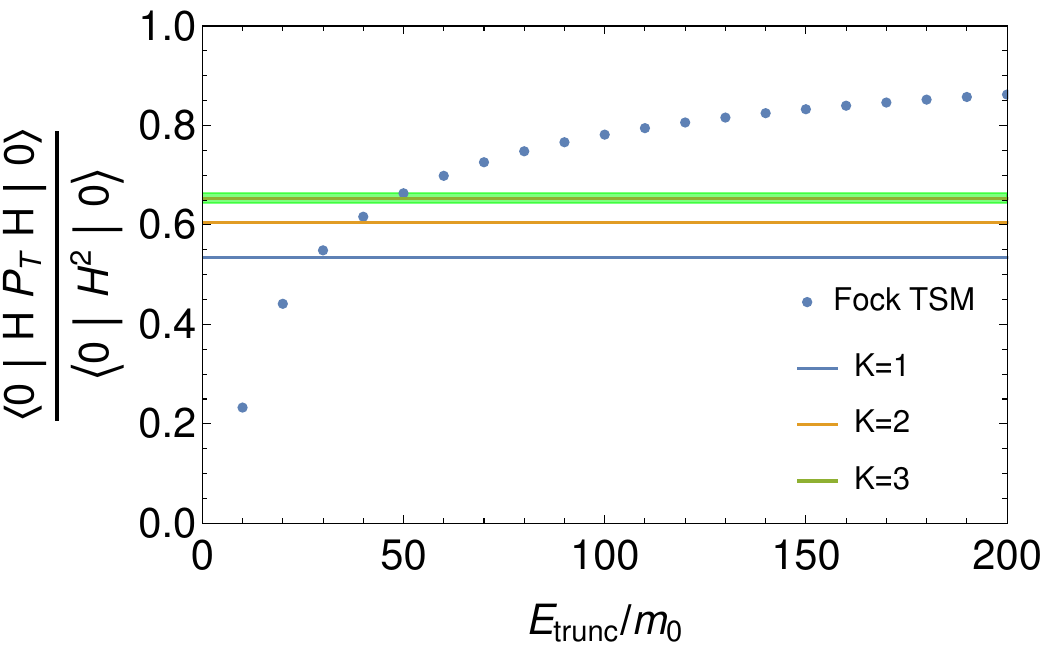}
\caption{Limitations of raw truncation (unbroken sector): squared norm of the projection of $H\ket{0}$ to a truncated Hilbert space, normalized by its squared norm. Here $\ket{0}$ is the unperturbed vacuum as a function of the TSM energy cutoff $E_{trunc}$, $L=10$, $g_4=1$. $\hat{\mathbf{P}}_T$ is a partial resolution of the identity in the truncated Hilbert space.  Also shown are the squared norms of the projections to the universal tail spaces of 1st (blue) 2nd (orange) and 3rd (green, with light green uncertainty) orders.}
\label{H2norm0}
\end{figure}
\begin{figure}[h!]
    \centering
    \subfloat[Norm of the residual error for the ground state as the function of the coupling $g$ in the unbroken phase. The results for the $K=1,2$ universal tails are shown with continuous blue and orange lines, respectively. Raw truncation with different energy cutoffs yield the set of dots at the top of the figure ($E_{cut}=5$: light gray, $E_{cut}=7$: gray, $E_{cut}=9$: black).]{\includegraphics[draft=false,width=\columnwidth]{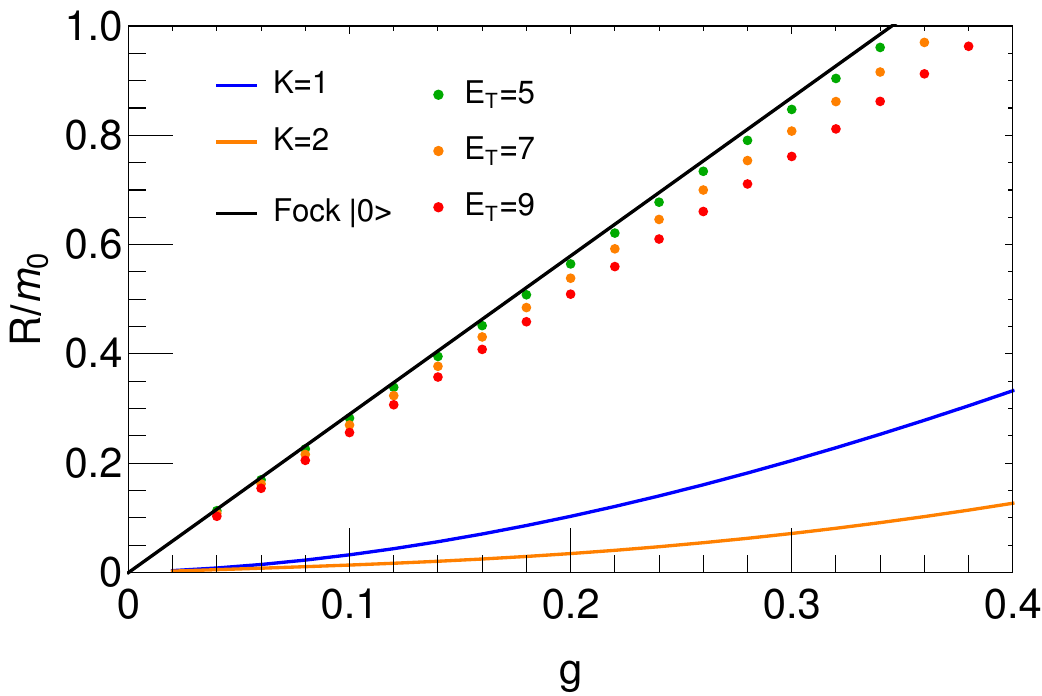}}\\
    \subfloat[Norm of the residual error as function of the Fock TSM energy cutoff, for various values of $g$. ]{\includegraphics[draft=false,width=\columnwidth]{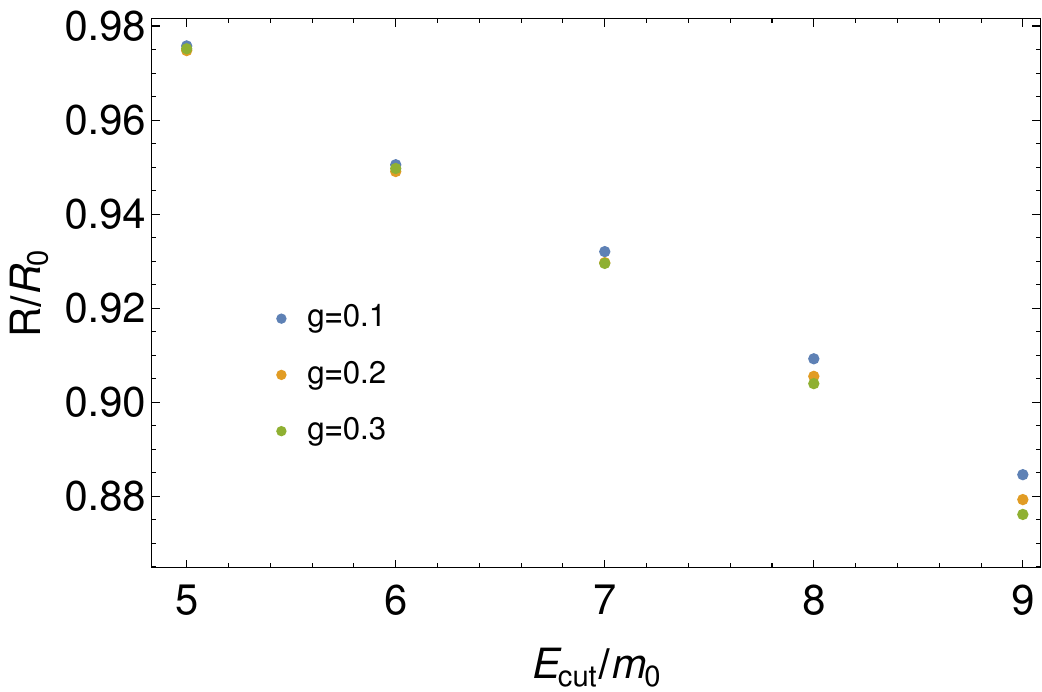}}

    \caption{Norm of the residual error for the ground state. We point out that the square of the \textit{full}, infinite-dimensional QFT Hamiltonian is sandwiched between the numeric eigenstates, making the computation difficult for large energy cutoffs. The largest raw energy cutoff corresponds to a $104$ dimensional truncated subspace.}
    \label{FigResNorm}
\end{figure}
In performing any numerical analysis, it is vital to have as much control on the error as possible. In truncated spectrum methods, a commonly adapted measure of precision is to compute the eigenvalues at special points where the results are known "exactly" (with much more precision) from independent approaches.  While this can give an idea on the expected performance of the method at large, it has a number of significant drawbacks:
\begin{enumerate}
    \item Information comes from special points, which are similar at best, but usually not identical to the actual point of interest.
    \item Moreover, it is generally hard to quantify the distance from the reference point in theory space, and make a connection with TSM precision.
    \item Finally, the error in the eigenvalues is a rather forgiving measure of the error, especially when actual eigenvectors are used to measure a  quantity of interest. 
\end{enumerate}

In the literature of iterative eigensolver methods, it is customary to introduce the norm $R$ of the residual error in the following way:
\be
R^2=\braket{\psi_{comp}|(\hat{\mathbf{H}}-E_{comp})^2|\psi_{comp}}.\label{reserrordef}
\ee
Note the square inside the expectation value. In eq. \eqref{reserrordef}, $\psi_{comp}$ is the computed (truncated) eigenvector, but $\hat{\mathbf{H}}$ is the entire Hamiltonian without truncation. Without the square, one would simply get zero, as $\braket{\psi_{comp}|\hat{\mathbf{H}}|\psi_{comp}}=E_{comp}$.

Inserting a partial resolution of the identity between the Hamiltonians results in a strong cutoff dependence, as depicted in Fig. \ref{H2norm0}. 
As the action of the operator $\hat{\mathbf{H}}$ changes particle number by at most $4$, it is easy to compute the projected matrix element to high energy cutoffs by introducing an additional particle number cutoff on the truncated space. (Here, for the raw truncation points, we do not separate the zero mode.) It is interesting to note the would-be dimensionality of the full Fock space with the corresponding energy cutoff $E_{cut}$. The required dimensionality would be $D=6.5*10^5$ for first order tails, about $D=2.3*10^7$ for second order tails, and $D\approx10^9$ for third order tails, giving a rough estimate of the raw truncation dimensionality needs to reproduce the precision of the Krylov tail numerics. This is impressive in itself, but still suffers from a significant deviation from the exact result.
Fortunately, our approach is naturally suited to calculate the matrix elements of $\hat{\mathbf{H}}^2$ directly, without any inherent truncation.

Let us note that when $\delta E= E_{comp}-E_{exact}$ is small, the correction obtained from substituting $E_{comp}$ with $E_{exact}$ in eq. \eqref{reserrordef} will be $O(\delta E^2)$, which is negligible compared to the effect coming from the square of the Hamiltonian. Intuitively, the residual error vector estimates the error resulting from acting with $\hat{\mathbf{H}}$ on the truncated eigenvector instead of its true eigenvector with eigenvalue $E_{exact}$. This measure can be particularly relevant in measurements which involve acting multiple powers of $\hat{\mathbf{H}}$ to an approximate eigenvector, something that happens in computations involving the real time dependence of observables.
At small coupling, the difference of the eigenvalues between the interacting and the free theory are indeed minuscule. Therefore, relying only on the precision of the eigenvalues is deceptive when the eigenvectors are needed. The residual norm provides a much more sensitive measure of the error.

We show the residual error corresponding to the ground state energy in Fig. \ref{FigResNorm}. A reference computation was also performed in a conventional massive Fock-space truncated space with an energy cutoff (with no separated zero mode). In that basis, we evaluated the matrix elements of $\hat{\mathbf{H}}^2$ by expressing the squared interaction terms with normal ordered ones and performing the nontrivial spatial integrals where necessary. In this case, the residual error decreases approximately linearly with the energy cutoff, at least for the small cutoffs examined. Extrapolating this linear tendency again gives an idea on the Fock space dimensionality required to reproduce the precision of the Krylov computation.

\section{Conclusion and Discussion}

In this paper we have implemented a Krylov-subspace method tailored to computing the low lying spectrum of quantum field theories, based on the iterative procedure originally described in \cite{PhysRevA.28.2151}. It can also be seen as a systematic improvement to the renormalization
group method of \cite{Hogervorst:2014rta,Rychkov:2014eea,Elias-Miro:2017tup}. We have demonstrated the method on the unbroken and broken sector of the two-dimensional $\phi^4$ model.

The method involves the computation of multi-point correlators in the ``unperturbed" theory. In the present case, this translates to the numerical calculation of Feynman diagrams. Therefore we see this method as a bridge between perturbative diagrammatics and truncated spectrum methods (TSMs).

Throughout this work we focused on the simplest spectral quantities: the ground state energy and the difference between the two lowest two energy levels. However, we remark that other physical quantities are accessible to the method by modest and often obvious modifications.
We succeeded in reproducing the bulk energy and the mass gap of the $\phi^4$ model in the unbroken sector in agreement with Borel resummation and other TSM results. Remarkably, in a wide range of couplings, a truncated Hamiltonian of only dimension four (per parity sector) was sufficient to provide a good estimate over a wide range of couplings. In the unbroken sector, this can be contrasted to Borel resummation, which utilizes very similar Feynman diagram computations in an entirely different way. We believe it is interesting that even though tail matrix elements suffer from the same asymptotic behavior as the perturbative series, the diagonalization of the Hamiltonian involving these same matrix element efficiently extracts convergent physics from these quantities, in analogy to Borel resummation.
In the unbroken section we have also reproduced the correct finite-volume behavior of the ground state energy and demonstrated a novel way to measure the error for TSM eigenvectors. 

Using our technique, we also analyzed the $\phi^4$ model in the broken sector.  We compared the bulk energy to raw TSM truncation and a 4-loop perturbative expansion and found reassuring agreement. We also used our code to provide a precise estimate of the critical point in the broken sector, $0.2645\pm0.002$.

While we have not pursued it here, the method indicates promising features for possible future real-time dynamics applications. Since the same universal tail basis provides a very good representation of lowest energy states over an entire range of couplings, it appears to be a natural basis to study quenches with both suddenly and gradually changing couplings.

This combination of Hamiltonian truncation and Feynman diagrammatics is a new method and our presentation leaves several issues that will require further elaboration upon in future work.  Perhaps the most pressing issue is the extrapolation of data through Krylov iteration orders. At the moment we lack a clear physical understanding to motivate our extrapolation functions.  We instead relied on the empirical quality of the fits and the consistency of the results with existing reference values from the literature.  We do however discuss convergence in Krylov-like methods in Appendix A.

Our method easily generalizes to linear sigma models in two dimensions. In the unbroken sector, this would entail trivial modifications of the Feynman integrals by symmetry factors. From the analysis of the broken symmetry phase at strong coupling, it is presumably possible to probe nonlinear sigma models as well.

This method is easily extendable to higher dimensions and will serve as the next test of the method.  For the $\phi^4$ theory in $2+1$d all that would be required would be to replace the form of the propagator.   It may however be advantageous to implement the Feynman integrals in momentum space, as the real space propagator suffers from stronger singularities in higher dimensions.

The transition to momentum space could cause certain complications in the present setup, where the partial cancellation of disconnected pieces is implemented at the level of the integrand, leading to better temporal convergence. 
Strictly speaking the explicit projections leading to the extra disconnected pieces are not necessary. We might have omitted them at the cost of overlaps between the low-energy basis states and the tails.  We opted for implementing the projectors in coordinate space as they act as a temporal cutoff, improving the convergence of matrix element integrals and decreasing their magnitude. Such cancellations in coordinate space do not obviously translate to momentum space.  Fortunately, the structure of the disconnected pieces is known and can be taken into account by other means even if one opts for keeping them.  

\section*{Acknowledgements}
We would like to thank Z. Bajnok, M. Serone, G. Takács, D. Horváth, A. Weichselbaum, N. Aryal, and S. Sarkar for useful discussions, and especially to J. Elias-Miro for providing TSM data related to \cite{Elias-Miro:2017tup}.
M.L. and R.M.K. was supported by the U.S. Department of Energy, Office of Basic Energy Sciences, under Contract No. DE-SC0012704.

\appendix

\section{Relation to conventional Krylov subspace methods} \label{OtherKrylovRel}

The original idea of Krylov subspace methods is to compress the essence of an $N\times N$ dimensional operator $\mathbf{A}$ into an $m\ll N$ dimensional operator $\mathbf{A}^{(K)}$ acting on $\mathcal{K}$. 
The Krylov subspace $\mathcal{K}$ is spanned by the set $\{\mathbf{A}^k\ket{r},\:0\leq k < m\}$. A block Krylov subspace of block size $n$ is obtained by the span of the set $\{\mathbf{A}^k\ket{r_i},\:0\leq k < m,1\leq i\leq n, n>1\}$ for linearly independent vectors $r_i,\:1\leq i \leq n$.
The solution to a linear system $\mathbf{A}\ket{x}=\ket{b}$ can be approximated by vectors ${\ket{x_K}+\ket{\delta x^{(m)}}}$ for which $\mathbf{A}\ket{\delta x^{(m)}}\perp \mathcal{K}$. Eigenpairs (eigenvalue-eigenvector pairs) of $\mathbf{A}$ are approximated by eigenpairs  of $\mathbf{A}^{(K)}$. 

The case relevant to us is when the operator $\mathbf{A}$ is diagonalizable, i.e. related to a symmetric matrix $\bar{\mathbf{A}}$ by a similarity transformation $\mathbf{A}=\mathbf{Q}^{-1}\bar{\mathbf{A}}\mathbf{Q}$. If $\bar{\mathbf{A}}$ is a bounded operator with a spectrum including at most a finite number $\nu$ of discrete negative eigenvalues, 
$\{\lambda^-_i\}^\nu_{i=1}$, 
the residual norm of the approximate solution after iteration $m$ is bounded by the formula \cite{SaadSchultz,Elman1982IterativeMF}
\bea
&&\|\mathbf{A}\delta x^{(m)}\|\leq\|\mathbf{Q}^{-1}\|\|\mathbf{Q}\ket{b}\|\cr\cr
&&\times\left(\frac{\lambda_{max}^+-\lambda_{min}^+}{\lambda_{max}^++\lambda_{min}^+}\right)^{m-\nu}\prod_{i=1}^\nu\frac{\lambda_{max}^+-\lambda_i^-}{|\lambda_i^-|},\label{genBoundGMRES}
\eea
where $\lambda_{max}^+$ and $\lambda_{min}^+$ are respectively the largest and smallest positive eigenvalues of $\bar{\mathbf{A}}$.

In our setup, the approximate tails are constructed iteratively from the linear system in eq. \eqref{IterLinEq}, and they are subsequently used as a variational basis for the determination of eigenvalues. Translating to the notation of the previous paragraph, in eq. \eqref{IterLinEq} we would replace $\mathbf{A}$ with $\boldsymbol{\mathcal{T}}\equiv1-\hat{\mathbf T}=(\hat{\mathbf{H}}_{0,hh}-E)^{-1}(\hat{\mathbf{H}}_{hh}-E)$ and take $\ket{b}\equiv{\ket{t_{l,0}}}$ and $\ket{x}\equiv\ket{T_l}$. One can then see the basis vectors in eq. \eqref{tildet} as linear combinations of vectors $\boldsymbol{\mathcal{T}}^i\ket{t_{l,0}}$, $0\leq i \leq k-1$. However unlike $\mathbf{A}$ above, in our case the operator $\boldsymbol{\mathcal{T}}$ is infinite dimensional and not bounded.

Due to the unbounded character of $\boldsymbol{\mathcal{T}}$, It is difficult to make general statements about the convergence properties of the Krylov iteration (although recently progress has been made \cite{Caruso:2019,Caruso:2020}). In particular, the bound \eqref{genBoundGMRES} formally gets ill-defined as there is no $\lambda_{max}^+$.  In any case it is reassuring that even in this case the Krylov subspace only becomes exactly degenerate in a finite number of steps if the exact solution has been found. Indeed, if one can write $\boldsymbol{\mathcal{T}}^n\ket{t_{l,0}}=\sum_{k=0}^{n-1}c_k\boldsymbol{\mathcal{T}}^k\ket{t_{l,0}}$ with some coefficients $c_k$, then the exact tail is given as $\ket{T_l}=c_0^{-1}(\boldsymbol{\mathcal{T}}^{n-1}-\sum_{k=0}^{n-2}c_{k+1}\boldsymbol{\mathcal{T}}^k)\ket{t_{l,0}}$. 
The operator $\boldsymbol{\mathcal{T}}$ is diagonalizable with $\mathbf{Q}=(\hat{\mathbf{H}}_{0,hh}-E)^{1/2}$.  We will assume that $E$ is strictly smaller than any eigenvalue of $\hat{\mathbf{H}}_{0,hh}$. It follows that $\mathbf{Q}$ is positive. Moreover, when the parameter $E$ is chosen to be the ground state energy in any symmetry sector respected by both $\hat{\mathbf{H}}$ and $\hat{\mathbf{H}}_0$, then the symmetrized operator $\bar{\boldsymbol{\mathcal{T}}}=\mathbf{Q}\boldsymbol{\mathcal{T}}\mathbf{Q}^{-1}$ is actually positive definite (in the given symmetry sector), as follows from the variational principle.

For concreteness, let us first consider the restricted version of eq. \eqref{IterLinEq} in the presence of both a momentum cutoff $\Lambda$ and a projection of the oscillator Hilbert space to a subspace spanned by Fock states with at most $N$ particles (the zero mode is not affected by this cutoff). Then we have a restricted problem
\be
\boldsymbol{\mathcal{T}}_{N,\Lambda}\ket{x_{N,\Lambda}}=\ket{b_\Lambda},
\ee
with a bounded operator $\boldsymbol{\mathcal{T}}_{N,\Lambda}$ (the truncated Hilbert space is finite dimensional). Two comments are in order. One is that in the limit $\Lambda\rightarrow\infty$, the operator  
$\boldsymbol{\mathcal{T}}_{N,\Lambda\rightarrow\infty}\equiv\boldsymbol{\mathcal{T}}_{N}$ is, according our numerics, bounded, with a numerical behavior $\lambda_{max}^{N,\Lambda}\rightarrow \lambda_{max}^{N}+O(\Lambda^{-2})$. The second is that the limit $\boldsymbol{\mathcal{T}}_{N}$ is also strictly positive definite. The latter can be seen as a consequence of the low-energy eigenvectors of $\mathbf{H}_{0,ll}$ having a finite weight in the ground state of $\mathbf{H}$, while this ground state is separated from the rest of the spectrum by a gap. 
In the following we assume the existence of the above limit and suppress the $\Lambda$ index. (In cases when the limit does not converge, it is still possible to reintroduce the cutoff. Then the statements below remain valid in the presence of a finite momentum cutoff.)

The exact solution to the equation $\boldsymbol{\mathcal{T}}_{N}\ket{T_{l,N}}=\ket{t_{l,0}}$ can be written as $\ket{T_{l,N}}=\ket{T_l}+\ket{\delta T_{l,N}}$, where $\ket{\delta T_{l,N}}$ is the difference between the regularized and the full ($N\rightarrow\infty$, cutoff $\Lambda$) solution. It is safe to assume that $\left\Vert \ket{\delta T_{l,N}}\right\Vert \rightarrow0$ for large N, as we expect to be able to approximate the eigenvectors arbitrarily well by systematically enlargening the truncated Hilbert space.

At iteration $m$, the vectors $\boldsymbol{\mathcal{T}}^{k}\ket{t_{l,0}}$, $0\leq k\leq m$ are identical to the vectors $\boldsymbol{\mathcal{T}}_{N=4\left(m+1\right)}^{k}\ket{t_{l,0}}$. This is because we build our tails on top of the oscillator vacuum, so $\ket{t_{l,0}}$ contains at most $4$-particle states; on the other hand, the application of $\boldsymbol{\mathcal{T}}$ changes the particle number by at most $4$. Therefore, at iteration $m$, the approximate solution $\ket{T_l^{\left(m\right)}}$ can be written as
\bea
\ket{T_l^{\left(m\right)}}&=&\ket{T_l}+\ket{\delta T_l^{\left(m\right)}}=\ket{T_{l,4(m+1)}}+\ket{\delta T_{l,4(m+1)}^{\left(m\right)}}\cr\cr
&=&\ket{T_l}+\ket{\delta T_{l,4(m+1)}}+\ket{\delta T_{l,4(m+1)}^{\left(m\right)}},
\eea
so $\ket{\delta T_l^{\left(m\right)}}=\ket{\delta T_{l,4(m+1)}}+\ket{\delta T_{l,4(m+1)}^{\left(m\right)}}$. In turn, we can bound the error of the full iteration by looking at the regularized problem.
Analogously to eq. \eqref{genBoundGMRES}, we obtain the bound
\bea
\|\boldsymbol{\mathcal{T}}\ket{\delta T_{l,N}^{(m)}}\|&\leq&\|\mathbf{Q}^{-1}\|\|\mathbf{Q}\ket{t_{l,0}}\|\left(1-\frac{f(m)}{m}\right)^m;\cr\cr
f(m)&=&2m\frac{\lambda_{min,m}}{\lambda_{min,m}+\lambda_{max,m}},\label{OurBound}
\eea
 where $\lambda_{min,m}$ and $\lambda_{max,m}$ are the extremal eigenvalues of the operator $\bar{\boldsymbol{\mathcal{T}}}_{N=4m+4}$with $\bar{\boldsymbol{\mathcal{T}}}_{N}=\mathbf{Q}\boldsymbol{\mathcal{T}}_{N}\mathbf{Q}^{-1}$. From the form of $f(m)$ in eq. \eqref{OurBound} it follows that $f(m)\leq m$. Depending on the large-$m$ behavior of $f(m)$, we can distinguish three cases:
 \begin{enumerate}
 \item $f(m)\rightarrow\infty$:  In this case the vanishing of the residual norm (with respect to $\boldsymbol{\mathcal{T}}$) is guaranteed, with rate given by eq. \eqref{OurBound}.  This is because $\|\mathbf{Q}^{-1}\|$ is bounded and $\|\mathbf{Q}\ket{t_{l,0}}\|=H_{1t,ll}^{(1)}$ as defined in eq. \eqref{GenHtails}. In particular, if $f(m)\sim \alpha log(m)$, this yields a power-law bound $m^{-\alpha}$.
 \item $f(m)\rightarrow C<\infty$: eq. \eqref{OurBound} yields a finite asymptotic bound, $\lim_{m\rightarrow\infty}\|\boldsymbol{\mathcal{T}}\delta T_{l,N}^{(m)}\|\leq\|\mathbf{Q}^{-1}\|\|\mathbf{Q}\ket{t_{l,0}}\|e^{-C}$.  
 \item $f(m)\rightarrow0$: in this case the bound is essentially empty, stating simply that $\|\boldsymbol{\mathcal{T}}\delta T_{l,N}^{(m)}\|\leq\mathbf{Q}^{-1}\|\|\mathbf{Q}\ket{t_{l,0}}\|$.
 \end{enumerate} 
 We note that the above bound provides a sufficient, but not necessary condition for convergence of the residual norm.
 
\begin{figure}[h]
    \centering
     \includegraphics[draft=false,width=\columnwidth]{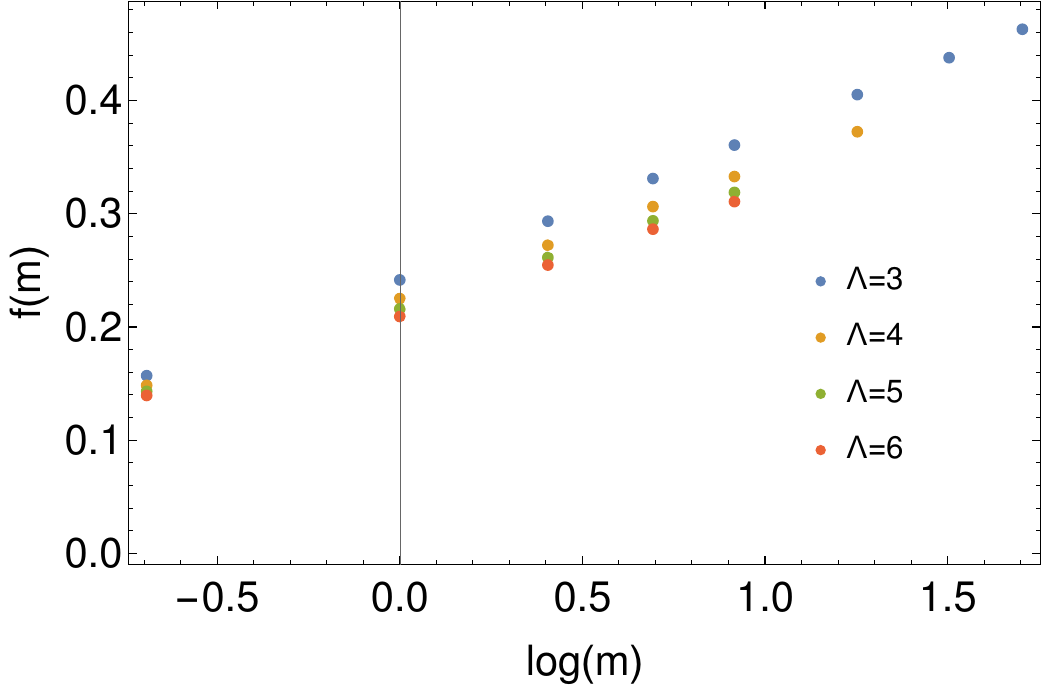}

    \caption{Numerical evaluation of the function $f(m)$ of eq. \eqref{OurBound} for various momentum cutoffs $\Lambda$, at parameters $g=1$, $L=10$, $E=-0.3941$. We used Fock space truncation in the even parity sector in the presence of $N_{ZM}=10$ kept zero mode states. }
    \label{fmfunc}
\end{figure}
To get an idea on the actual behavior of $f(m)$ in the $\phi^4$ model, we computed the extremal eigenvalues of $\bar{\boldsymbol{\mathcal{T}}}$ in a Fock space truncation for various momentum and particle number cutoffs. The results shown on Figure \ref{fmfunc} indicate that we are consistent with case 1, at least numerically. Note that we calculate the eigenvalues for various particle number cutoffs $N$ and use $m=(N-4)/4$ and we include fractional values of $m$ to increase the density of measurement points. 

 Many standard incarnations of Krylov subspace methods involve a step-by-step orthogonalization of the Krylov subspace. In higher orders, this orthonormalization procedure ensures numerical stability. In our setup, this orthogonalization is omitted. This is because we only consider low orders of the iteration and our matrix elements are provided with finite errors. As different matrix elements involve integrals of different dimensionality, the only way to perform a Gram-Schmidt-like procedure is to integrate first and propagate the error into the linear combinations, which is thus expected to increase. Since we deal with the generalized eigenvalue problem directly, in our case the explicit orthogonalization appears to be an unnecessary extra step.

We note in passing that there are alternative Krylov subspace methods that could be examined. In particular, we could have started from the original eigenvalue problem, and set up the Krylov subspace using powers of the full Hamiltonian $\hat{\mathbf H}$ acting on eigenstates of the unperturbed part $\hat{\mathbf H}_{0}$. The eigenvalues could be approximated by a Lanczos method \cite{Lanczos:1950zz,Ojalvo:1970aa,Paige:1972aa}, that could be implemented in an approach very similar to ours. Note that the matrix elements of powers of $\hat{\mathbf H}$ numerically grow significantly faster without the energy denominators inserted. The operator $\hat{\mathbf{H}}$ is evidently unbounded in the $\Lambda\rightarrow\infty$ limit, even if there is a projection to finite particle numbers.
Alternatively, we could include the energy denominator as a preconditioner to the aforementioned Lanczos method. We do not cover these possibilities in the present work, but it may be useful to compare the performance of these alternatives on different models. 

\section{Normal ordering schemes}  \label{SubsecNormOrdSchemes}

In this Appendix we develop expressions where we connect the Hamiltonian where normal ordering for infinite volume $L=\infty$ to normal ordering schemes in general.  We first consider the problem in full generality.
According to Wick's theorem, we can write for any two masses $m$ and $\mu$,
\bea
\varphi(x)^n &=& \sum_{k=0}^{\left\lfloor n/2 \right\rfloor} \frac{n! :\varphi(x)^{n-2k}:_m}{2^k k!\cdot(n-2k)!}\left\langle 0_m \left| \varphi(x)^2 \right| 0_m \right\rangle^k\cr\cr
&=&\sum_{k=0}^{\left\lfloor n/2 \right\rfloor} \frac{n!:\varphi(x)^{n-2k}:_\mu}{2^k k!\cdot(n-2k)!}\left\langle 0_\mu \left| \varphi(x)^2 \right| 0_\mu \right\rangle^k \label{NormOrdFund}
\eea
where the quadratic expectation values are simply calculated. In finite volume we have,
\bea
\left\langle 0_m \left| \varphi(x)^2 \right| 0_m \right\rangle &\equiv& Z(m,L,\Lambda)=\left\langle 0_m \left| \left[\varphi_+,\varphi_-\right] \right| 0_m \right\rangle\cr\cr 
&=&\frac{1}{2L}\sum_{n=-\frac{L\Lambda}{2\pi}}^{\frac{L\Lambda}{2\pi}}\frac{1}{\omega_n},
\eea
while in infinite volume, $Z$, becomes
\be
Z(m,\infty,\Lambda)=\intop_{-\Lambda}^{\Lambda}\frac{1}{\omega(k)}dk.
\ee

Let us introduce the vector with components
\be
\psi_i^{(m)}(x)=:\varphi(x)^{r+2i}:_m,\quad0\leq i\leq\left\lfloor n/2 \right\rfloor
\ee
where $r=1$ if $n$ is odd and $r=0$ otherwise. In this notation, eq. \eqref{NormOrdFund}
can be rewritten as
\be
A_{ij}^{(m)}\psi_j^{(m)}(x)=A_{ij}^{(\mu)}\psi_j^{(\mu)}(x),
\ee
where
\bea
&&A_{ij}^{(m)}\equiv A_{ij}(Z(m,L,\Lambda))\nonumber\\
&&=
\begin{cases}
\frac{(2i)!}{2^{i-j}\cdot(2j)!\cdot(i-j)!}Z(m,L,\Lambda)^{i-j},& \text{for}\quad i\geq j\\
0. & \text{for}\quad i < j
\end{cases}
\eea
The matrices $A_{ij}$ are of lower triangular type with unit diagonal. They constitute a one-parameter group with the multiplication law
\be
A(Z_1)A(Z_2)=A(Z_1+Z_2).
\ee
Therefore the relation between different normal ordered expressions can be written compactly as
\bea
\psi_j^{(m,L_1)}(x) &=& \cr\cr 
&& \hskip -.85in A_{jk}(Z(\mu,L_2,\Lambda)-Z(m,L_1,\Lambda))\psi_k^{(\mu,L_2)}(x) .\label{NOrdMaster}
\eea
\subsection{Massive oscillators}
We now consider the case where our oscillator modes are massive, $m\neq 0$.
The explicit form of the relation for normal ordering with mass $m$
between finite volume $L$ and infinite volume takes the form
\bea
:\varphi^2:_{m,\infty}&=&:\varphi^2:_{m,L}+\Delta Z\cr\cr
:\varphi^3:_{m,\infty}&=&:\varphi^3:_{m,L}+\Delta Z \varphi\cr\cr
:\varphi^4:_{m,\infty}&=&:\varphi^4:_{m,L}+6\Delta Z:\varphi^2:_{m,L}+3\Delta Z^2,
\eea
where $\Delta Z$ is a correction exponentially small for large volumes
with the explicit form
\be
\Delta Z(mL)=\frac{1}{2\pi}\intop_{-\infty}^{\infty}\frac{du}{e^{mL\cosh u}-1} .\label{zfuncdef}
\ee

\begin{figure}[h]
    \centering
     \subfloat[]{\includegraphics[draft=false,width=0.5\columnwidth]{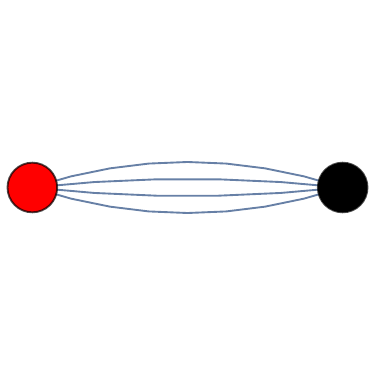}}\\
 \subfloat[]{\includegraphics[draft=false,width=0.5\columnwidth]{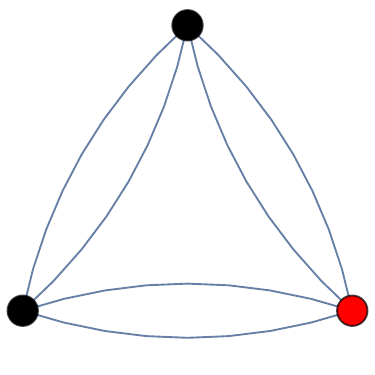}}%

    \caption{Lowest order Feynman diagrams contributing to the ground state energy: $O(g_4^2)$ (top), $O(g_4^3)$ (bottom). The vertex with the zero time argument is distinguished by being colored red. }
    \label{FigGSFeyns1}
\end{figure}

In finite volume, the kinetic term obeys the relation
\bea
\intop_0^L\!\frac{dx}{2}((\partial_x \varphi)^2+\pi(x)^2) &=& \intop_0^L \! \frac{dx}{2}:(\partial_x \varphi)^2+\pi(x)^2:_{m,L}\cr\cr 
&&\hskip -.5in +
\frac{1}{2L}\sum_{n=-\frac{L\Lambda}{2\pi}}^{\frac{L\Lambda}{2\pi}}\frac{1}{\omega_n}(k_n^2+\omega_n^2),
\eea
while in infinite volume, we have
\bea
\intop_{-\infty}^\infty\!\frac{dx}{2}((\partial_x \varphi)^2+\pi(x)^2) \!&=&\!
\intop_{-\infty}^\infty\!\frac{dx}{2}:(\partial_x \varphi)^2+\pi(x)^2:_{m,\infty}\cr\cr 
&&\hskip -.5in +
\frac{1}{4\pi}\intop_{-\Lambda}^{\Lambda}\frac{1}{\omega(k)}(k^2+\omega(k)^2),
\eea
which leads to the relation
\bea
\intop\!\frac{dx}{2}\!:(\partial_x \varphi)^2+\pi(x)^2:_{m,\infty}
\!&=&\!\intop\!\frac{dx}{2}\! :(\partial_x \varphi)^2+\pi(x)^2:_{m,L}\cr\cr 
&&\hskip -.75in +me_0(mL)-\frac{m^2L}{2}\Delta Z(mL)
\eea
where
\be
e_0(mL)=\intop_{-\infty}^{\infty} \frac{d\theta}{2\pi}\cosh\theta\log(1-e^{-mL\cosh\theta}). \label{e0funcdef}
\ee
The explicit form of the Hamiltonian, expressed with finite volume normal ordering,
reads
\bea
H&=&me_0(mL)+L\left(g_2-\frac{m^2}{2}\right)\Delta Z(mL)+ 3Lg_4(\Delta Z)^2\nonumber\\
&+&\intop dx :\frac{\pi^2}{2}+\frac{(\partial_x \phi)^2}{2}:_{m,L}\nonumber\\
&+&\intop dx:(g_2+6g_4\Delta Z) \phi(x)^2+g_4 \phi(x)^4:_{m,L}.
\eea
At this point the zero mode can be separated
\be
\varphi(x)=\varphi_0+\tilde{\varphi}(x),\quad \pi=\pi_0+\tilde{\pi}(x)
\ee
resulting in the following form
\bea
H&=&me_0(mL)+L\left(g_2-\frac{m^2}{2}\right)\Delta Z(mL)+ 3Lg_4(\Delta Z)^2\cr\cr
&+&m a_0^\dagger a_0+\left(g_2-\frac{m^2}{2}+6g_4\Delta Z\right)L :\phi_0^2:_m\cr\cr
&+&g_4 L :\phi_0^4:_m+H_{0,osc}^{(m)}\cr\cr 
&+&\intop dx :\left(g_2-\frac{m^2}{2}+6g_4\Delta Z\right) \tilde\phi(x)^2:_{m,L}\cr\cr 
&+& g_4 \intop dx :(\phi(x)^4-\phi_0^4):_{m,L},
\eea
where
\be
H_{0,osc}^{(m)}=\intop dx :\frac{\tilde\pi^2}{2}+\frac{(\partial_x \tilde\phi)^2}{2}+\frac{m^2}{2}\phi^2:_{m,L}=\sum_{n\neq 0} \omega_n a_n^\dagger a_n.
\ee
\subsection{Massless oscillators}
When the massless oscillator basis is used, another renormal-ordering is necessary
for the powers of $\tilde\phi$:
\bea
\intop dx :\frac{\tilde\pi^2}{2}+\frac{(\partial_x \tilde\phi)^2}{2}:_{m,L}&=&\intop dx :\frac{\tilde\pi^2}{2}+\frac{(\partial_x \tilde\phi)^2}{2}:_{0,L}\nonumber\\
&&\hskip -.5in + \sum_{n\neq 0} \left(\frac{|k_n|}{2}-\frac{\omega_n}{2}+\frac{m^2}{4\omega_n}\right).
\eea
The conditionally convergent sum is understood in the presence of some cutoff $|n|\leq n_{max}$, and $n_{max}$ is subsequently taken to infinity. This results is a finite limit.
The extra term on the RHS can be written in the following integral representation
\bea
&&\sum_{n\neq 0} \frac{|k_n|}{2}-\frac{\omega_n}{2}+\frac{m^2}{4\omega}=\nonumber\\
&&-\frac{\pi}{6L}-\frac{m^2}{8\pi}L+\frac{m}{4}-me_0(mL)+\frac{m^2}{2}L\Delta Z.
\eea
Furthermore, using the appropriately modified version of eq. \eqref{NOrdMaster} for the powers of $\tilde\phi$ with $m_1=m$, $m_2=0$, $L_1=L_2=L$,
we obtain
\bea
:\tilde\varphi^2:_{m,L}&=&:\tilde\varphi^2:_{0,L}+\tilde\Delta\cr\cr
:\tilde\varphi^3:_{m,L}&=&:\tilde\varphi^3:_{0,L}+\tilde\Delta \varphi\cr\cr
:\tilde\varphi^4:_{m,L}&=&:\tilde\varphi^4:_{0,L}+6\tilde\Delta:\tilde\varphi^2:_{0,L}+3\tilde\Delta^2,
\eea
where
\be
\tilde\Delta= -\Delta Z(mL)+\frac{1}{2 m L}+\frac{\gamma_E}{2\pi}+\frac{1}{2\pi}\log\left(\frac{mL}{4\pi}\right)
\ee
Expressing the Hamiltonian with the $::_{0,L}$ normal ordered operators, we obtain
($\hat\Delta=\tilde\Delta+\Delta Z$):
\bea
H&=&-\frac{\pi}{6L}+\frac{m}{4}-\frac{m^2}{8\pi}L+L g_2\hat\Delta  + 3Lg_4(\hat\Delta)^2\nonumber\\
&+&m a_0^\dagger a_0+\left(g_2-\frac{m^2}{2}+6g_4\hat\Delta \right)L :\phi_0^2:_m+g_4 L :\phi_0^4:_m\nonumber\\
&+&H_{0,osc}^{(0)}+\intop dx \left[\left(g_2+6g_4\hat\Delta\right) :\tilde\phi(x)^2:_{0,L}\right.\nonumber\\
&+&g_4 :\widetilde{\phi}(x)^4:_{0,L}+4g_4:\widetilde{\phi}(x)^3:_{0,L}\phi_0\cr\cr
&+&\left.6g_4:\widetilde{\phi}(x)^2:_{0,L}:\phi_0^2:_{m}\right]
\eea
In the above the zero mode remains normal ordered w.r.t. to the massive $a_0$ while the oscillators are normal ordered w.r.t. to massless finite $q$ modes, $a_{q\neq0}$.

\section{TSM details of the ``conventional" tail state approach}\label{AppZMExDetails} 
\subsection{Matrix elements}
Tthe potential $\hat{\mathbf{V}}'$ in \eqref{V1OrigChoice} to be integrated out
has the form
\be
\hat{\mathbf{ V'}} = \sum_{n=2}^4\mathbf{g}_n\intop dx  :\widetilde{\boldsymbol{\phi}}(x)^n:_{m_0,L} 
\ee

Note that the projection into the zero mode $\hat{\mathbf P}:\widetilde{\boldsymbol{\phi}}(x)^n: \hat{\mathbf P}$ is zero for all $n>0$.
The matrix elements between the emphasized subspace and the first order tails can be written as
\bea
(\Hlt{1})_{m\tilde t_{m'1}}&=&-\braket{m|\hat{\mathbf{V}}'(\tau)\hat{\mathbf{P}}_\perp\hat{\mathbf{V}}'(0)|\tilde t_{m'1}}\nonumber\\
&=&\intop_0^Ldx\intop_0^{\infty}dt\sum_{\alpha_1,\alpha_2=2}^4\braket{m|\mathbf{g}_{\alpha_1}(\tau)\mathbf{g}_{\alpha_2}(0)|m^\prime}\cdot\nonumber\\
&&V_{\alpha_1\alpha_2}(\tau,0)e^{(E_*-E_m^{ZM})\tau}\label{H1tappeq}
\eea

where the two-point function $V_{\alpha_1\alpha_2}$ was defined in eq. \eqref{multiV}.
Since $V_{\alpha_1\alpha_2}=0$ for $\alpha_1\neq \alpha_2$, the sums in eq. \eqref{H1tappeq} can be restricted to $\alpha_1=\alpha_2$.

\begin{figure*}[t]
\centering
\subfloat[]{\includegraphics[draft=false,width=0.5\columnwidth]{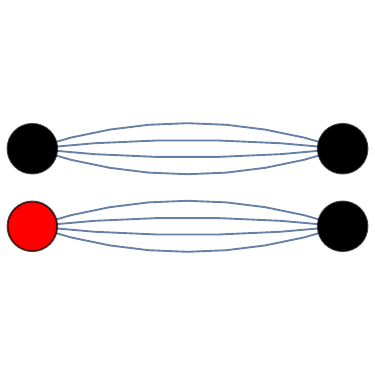}} 
 \subfloat[]{\includegraphics[draft=false,width=0.5\columnwidth]{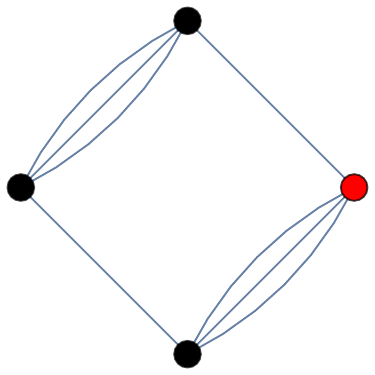}}%
  \subfloat[]{\includegraphics[draft=false,width=0.5\columnwidth]{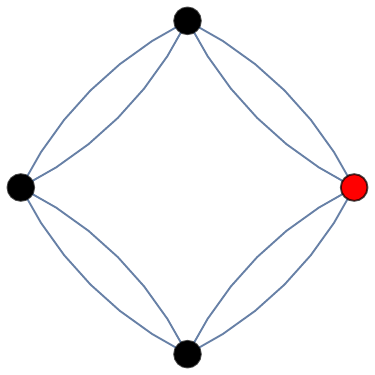}} 
 \subfloat[]{\includegraphics[draft=false,width=0.5\columnwidth]{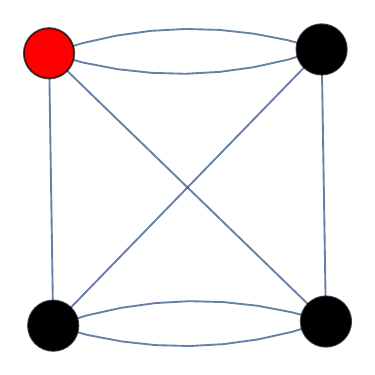}}%
    \caption{Feynman diagrams contributing to the ground state energy at $O(g_4^4)$.  The multiplicities of the diagrams in (a), (b), (c), and (d), are $3$, $6$, $3$, and $3$, respectively. The vertex with zero time argument is distinguished by being colored red.} 
    \label{FigGSFeyns2}
\end{figure*}
Going beyond the leading order, we get for the matrix elements (here and in the following, $T_n=\sum_{j=1}^n \tau_n$, and we use the notation $d^n\tau=\prod_{k=1}^{n} \intop_{0}^{\infty} d\tau_k$)

\bea
&&(\Hlt{n-1})_{m\tilde t_{m',n-1}}=(-1)^{n-1} \intop_0^\infty d^{n-1}\tau \sum_{\alpha_1,\dots\alpha_n=2}^4 \cr\cr
&&I_{n}^{\alpha_1\dots \alpha_n}(\{T(\{\tau\})\})e^{(E_*-E_m^{ZM})T_{n-1}}\nonumber\\
&&~~\times\braket{m|\mathbf{g}_{\alpha_1}(T_{n-1})\dots\mathbf{g_{\alpha_n}}(0)|m'};\label{H1tappGeneq}\\
&&(G^{(ij)})_{\tilde{t}_{mi},\tilde{t}_{m'j}}=(-1)^{(i+j)}\intop_0^{\infty}d^{i+j-1}\tau\sum_{\alpha_1,\dots\alpha_{i+j}=2}^4\cr\cr
&&{\color{red}\tau_{j}} I_{i+j}^{\alpha_1\dots \alpha_{i+j}}(\{T(\{\tau\})\})e^{(E_*-E_m^{ZM})T_{i+j-1}}\cr\cr 
&&\times\braket{m|:\mathbf{g}_{\alpha_1}(T_{i+j-1})\dots\mathbf{g}_{\alpha_{i+j}}(0)|m'}.\label{GijappGeneq}
\eea
We give the explicit forms of the objects $I_n^{k_1,\dots k_n}$ below:
\bea\label{deltaHDiscExampleApp}
I_3^{ijk}(\{T\})&=& V_{ijk}(T_2,T_1,0); \cr\cr
I_4^{ijkl}(\{T\}) &=& V_{ijkl}(T_3,T_2,T_1,0) \cr\cr 
&&~~~~-V_{ij}(T_3,T_2)V_{kl}(T_1,0);\cr\cr
I_5^{ijklm}(\{T\}) &=& V_{ijklm}(T_4,\dots,T_1,0) ;\cr\cr
&&\hskip -.5in-V_{ijk}(T_4,T_3,T_2)V_{lm}(T_1,0)\cr\cr 
&&\hskip -.5in-V_{ij}(T_4,T_3)V_{klm}(T_2,T_1,0)\label{I5eq}\cr\cr
I_6^{ijklmn}(\{T\}) &=& V_{ijklmn}(T_5,\dots,T_1,0) \cr\cr 
&&\hskip -.5in-V_{ijkl}(T_5,T_4,T_3,T_2)V_{mn}(T_1,0) \cr\cr 
&&\hskip -.5in-V_{ijk}(T_5,T_4,T_3)V_{lmn}(T_2,T_1,0)  \cr\cr 
&&\hskip -.5in-V_{ij}(T_5,T_4)V_{klmn}(T_3,T_2,T_1,0)  \cr\cr 
&&\hskip -.5in+V_{ij}(T_5,T_4)V_{kl}(T_3,T_2)V_{mn}(T_1,0) ;\cr\cr 
I_7^{ijklmno}(\{T\}) &=& V_{ijklmno}(T_6,\dots,T_1,0)  \cr\cr 
&&\hskip -.5in - V_{ijklm}(T_6,T_5,T_4,T_3,T_2)V_{no}(T_1,0) \cr\cr 
&&\hskip -.5in- V_{ijkl}(T_6,T_5,T_4,T_3)V_{mno}(T_2,T_1,0)  \cr\cr 
&&\hskip -.5in-V_{ijk}(T_6,T_5,T_4)V_{lmno}(T_3,T_2,T_1,0)  \cr\cr 
&&\hskip -.5in-V_{ij}(T_6,T_5)V_{klmno}(T_4,T_3,T_2,T_1,0) \cr\cr 
&&\hskip -.5in+ V_{ijk}(T_6,T_5,T_4)V_{lm}(T_3,T_2)V_{no}(T_1,0)  \cr\cr 
&&\hskip -.5in+V_{ij}(T_6,T_5)V_{klm}(T_4,T_3,T_2)V_{no}(T_1,0) \cr\cr 
&&\hskip -.5in+ V_{ij}(T_6,T_5)V_{kl}(T_4,T_3)V_{mno}(T_2,T_1,0). \label{I7eq}
\eea
Similarly to perturbation theory, the integrands in eqs. \eqref{H1tappGeneq}-\eqref{GijappGeneq} simplify considerably by introducing explicit time ordering and exploiting the resulting permutation symmetry. However, additional care is needed in symmetrizing the intrgration domain. Since the interaction consists of multiple types of vertices (quadratic, cubic and quartic terms), it turned out convenient to keep the $\tau=0$ operator distinguished, while the remaining operators are assumed equivalent.
In particular, we can write
\bea
&&(\Hlt{3})_{m\tilde t_{m',3}}=- \intop_0^\infty d^{3}T   \cr\cr
&&\left(\sum_{2\leq\alpha_1\leq\alpha_2\leq\alpha_3\leq\alpha_4}^4\frac{3!}{n_2! n_3! n_4!}V^{\mathrm{comp}}_{\alpha_1\alpha_2\alpha_3\alpha_4}(\{T\})\right.\cr\cr
&&~~\times \braket{m|\hat{\mathcal{T}}\mathbf{g}_{\alpha_1}(T_{3})\dots\mathbf{g_{\alpha_n}}(0)|m'}\Biggr)\cr\cr
&&+\frac12\Theta(T_1-\min(T_2,T_3))\Biggr(\sum_{\alpha_1,\alpha_2=2}^4\cr\cr
&&V_{\alpha_1\alpha_1}(T_3,T_2)V_{\alpha_2\alpha_2}(T_1,0)\cr\cr
&&\braket{m|\hat{\mathcal{T}}\mathbf{g}_{\alpha_1}(T_3)\mathbf{g}_{\alpha_1}(T_2)\mathbf{g_{\alpha_2}}(T_1)\mathbf{g_{\alpha_2}}(0)|m'}\Biggr)
\eea
where $\hat{\mathcal{T}}$ denotes time ordering so that larger time arguments are arranged to the left. The variables $n_2$, $n_3$, $n_4$ refer to the number of quadratic, cubic and quartic vertices excluding the one with zero argument, while $\Theta(T)$ is the Heaviside unit step function. The compressed correlator $V^{comp}_{\alpha_1,\dots \alpha_n}$ is obtainted from $V_{\alpha_1,\dots \alpha_n}$ in the following way. First we remove any disconnected Feynman diagrams appearing in $V_{\alpha_1,\dots \alpha_n}$. Second, we distribute diagrams which can be transformed into each other by a permutation of the first $n-1$ vertices into equivalence classes. We choose a representative diagram from each class so that $\alpha_1\leq \alpha_2\leq \dots\leq \alpha_{n-1}$, and multiply it with the number of diagrams in the class. Thus extra symmetry factors corresponding to this partial symmetrization are obtained by an explicit counting of equivalent diagrams.

The subset of Feynman diagrams containing only quartic vertices, for the lowest few orders are pictured in Figures \ref{FigGSFeyns1} and \ref{FigGSFeyns2}. While restricting the evaluation to topologically different diagrams greatly reduces the complexity of evaluations, one has to keep track of the time arguments in the zero mode correlator, which have to be arranged ``manually" into the correct time order in their numerical evaluation.

\begin{figure}
\centering
\subfloat[Partial cancellation for the lowest order disconnected diagrams. Here $T_3\geq T_2 \geq T_1 \geq 0$. The top diagram is cancelled by the explicit subtraction in $I_4^{ijkl}(\{\tau\})$, but the other two are not.]{\includegraphics[width=0.45\textwidth]{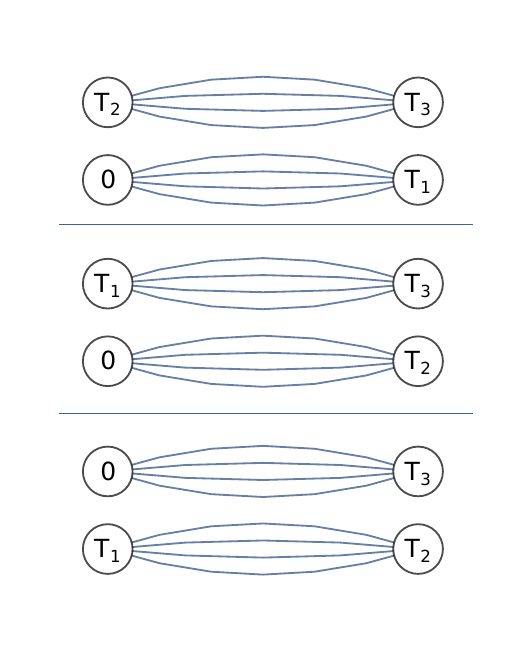}}\\
\subfloat[An example of higher orders (arising in $I_7$): disconnected diagrams with disjoint argument configurations (like the one depicted) cancel. Here time increases from left to right (in particular, $T_2 < T_1$.) The correlator $\langle V(T_6)V(T_5)\rangle$ is completely separated in its arguments from the other correlators which results in the cancellation by $V(T_6,T_5)V(T_4,T_3,T_2,T_1,0)$.]{\includegraphics[width=0.45\textwidth]{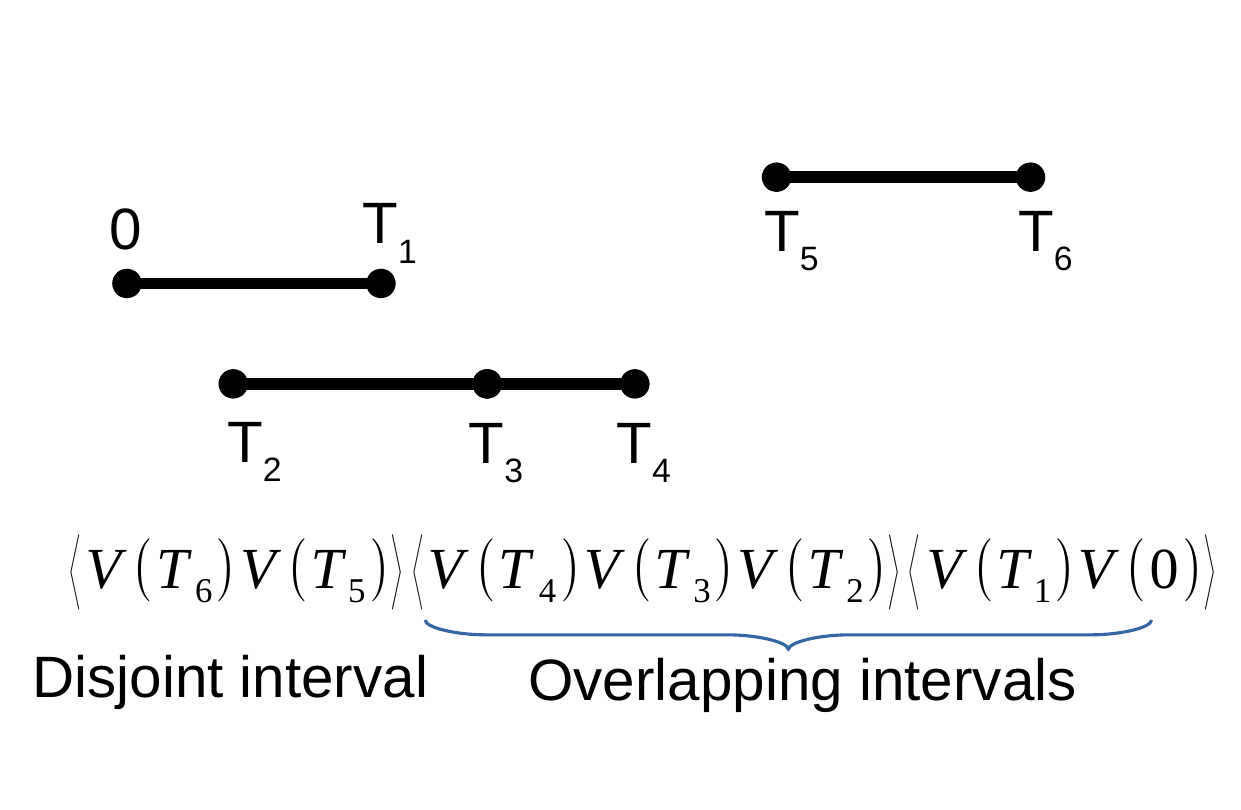}}
\caption{Partial cancellation of disconnected terms}
%\ref{tablech} .) }
\label{FigIntervals}
\end{figure}

From the viewpoint of Feynman diagrams, the $n$-point functions contain disconnected pieces. (see e.g., Fig. \ref{FigGSFeyns2} (a)). The explicit disconnected terms of $I_{n}^{k_1\dots k_n}$ only partially cancel these terms, as the time ordering of the latter is partially fixed. Nonetheless, these terms combine in a nice way. Cancellation occurs whenever the disconnected correlator factors can be arranged into at least two subsets, so that the narrowest time interval encompassing all time arguments inside a single subset is disjoint from the corresponding intervals of the other subsets (Figure \ref{FigIntervals}).
In other words, there is a cutoff in the separation of disconnected pieces. Together  with the exponential decay of the propagator, this feature ensures that the dominant contribution to the $\tau$-integral comes from the the region with all $\tau$ variables being small. When the $\tau$ integrands are symmetrized under $\tau$ ordering, the regions where the disconnected terms give a contribution needs to be symmetrized explicitly.

The above method is suitable for calculating the ground state and zero-momentum one-particle states
of the interacting theory. 

Things become more complicated if one is to calculate general excited
states. One could in principle include nontrivial oscillator states. When states having non-trivial oscillator content are included among $\mathcal{H}_1$ states, one obtains Feynman diagrams with external lines, which adds to the complexity of calculating the associated matrix elements. 

\subsection{Choice of $E_*$}

All `conventional' tail states depend on the energy parameter $E_*$. Technically $E_*$ appears as the (a priori unknown) exact energy of the eigenvalue that we seek. In Fig. \ref{unapproximated_tails}, the emphasis was on the numerical convergence in $N_{ZM}$ and so we took $E_*$ from reference data originally obtained with the method described in \cite{Elias-Miro:2017tup}. In general it is tempting to perform a self-consistent iteration to fix this energy parameter. However, we argue that such an iteration is actually not favorable in general. The stage of the derivation in which $E_*$ is identified with the exact energy eigenvalue involves the full, non-truncated Hamiltonian. There is no reason why an iteration in a restricted Hilbert space should be optimal in obtaining the eigenvalue. A better point of view may be to consider $E_*$ as a variational parameter, and look for the optimal set of tails by minimizing the ground state energy with respect to $E_*$.

\subsection{Feynman diagrams} \label{FeynApp}
Practically, we generate the Feynman diagrams as follows. In our setup the relevant diagrams have no external legs, so we will restrict our attention to these ``vacuum" diagrams.
\begin{enumerate}
\item For a given number of vertices $n$, generate all distinct ordered sets
of vertex ranks (no. of legs) from which at least one diagram can be assembled.
For the $\phi^4$ model, $2\leq\text{rank}\leq4$ for each vertex and the sum of all ranks has to be even.
For example, there are six relevant ordered sets for $n=3$.
Henceforth we introduce the notation $(r_1,r_2,\dots,r_n)$ for the set of diagrams
having $n$ vertices with ranks $r_1$,\dots, $r_n$.

\item To each collection of vertices, e.g. $\left(2,2,4,4\right)$, construct
a list of all possible Feynman diagrams by connecting the vertices
in all possible ways, such that no line starts and ends on the same
vertex. Disconnected diagrams
are also kept. All possibilities are easily generated in Mathematica
as long as the number of vertices is not too large.

\item Distribute the appropriate symmetry factors to the graphs. Each graph
is multiplied by

\be
S(\textit{graph}) =\left(\frac{\prod_{i\in\textit{vertices}}\mathcal{R}_i!}{\prod_{i,j\in\textit{vertices};\:j>i}\mathcal{P}_{ij}!}\right)
\ee
where $\mathcal{R}_i$ is the number of legs of vertex $i$, and $\mathcal{P}_{ij }$ is the number of propagators between vertices $i$ and $j$.
\item If $V^{\mathrm{comp}}$ is calculated,  we select the connected Feynman diagrams appearing in $V_{\alpha_1,\dots \alpha_n}$. We distribute diagrams which can be transformed into each other by a permutation of the first $n-1$ vertices into equivalence classes. We choose a representative diagram from each class so that $\alpha_1\leq \alpha_2\leq \dots\leq \alpha_{n-1}$, and multiply it with the number of diagrams in the class.
\item
Finally, the integrand corresponding to a given set of graphs is generated
in real (Euclidean) space.
The formula is then exported to $C++$ syntax using Mathematica's \texttt{CForm} function.
When the number of diagrams approaches the order of $10^4$, the corresponding
source code is in the order of megabytes, and it becomes convenient to
divide the integrand into a number of separate $C++$ functions and compile a static library.
This greatly reduces compilation time.

\end{enumerate}

\subsection{Feynman diagram data}
\subsubsection{First Krylov order (NLO): $2$ and $3$ vertices}
At this order, the lowest order contributing diagrams have vertex
content $(2,2)$, $(3,3)$ and $(4,4)$, respectively.
There are six relevant ordered sets involving three vertices.
A single possible Feynman diagram corresponds to each set. They
are depicted in Figure \ref{FigFeynNLO}.

\subsubsection{Second Krylov order (2NLO): $4$ and $5$ vertices}
For $4$ vertices, there are $9$ different ordered rank sets.
The corresponding numbers of diagrams are shown in Table \ref{tab:tab45vert}.
For $5$ vertices, there are $12$ different ordered rank sets. The respective numbers of diagrams are depicted in Table \ref{tab:tab45vert}.

\begin{table}[]
\centering
\begin{tabular}{|c|c|c|c|}
\hline
4-Vertex set & \# diagrams & 5-Vertex set & \# diagrams \\    \hline
(2,2,2,2)  & 6 & (2,2,2,2,2)  & 22 \\
(2,2,2,4)  & 3 & (2,2,2,2,4)  & 21 \\
(2,2,3,3)  & 6 & (2,2,2,3,3)  & 29 \\
(2,2,4,4)  & 6 & (2,2,2,4,4)  & 29 \\
(2,3,3,4)  & 6 & (2,2,3,3,4)  & 37 \\
(2,4,4,4)  & 6 & (2,2,4,4,4)  & 46 \\
(3,3,3,3)  & 10 & (2,3,3,3,3)  & 48 \\
(3,3,4,4)  & 10 & (2,3,3,4,4)  & 58 \\
(4,4,4,4)  & 15 & (2,4,4,4,4)  & 84 \\ 
& & (3,3,3,3,4)  & 72 \\ 
& & (3,3,4,4,4)  & 101 \\
& & (4,4,4,4,4)  & 158 \\   \hline
\end{tabular}
\caption{Number of diagrams with $4$ and $5$ vertices before symmetrization}
\label{tab:tab45vert}
\end{table}
\subsubsection{Third Krylov order (3NLO): $6$ and $7$ vertices}
For $6$ vertices, there are $16$ different ordered rank sets.
The corresponding numbers of diagrams are shown on Table \ref{tab:tab67vert}.
For $7$ vertices, there are $20$ different ordered rank sets.
The corresponding numbers of diagrams are shown on Table \ref{tab:tab67vert2}.

\begin{table}[]
\centering
\begin{tabular}{|c|c|c|c|}
\hline
6-Vertex set & \# diagrams & \# connected & \# compressed \\ \hline
(2,2,2,2,2,2)  & 130 & 60 & 1\\
(2,2,2,2,2,4)  & 130 & 90 & 2 \\
(2,2,2,2,3,3)  & 190 & 150 & 10 \\
(2,2,2,2,4,4)  & 209 & 163 & 12 \\     
(2,2,2,3,3,4)  & 262 & 237 & 29 \\
(2,2,2,4,4,4)  & 346 & 309 & 37 \\
(2,2,3,3,3,3)  & 352 & 306 & 33 \\
(2,2,3,3,4,4)  & 449 & 425 & 123 \\
(2,2,4,4,4,4)  & 691 & 636 & 68\\
(2,3,3,3,3,4)  & 574 & 532 & 32\\
(2,3,3,4,4,4)  & 856 & 828 & 233\\
(2,4,4,4,4,4)  & 1430 & 1360 & 74 \\
(3,3,3,3,3,3)  & 760 &  640 & 10\\
(3,3,3,3,4,4)  & 1093 & 1023 & 60 \\
(3,3,4,4,4,4)  & 1819 & 1746 & 175\\
(4,4,4,4,4,4)  & 3355 & 3150 & 42 \\ 
\hline
\end{tabular}
\caption{Number of diagrams with $6$ vertices before and after symmetrization. The "compressed" column indicates the number of inqeuivalent connected diagram classes}
\label{tab:tab67vert}
\end{table}

\begin{table}[]
\centering
\begin{tabular}{|c|c|c|c|}
\hline
7-Vertex set & \# diagrams & \# connected & \# compressed \\ \hline
(2,2,2,2,2,2,2) & 822 & 360 & 1\\
 (2,2,2,2,2,2,4) & 1005 & 630 & 3\\
 (2,2,2,2,2,3,3) & 1402 & 1050 & 15\\
 (2,2,2,2,2,4,4) & 1662 & 1250 & 19\\     
 (2,2,2,2,3,3,4) & 2159 & 1856 & 61 \\
(2,2,2,2,4,4,4) & 3093 & 2622 & 88\\
 (2,2,2,3,3,3,3) & 2878 & 2484 & 94 \\
(2,2,2,3,3,4,4) & 3975 & 3669 & 382\\
 (2,2,2,4,4,4,4) & 6453 & 5862 & 216\\
 (2,2,3,3,3,3,4) & 5140 & 4746 & 139\\
 (2,2,3,3,4,4,4) & 8188 & 7842 & 1129\\
 (2,2,4,4,4,4,4) & 14613 & 13790 & 372\\
 (2,3,3,3,3,3,3)  & 6720 & 5940 & 70\\
 (2,3,3,3,3,4,4)  & 10466 & 9946 & 503\\
 (2,3,3,4,4,4,4)  & 18497 & 17960 & 1659\\
(2,4,4,4,4,4,4)  & 35865 & 34350 & 361 \\ 
 (3,3,3,3,3,3,4)  & 13440 & 12300 & 35\\
 (3,3,3,3,4,4,4)  & 23453 & 22483 & 608         \\
 (3,3,4,4,4,4,4)  & 45103 & 43810 & 1103         \\
(4,4,4,4,4,4,4)  & 93708 & 90075 & 195 \\ \hline
\end{tabular}
\caption{Number of diagrams with $6$ vertices before and after symmetrization. The "compressed" column indicates the number of inqeuivalent connected diagram classes}
\label{tab:tab67vert2}
\end{table}

\section{Matrix elements for universal tails} \label{AppExplicitDiscTerms}

Using the notations from Appendix \ref{AppZMExDetails}, the matrix elements of the Hamiltonian and the Gram matrix between the universal tail states can be written as ($T_n=\sum_{j=1}^n \tau_n$, and we use the notation $d^n\tau=\prod_{k=1}^{n} \intop_{0}^{\infty} d\tau_k$))
\begin{eqnarray}
\braket{\widetilde{t}_{i_1,\cdots,i_m}|\widetilde{\mathbf{V}}_j|\widetilde{t}_{k_1,\cdots,k_n}} &=& (-1)^{m+n}\intop d^{m+n}\tau\cr\cr 
&&\hskip -1.in \times I_{m+n+1}^{i_1,\cdots,i_m,j,k_1,\cdots,k_n}(\{T(\{\tau\})\})\cr\cr
\braket{\widetilde{t}_{i_1,\cdots,i_m}|\widetilde{\mathbf{H}}_{0,osc}^{(m)}|\widetilde{t}_{k_1,\cdots,k_n}} &=& -\braket{\widetilde{t}_{i_1,\cdots,i_m}|\widetilde{\mathbf{V}}_{k_1}|\widetilde{t}_{k_2,\cdots,k_n}}\cr\cr
\braket{\widetilde{t}_{i_1,\cdots,i_m}|\widetilde{t}_{k_1,\cdots,k_n}} &=& (-1)^{m+n}\intop d^{m+n-1}\tau\cr\cr 
&&\hskip -1.in \times \tau_{n}I_{m+n}^{i_1,\cdots,i_m,k_1,\cdots,k_n}(\{T(\{\tau\})\})
\end{eqnarray}
In the above, the quantities $I_{n}^{k_1,\dots k_n}$ were defined in eqs. \eqref{deltaHDiscExampleApp}.

Note that the time arguments are in explicit time order and so the disconnected pieces do not completely cancel the ones implicit in the $n$-point functions $V_{ijk\dots}$. 
Since our $K=3$ basis spans $40$ states in the oscillator subspace, in principle we would need to calculate a total of $4100$ matrix elements corresponding to the operators $\widetilde{\mathbf{V}}_2$ $\widetilde{\mathbf{V}}_3$, $\widetilde{\mathbf{V}}_4$, $\widetilde{\mathbf{H}}_{0,osc}^{(m)}$ and the Gram matrix $\widetilde{\mathbf{G}}$ (their representations are symmetric matrices). However, due to their structure, it is sufficient to compute a total of $1312$ matrix elements, $574$ of which are $12$ dimensional integrals.

\section{Euclidean finite volume propagator} \label{AppFVprop}
In this appendix we compute the finite volume propagators.
For the massless case, the finite volume two-point function is defined as
\be
G_0(t,x)=\left\langle\left|\left[\tilde\varphi_+(t,x),\tilde\varphi_-(0,0)-\right]\right|\right\rangle
=\frac{1}{2L}\sum_{n\neq 0} \frac{e^{i k_n x- |k_n| t}}{|k_n|},
\ee
which can be summed up analytically to yield
\be
G_0(t,x)=-\frac{1}{4\pi}\log\left[1+e^{\frac{-4\pi}{L} t}-2e^{\frac{-2\pi}{L} t}\cos\left(\frac{2\pi}{L} x\right)\right].
\ee

In the massive case, the finite volume propagator $G_m(t,x)$ for a free boson of mass $m$ is calculated
in the following way. It follows from the massive mode expansion that $G_m(t,x)$ can be written as
\be
G_m(t,x)= \frac{1}{2L}\sum_{n\neq 0} \frac{e^{i k_n x- \omega_n t}}{\omega_n} .\label{massivePropMode}
\ee
In the following it will be convenient to explicitly subtract the zero mode contribution
\be
G_m(t,x)=\frac{1}{2L}\sum_{n\in\mathbb{Z}} \frac{e^{i k_n x- \omega_n t}}{\omega_n}-\frac{1}{2Lm}e^{-mt},
\ee
which allows us to write $G_m(t,x)$ in terms of a contour integral
\bea
&&G_m(t,x)=-\frac{e^{-mt}}{2mL}\nonumber\\
&&+\frac{1}{8\pi}\intop_C \frac{e^{ipL}}{e^{ipL}-1}(e^{ipx}+e^{-ipx})\frac{e^{-t\sqrt{p^2+m^2}}}{\sqrt{p^2+m^2}}dp,
\eea
where we have also exploited the $x\rightarrow -x$ symmetry of the sum. The contour $C$ consists
of many disconnected pieces, encircling all poles on the real axis in the positive direction.
This contour is first deformed into two horizontal sections. One section goes slightly
below the real axis from $-\infty-i\epsilon$ towards $+\infty-i\epsilon$ ($\epsilon>0$),
while the other goes above the real axis from $+\infty+\epsilon$ to $-\infty+\epsilon$.
From the contour, we subtract (and then add explicitly) the large volume approximation of the sum,
\be
G_{m,\infty}=\frac{1}{4\pi}\intop_{-\infty}^{\infty}dp\frac{\cos(px)e^{-t\sqrt{p^2+m^2}}}{\sqrt{p^2+m^2}},
\ee
after which the contours can be transformed into two large semicircles, avoiding
the vertical square root branch cuts above and below the real axis.

In the following we will rely on the identity
\bea
&&K_0(\sqrt{\rho^2+\tau^2})=\intop_0^\infty e^{-\sqrt{\rho^2+\tau^2}\cosh\theta}d\theta\nonumber\\
&&=\frac{1}{2}\intop_{-\infty}^{\infty}e^{-r\cosh u \cosh\theta}d\theta\cr\cr
&&=\frac{1}{2}\intop_{-\infty}^{\infty}e^{-r\cosh^2 u \cosh\theta-r\cosh u\sinh u\sinh\theta}d\theta \label{mpropint}.
\eea
where $K_0(x)$ is the modified Bessel function of the second kind.

Let us first study the large-volume asymptotic $G_{m,\infty}(t,x)$.
Changing the integration variable to $p=m\sinh\theta$, we obtain
\be
G_{m,\infty}=\frac{1}{4\pi}\intop_{-\infty}^{\infty}\cos(mx\sinh\theta)e^{-mt\cosh\theta}.
\ee
Comparing this form to the integral eq. \eqref{mpropint}, we obtain
\be
G_{m,\infty}=\frac{1}{2\pi}K_0\left(m\sqrt{t^2+x^2}\right).
\ee
The contribution of the contours tightened around the branch cuts can be written
in the form
\be
G_{m,\Delta}= \frac{1}{4\pi}\intop_0^{\infty}\frac{4\cos(m t \sinh\theta)\cosh(m x \cosh\theta)}{e^{mL \cosh\theta}-1}.
\ee
Expanding the denominator in the small parameter $e^{-mL\cosh\theta}$, we obtain an
infinite series of terms of the form \eqref{mpropint}. In this way we get ($0<x<L$)
\begin{eqnarray}\label{FVMassivePropApp}
G_{m}(t,x) &=& G_{m,\infty}+G_{m,\Delta}\cr\cr
&=&\frac{1}{2\pi}K_0\left(m\sqrt{x^2+t^2}\right)\cr\cr
&&+\frac{1}{2\pi}K_0\left(m\sqrt{(L-x)^2+t^2}\right)+\sum_{n=1}^{\infty}\delta G_n,\cr\cr
&&\delta G_n=\frac{1}{2\pi}K_0\left(m\sqrt{(nL+x)^2+t^2}\right)\cr\cr 
&&+\frac{1}{2\pi}K_0\left(m\sqrt{((n+1)L-x)^2+t^2}\right).
\end{eqnarray}
This form is manifestly periodic in $x$ with period $L$.

\vskip 10pt

\section{Evaluating the integrals}

To obtain the numerical matrix elements, the integration was performed
using the globally adaptive \texttt{Cuhre} algorithm of the \textsc{Cuba} package.
This is a very competitive algorithm for multidimensional integrals of moderate
dimensions, with the error generally reducing as $N_{\mathrm{eval}}^{-1}$
instead of $N_{\mathrm{eval}}^{-1/2}$ common in Monte Carlo methods.

The integration used around $50$ million function
evaluations for each matrix element. This results in a relative precision of order
$10^{-3}$ for matrix elements of the third order tail states.

For the actual computations it was advantageous to use polar
coordinates in the time subspace for $4>2$ dimensional integrals. The parameterization that we used
is as follows:
\begin{align*}
t_{1} & =r\cos\varphi_{1}\\
t_{2} & =t_{1}\tan\varphi_{1}\cos\varphi_{2}\\
\vdots\\
t_{D/2-1} & =t_{D/2-2}\tan\varphi_{D/2-2}\cos_{D/2-1}\\
t_{D/2} & =t_{D/2-1}\tan\varphi_{D/2-1}
\end{align*}
and where we introduce $r=R_0 \rho$, with a rescale parameter $R_0=40$.
%$r=\tanh^{-1}\rho$.

For the numerical integration, we have to transform the integral
into the unit hypercube. The angles are naturally bounded as
\be
\left\{ \varphi_{1},\dots,\varphi_{D/2}\right\}  \quad \in\left[0,\frac{\pi}{2}\right],
\ee
while the space coordinates are bounded as
\be
\left\{ x_{1},\dots,x_{D/2}\right\} \quad \in\left[0,L\right].
\ee
We also restrict $\rho$ to the interval $[0,1]$, so $R_0$ amounts to a temporal cutoff. It is chosen so as the results are independent of it.
For the integration, these variables are rescaled to their barred
counterparts $\bar{\varphi}_{i},\bar{x_i}$, so that they
are bounded in the $\left[0,1\right]$ interval. Thus the final Jacobi
determinant takes the form
\begin{equation}
J=R_0^{D/2-1}\rho^{D/2-1}L^{D/2}\left(\frac{\pi}{2}\right)^{D/2-1}\prod_{k=1}^{D/2-2}\left(\sin\varphi_{k}\right)^{\frac{D}{2}-1-k},\label{eq:Jacobi}
\end{equation}
(note that \eqref{eq:Jacobi} is expressed using the non-rescaled
angles, but the integration is understood with respect to the barred
ones, hence the extra numerical prefactor.)

For the modified Bessel functions, the C++ code uses the implementation
of the \texttt{alglib} package. A technical point is that $\mathrm{K}_{0}\left(x\right)$
behaves as $-\ln x+\mathrm{const}$ in the $x\rightarrow0$ limit.
Therefore, it was necessary to define a regularized function with a momentum cutoff $\delta_{\text{cut}}^{-1}$
\begin{align*}
\tilde{\mathrm{K}}_{0}\left(x\right) & =
\begin{cases}
\mathrm{K}_{0}\left(x\right), & x>\delta_{\text{cut}}\\
\mathrm{K}_{0}\left(\delta_{\text{cut}}\right), & x<=\delta_{\text{cut}}
\end{cases}
\end{align*}
and $\tilde{\mathrm{K}}_{0}$ is used instead of $\mathrm{K}_{0}$
in the numerical computations.

\bibliography{rgeffpot.bib}

\end{document}